\documentstyle[12pt,epsf]{article}
\begin{document}

\begin{titlepage}

\begin{flushright}
\vspace{-2.0cm}{\normalsize UTHEP-334\\
UTCCP-P-10\\
May 1996\\}
\end{flushright}

\vspace*{0.8cm}

\begin{centering}
{\LARGE
Finite Temperature Transitions\\
\vspace*{0.3cm}
in Lattice QCD with Wilson Quarks\\
\vspace*{0.3cm}
--- Chiral Transitions and the\\
\vspace*{0.3cm}
Influence of the Strange Quark ---
}

\vspace{1.6cm}

{\large 
Y.\ Iwasaki\rlap,$^{\rm a}$ K.\ Kanaya\rlap,$^{\rm a}$ 
S.\ Kaya\rlap,$^{\rm a}$ S.\ Sakai\rlap,$^{\rm b}$ 
and T.\ Yoshi\'e$^{\rm a}$
}

\vspace{0.3cm}

{\it
$^{\rm a}$
Institute of Physics and Center for Computational Physics, \\
University of Tsukuba, Ibaraki 305, Japan\\
$^{\rm b}$
Faculty of Education, Yamagata University,
Yamagata 990, Japan
}

\end{centering}

\vspace{1.6cm}\noindent

{\baselineskip=.8cm
The nature of finite temperature transitions in lattice QCD
with Wilson quarks
is studied near the chiral limit 
for the cases of 2, 3, and 6 flavors of degenerate quarks
($N_F=2$, 3, and 6)
and also for the case of massless up and down quarks and a 
light strange quark ($N_F=2+1$). 
Our simulations mainly performed on lattices 
with the temporal direction extension $N_t=4$ indicate
that the finite temperature transition in the chiral limit 
(chiral transition) is continuous for $N_F=2$, 
while it is of first order for $N_F=3$ and 6.
We find that the transition is of first order 
for the case of massless up and down quarks and the 
physical strange quark 
where we obtain a value of $m_\phi/m_\rho$ 
consistent with the physical value.
We also discuss the phase structure at zero temperature 
as well as that at finite temperatures.
}

\vfill \noindent

\end{titlepage}


\section{Introduction}
\label{sect:introduction}

One of major goals of numerical studies in lattice QCD 
is to determine the nature
of the transition from the high temperature
quark-gluon-plasma phase to the low temperature hadron 
phase, which is supposed to occur at the early stage of 
the Universe and possibly at heavy ion collisions.
It is, in particular, crucial to know
whether the transition is a first order phase transition
or a smooth transition 
(second order phase transition or crossover)
to understand the evolution of the Universe. 

Determination of the order of the transition 
for the case of degenerate $N_F$ flavors, 
is an important 
step toward the understanding of the nature of the QCD 
transition in the real world. 
We can compare the numerical results for 
various number of flavors with theoretical predictions
based on the study of the effective $\sigma$ model 
\cite{Pisarski,Wilczek}. 
In order to investigate what really happens in the nature,
we have to ultimately study 
the effect of the strange quark together with those of
almost massless up and down quarks,
because the critical temperature is of the same order 
of magnitude as the strange quark mass.

In this article we investigate 
finite temperature transitions in lattice QCD
using the Wilson formalism for quarks
for various numbers of flavors ($N_F=2$, 3, and 6)
near the chiral limit 
and also for the case of massless up and down quarks and a 
light strange quark ($N_F=2+1$).
Most simulations of finite temperature QCD 
were performed with staggered quarks.
However, because the Wilson formalism of fermions on the lattice
is the only known formalism which possesses a local action 
for any number of flavors, it is important to investigate 
the finite temperature transition
with Wilson quarks and compare the results with those 
for staggered quarks.

In Sec.~\ref{sect:action}, we define our action and 
coupling parameters. 
Because chiral symmetry is explicitly broken on the lattice 
in the Wilson formalism, 
we first define the chiral limit for Wilson quarks 
and give a brief survey of the phase structure 
in Sec.~\ref{sect:survey}. 
Our simulation parameters are summarized 
in Sec.~\ref{sect:simulations}.
Numerical results for the chiral limit are summarized 
in Sec.~\ref{sect:numericalKc}.
We then discuss, in Sec.~\ref{sect:problems}, problems 
and caveats which appear
in a study of the finite temperature transition 
with Wilson quarks when performed on lattices available with 
the present power of computers. 
Sec.~\ref{sect:chiral} deals with the transition in the 
chiral limit (chiral transition) in 
the degenerate cases of $N_F=2$, 3 and 6.
In Sec.~\ref{sect:strange}, we study the influence of the strange 
quark on the QCD transition both in the degenerate 
$N_F=3$ case and in a more realistic case of the massless 
up and down quarks with a massive strange quark, $N_F=2+1$.
We finally conclude in Sec.~\ref{sect:conclusions}.
Preliminary reports are given in 
\cite{ourOldLatC,preliminaryC,preliminaryS}.

\section{Action and coupling parameters}
\label{sect:action}

We use the standard one-plaquette gauge action
\begin{equation}
S_g = {2\over g^2 } \sum_{P} {\rm Re} \, {\rm Tr} (U_P)
\end{equation}
and the Wilson quark action \cite{WilsonF}
\begin{equation}
S_q = - \sum_{f=1}^{N_F} 
\sum_{n,m} \bar\psi_f(n) \, D(K_f,n,m) \, \psi_f(m),
\label{eq:WilsonF}
\end{equation}
\begin{equation}
D(K,n,m) = \delta_{n,m} - K \sum_{\mu} {
\{ (1-\gamma_{\mu})\,U_{n,\mu}\delta_{n+\mu,m} +
(1+\gamma_{\mu})\,U^{\dagger}_{m,\mu}\delta_{m+\mu,n}  \} },
\label{eq:WilsonD}
\end{equation}
where $g$ is the bare coupling constant and
$K$ is the hopping parameter.
In the case of degenerate $N_F$ flavors, 
lattice QCD contains two parameters:
the gauge coupling constant $\beta =6/g^2$ 
and the hopping parameter $K$. In the non-degenerate case,
the number of the hopping parameters is $N_F$.

We denote the linear extension of a lattice 
in the temporal direction by $N_t$ 
and the lattice spacing by $a$.

\section{Brief survey of phase structure}
\label{sect:survey}

In the Wilson formalism of fermions on the lattice,
chiral symmetry is explicitly broken by the Wilson term
even for vanishing bare quark mass \cite{WilsonF}.
The lack of chiral symmetry of chiral symmetry causes much 
conceptual and technical difficulties in numerical simulations 
and physics interpretation
of data.
Therefore before going into discussion of details of data 
and analyses, we give a brief survey of the phase
structure at zero temperature 
as well as that at finite temperatures
 \cite{IwasakiLat94,KanayaLat95},
including the results presented in this article.

\subsection{Quark mass and PCAC relation}
\label{subsect:quarkmass}

We first define the quark mass 
through an axial-vector Ward identity
\cite{Bo,ItohNP}.
\begin{equation}
2 m_q \, \langle\,0\,|\,P\,|\,\pi(\vec{p}=0)\,\rangle
= - m_\pi \, \langle\,0\,|\,A_4\,|\,\pi(\vec{p}=0)\,\rangle
\label{eq:mq}
\end{equation}
where $P$ is the pseudoscalar density and $A_4$ the fourth
component of the local axial vector current.
[Note that we have absorbed a multiplicative normalization
factor into the definition of the quark mass $m_q$,
because this convention is sufficient for our later study.
We also note that there is an alternative definition 
of the quark mass
replacing $m_\pi $ with e.g. $(\exp(-m_\pi a) - 1)/a$, 
which gives the quark
mass identical with the above within order of $a$.]

\begin{figure}
\centerline{ \epsfxsize=14cm \epsfbox{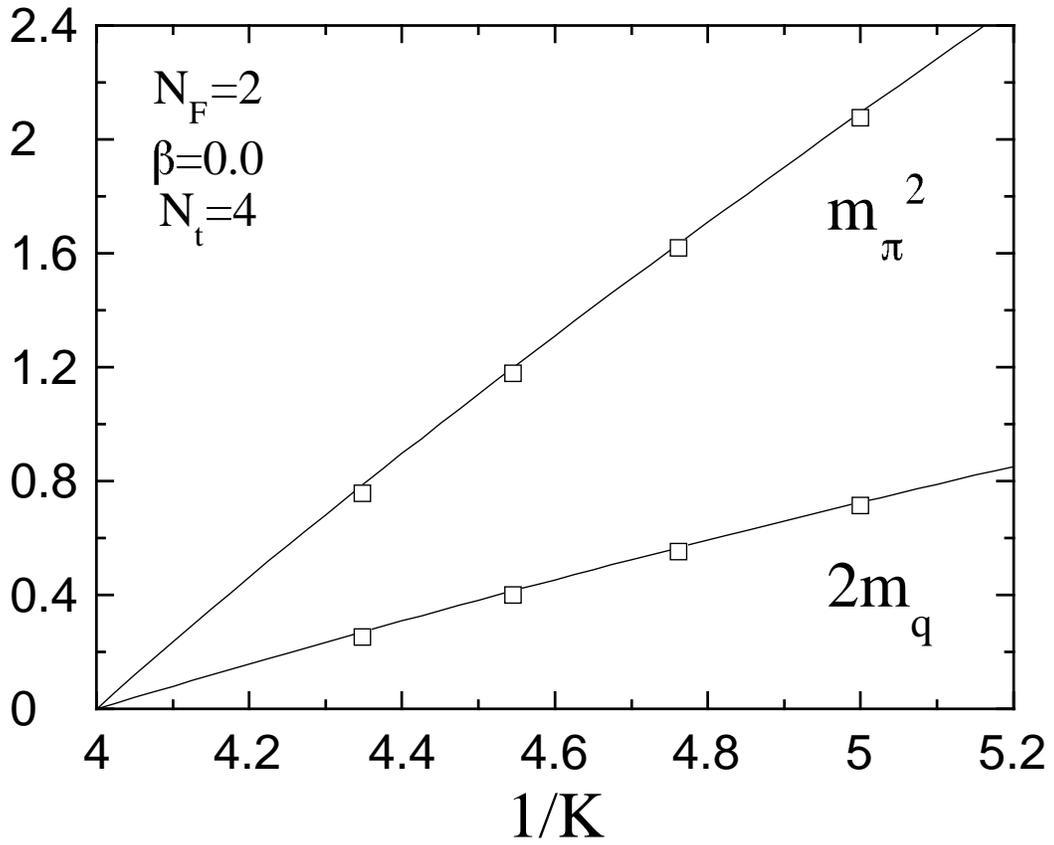} }
\caption{\baselineskip=.8cm
Squared pion screening mass $m_\pi^2 a^2$ and 
twice the quark mass $2m_q a$ for $N_F=2$ 
at $\beta=0$ on an $8^2\times10\times4$ lattice. 
Errors are smaller than the size of symbols. 
Solid curves are the results of a strong coupling calculation, 
Eq.(\protect\ref{eq:strongc}).
}
\label{fig:F2B0.0}
\end{figure}

With this definition of quark mass, 
the PCAC relation, 
\begin{equation}
m_\pi^2 \propto m_q, 
\label{eq:pcac}
\end{equation}
which is expected to be satisfied near the continuum limit,
was numerically first verified within numerical uncertainties 
for the quenched QCD at zero temperature in \cite{ItohNP,mm} 
and subsequently for various cases including QCD with $N_F=2$
in \cite{ourOldLatC,ourStrong,ChNf2,ChQ,DanielGKPS92,MILC4}.
It should be noted that 
the PCAC relation is satisfied not only in the continuum limit, 
$\beta=\infty$, 
but also even in the strong coupling limit, $\beta=0$:
The result of the strong coupling expansion without quark loops 
\cite{ourStrong},
\begin{eqnarray}
\cosh(m_\pi a)
& = & 1 +
\frac{(1-16K^2)(1-4K^2)}{4K^2 \, (2-12K^2)} \nonumber \\
2 m_q a
& = &
m_\pi a\,\frac{4K^2\sinh(m_\pi a)}{1-4K^2\cosh(m_\pi a)},
\label{eq:strongc}
\end{eqnarray}
gives the relation $m_\pi^2 \propto m_q$ at small $m_q$.
Our numerical data for $N_F=2$ at $\beta=0$ agrees well with
these formulae within errors as shown in Fig.~\ref{fig:F2B0.0}.%
\footnote{In Ref.\cite{ourStrong}, agreement between
Eq.(\ref{eq:strongc})
and numerical data in the confining phase is shown
also for the case $N_F=18$.
The rho meson mass, the nucleon mass,
and the delta mass also agree with corresponding
strong coupling mass formulae.}

We note that, if 
\begin{equation}
\langle\,0\,|\,A_4\,|\,\pi(\vec{p}=0)\,\rangle \propto m_\pi
\label{eq:a4pi}
\end{equation}
is satisfied for small $m_q$ as is the case both for $\beta=0$ and 
$\beta=\infty$,
then the definition (\ref{eq:mq}) implies that the PCAC relation 
(\ref{eq:pcac}) is exact. 
It should be also noted that Eq.(\ref{eq:a4pi}) holds 
when Euclidean invariance is recovered \cite{ItohNP}.

Eq.(\ref{eq:mq}) implies that when $m_q=0$, 
either $m_\pi=0$ or 
$\langle\,0\,|\,A_4\,|\,\pi(\vec{p}=0)\,\rangle=0$.
This further implies,
when we define the pion decay constant $f_\pi$ by 
\begin{equation}
\langle\,0\,|\,A_4\,|\,\pi(\vec{p}=0)\,\rangle = m_\pi f_\pi,
\label{eq:fpi}
\end{equation}
that when $m_q=0$, 
either $m_\pi=0$ or $f_\pi=0$ is satisfied.
Note that $f_\pi=0$ is the relation which should be satisfied 
when chiral symmetry is restored,
and that $m_\pi=0$ is the relation when
chiral symmetry is spontaneously broken, both in the chiral limit.
It might be emphasized that although the action does not possess
chiral symmetry, either relation of $m_\pi=0$ or $f_\pi=0$
holds
in the massless quark limit when the quark mass is defined
by Eq.(\ref{eq:mq}). In particular, in the confining phase, 
$m_\pi=0$ when $m_q=0$ and {\it vice versa}.

\subsection{Definition of chiral limit and phase structure 
at zero temperature}
\label{subsect:chirallim}

We identify the chiral limit as the limit where the quark mass 
vanishes at zero temperature.
This defines a chiral limit line $K_c$ in the $(\beta,K)$ plane,
which is a curve from $K \simeq 1/4$ at $\beta=0$ to 
$K=1/8$ at $\beta = \infty$. 
See Fig.~\ref{fig:PDfiniteT}.
In the following we also discuss alternative identifications of
the chiral limit.
When clear specification is required, 
we denote this $K_c$ as $K_c(m_q)$. 

Let us denote a line  
where the pion mass vanishes at zero temperature by $K_c(m_\pi^2)$. 
This line is the critical line of the theory 
because the partition function has singularities there. 
As discussed in the previous subsection, we expect that 
$K_c(m_q)$ and $K_c(m_\pi^2)$ are identical for small $N_F$. 
It should be, however, noted that the $K_c(m_q)$ line 
is conceptually different from the $K_c(m_\pi^2)$ line: 
If quarks are not confined and chiral symmetry is not spontaneously
broken, there is no $K_c(m_\pi^2)$ line. 
In fact, for the case of $N_F \ge 7$,  
the $K_c(m_q)$ line belongs to the deconfining phase 
and $m_\pi$ remains nonzero there --- 
i.e.\ there is no $K_c(m_\pi^2)$
line around the $K_c(m_q)$ line, at least for small 
$\beta$ \cite{ourStrong}.

As a statistical system on the lattice, 
QCD with Wilson quarks is well-defined 
also in the region above the $K_c$ line. 
Some time ago, S.\ Aoki \cite{aoki} proposed and numerically 
verified that
the critical line $K_c(m_\pi^2)$ (for small $N_F$) can be 
interpreted as a second order phase transition line 
between the parity conserving phase and a parity violating phase.
This interpretation is useful in understanding the 
existence of singularities of the partition function.
Once its existence is established, 
various properties of hadrons
can be investigated in the parity conserving phase.
In particular, 
even with the Wilson term, 
various amplitudes near the chiral limit do satisfy Ward-Takahashi 
identities derived from chiral symmetry 
to the corrections of $O(a)$ \cite{Bo}.\footnote{ 
In the particular form of Eq.(\ref{eq:mq}), we have absorbed 
these $O(a)$ corrections in the definition of $m_q$, 
or, equivalently, in the value of $K_c$.
}
Therefore, although the action does not have chiral symmetry, 
the concept of the spontaneous breakdown of chiral symmetry is
phenomenologically very useful.
Because our main interest is to study the physical properties of 
hadrons in the continuum limit, 
it is important to study these axial Ward-Takahashi identities 
and estimate the magnitude of the $O(a)$ corrections
from the Wilson term in the physical quantities. 

We have defined the $K_c$ line by the vanishing point of $m_q$ 
at zero temperature, 
because this line corresponds to massless QCD.
In this connection, however, it should be noted that 
there necessarily are ambiguities of $O(a)$ off the continuum limit
for lines in the $(\beta,K)$ plane which give the same theory
in the continuum limit.
This is true also for massless QCD: 
Instead of the condition 
$m_\pi=0$, we may fix other quantities such as $m_\rho/m_N$, 
which will lead to a line different from the $K_c$ line. 
Of course, the continuum limit is not affected by these 
$O(a)$ ambiguities.
We, however, would like to stress that 
the definition we have taken for
the $K_c$ is conceptually natural and useful for the 
reasons given in Sec.~\ref{subsect:quarkmass}.

\subsection{Phase structure at finite temperatures}
\label{subsect:finitetemp}

The temperature on a lattice with the linear extension in the
temporal direction $N_t$ is given by $T=1/{N_t a}$.
On a lattice with a fixed $N_t$, 
finite temperature transition or crossover from 
the low temperature regime
to the high temperature regime occurs at some hopping parameter 
when $\beta$ is fixed. This defines a curve $K_t$ 
in the $(\beta, K)$ plane. 
In this paper, for simplicity, we use the term ``transition'' 
for both genuine phase transitions and sharp crossovers, 
unless explicitly specified. 
At finite temperatures we denote the screening pion mass by $m_\pi$
and sometimes we call it simply the pion mass, and similarly
for other hadron screening masses.
Quark mass at finite temperatures is defined through 
Eq.(\ref{eq:mq}) with $m_\pi$ the screening pion mass,
and similarly for $f_\pi$ through Eq.(\ref{eq:fpi}).
Note that, with these definitions of $m_\pi$ and $f_\pi$, 
the discussions given in Sec.~\ref{subsect:quarkmass} hold
also at finite temperatures.

One of fundamental problems is whether the finite temperature 
transition line $K_t$ does cross the chiral limit line $K_c$, where
we define the $K_c$ line by the vanishing point of $m_q$ 
{\em at zero temperature} (cf.\ Sec.~\ref{subsect:chirallim}).
If the $K_t$ line does not cross the $K_c$ line, it means 
that there is no chiral limit in the low temperature confining phase.
Therefore it is natural to expect that it does cross.  
However, as first noted by Fukugita {\it et al.} \cite{FOU}, 
it is not easy to confirm this: The $K_t$
line creeps deep into the strong coupling region.
In this paper we show that
the $K_t$ line indeed crosses the chiral line $K_c$
at $\beta \sim 3.9$ --- 4.0 at $N_t=4$ 
and $\beta \sim 4.0$ --- 4.2 at
$N_t=6$ for the case of $N_F=2$. 
(For previous reports see Refs.\cite{ourOldLatC,preliminaryC}.) 

Because the $K_c$ line describes the massless QCD, 
we identify the crossing 
point of the $K_c$ and $K_t$ lines as
the point of the finite temperature
transition of the massless QCD, i.e.\ the chiral transition point. 
(We will discuss later $O(a)$ ambiguities 
in the definition of the chiral limit at finite temperatures 
which come from the lack of chiral symmetry.)

Numerical studies show that, in the confining phase, 
the pion mass vanishes, for a fixed $\beta$, 
at the hopping parameter which approximately
equals the chiral limit $K_c$.
On the other hand, in the deconfining phase, 
the pion mass is of order of 
twice the lowest Matsubara frequency $2\pi/N_t$ in the chiral limit.
Therefore, in the deconfining phase, the system is not singular
even on the $K_c$ line.

\begin{figure}
\centerline{ \epsfxsize=14cm \epsfbox{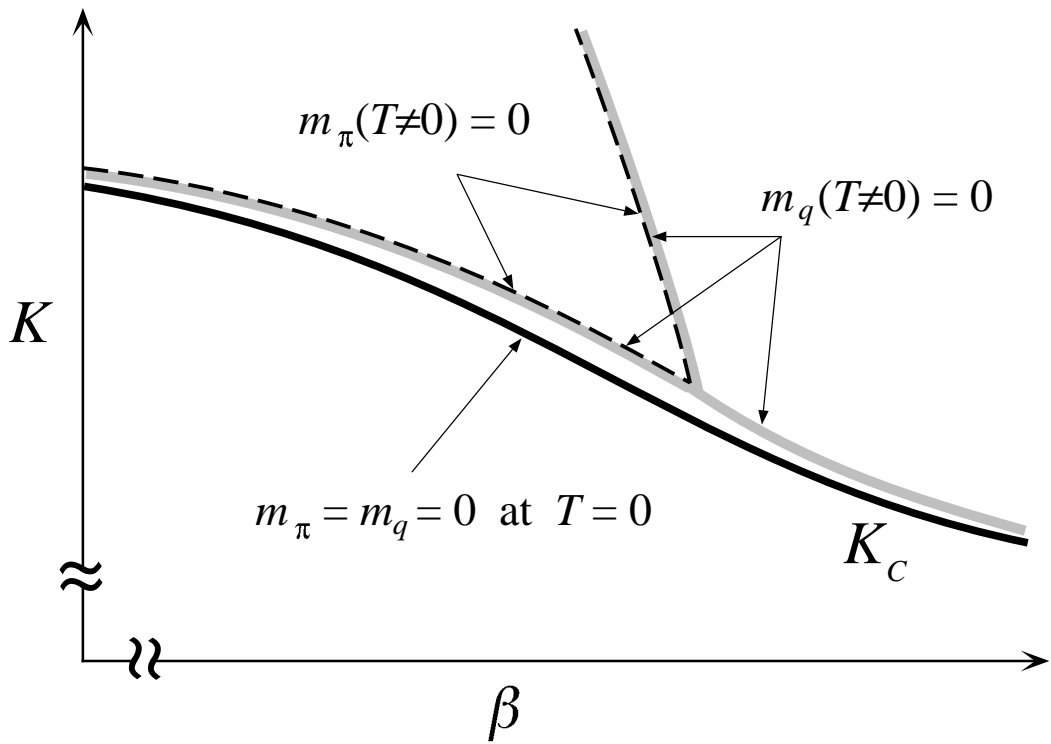} }
\caption{\baselineskip=.8cm
Schematic graph for
the chiral limit line $K_c$ defined by $m_q=0$ or $m_\pi=0$ 
at $T=0$ in the coupling parameter space $(\beta,K)$. 
Also plotted are the curves where $m_q=0$ and $m_\pi=0$ 
at finite $N_t$. 
See text for discussions.
}
\label{fig:PDfiniteT}
\end{figure}

Recently, Aoki {\it et al.} \cite{aokiukawaumemura} investigated
a critical line where the screening pion mass vanishes 
at finite temperatures, which we denote by $K_c(m_\pi^2; T\neq 0)$.
Based on analytic studies of the 2d Gross-Neveu model 
and numerical results
in lattice QCD with $N_F=2$, they showed that 
the $K_c(m_\pi^2; T\neq 0)$ line 
starting from $K \simeq 1/4$ at $\beta=0$
sharply turns back upwards (to larger $K$ region) at finite $\beta$. 
The lower part of the $K_c(m_\pi^2; T\neq 0)$ line is 
almost identical with the $K_c(m_\pi^2)$ line
up to the sharp turning point, 
while the analytic results of the 2d Gross-Neveu model suggest that 
they slightly differ from each other, 
probably with $O(a)$. 
See Fig.~\ref{fig:PDfiniteT}.

The non-existence of the $K_c(m_\pi^2; T \neq 0)$ line 
in the large $\beta$ region is consistent with 
the previous results that $m_\pi$ does not vanish 
in the deconfining phase along the chiral line $K_c$.
The slight shift of the $K_c(m_\pi^2; T\neq 0)$ line from 
the $K_c(m_\pi^2)$ line in the confining phase was observed 
also in our previous study \cite{ourOldLatC,preliminaryC} 
(see also Sec.~\ref{sect:numericalKc}).
This slight shift of the $K_c(m_\pi^2; T\neq 0)$ line means that 
$m_\pi$ is not rigorously zero on the $K_c(m_\pi^2)$ line 
in the confining phase at finite temperatures.
This small pion mass on the $K_c$ line in the confining phase
is caused by the chiral symmetry violation due to the Wilson term 
and should be of $O(a)$.

Similarly to the $K_c(m_\pi^2; T\neq 0)$ line, we define 
the line $K_c(m_q; T\neq 0)$ where the quark mass vanishes 
at finite temperatures. 
When we follow the line $K_c(m_q; T\neq 0)$ from $\beta=0$,
it is first identical with the $K_c(m_\pi^2; T\neq 0)$ line.
The line $K_c(m_q; T\neq 0)$ passes through the turning point
of the $K_c(m_\pi^2; T\neq 0)$ line and
runs into the larger $\beta$ region, where 
$f_\pi$ starts to vanish instead of $m_\pi$ 
on the $K_c(m_q; T\neq 0)$ line. 
See Fig.~\ref{fig:PDfiniteT}. 
This suggests that the turning point
which is the boundary between $f_\pi=0$ and $m_\pi=0$
is the finite temperature transition point.
This further implies
that the transition line $K_t$ 
touches the turning point
of the $K_c(m_\pi^2; T\neq 0)$ line and moves upwards in the 
$(\beta,K)$ plane.
This observation is not in accord with the argument
by Aoki {\it et al.} \cite{aokiukawaumemura} that
there is a small gap between 
the $K_c(m_\pi^2; T\neq 0)$ and $K_t$ lines.

We have identified the crossing 
point of the $K_c$ and $K_t$ lines as
the chiral transition point.  
In connection with the $O(a)$ ambiguities
of the line for massless QCD in the coupling parameter space 
mentioned in Sec.~\ref{subsect:chirallim}, 
there are $O(a)$ ambiguities also in the definition
of the chiral transition. 
Therefore, one may alternatively identify the sharp turning 
point of the $K_c(m_\pi^2; T\neq 0)$ line as the chiral 
transition point.
The property of the chiral transition in the continuum limit 
is, of course, not affected by these $O(a)$ ambiguities.

\subsection{Characteristics for Wilson quarks}
\label{subsect:wilsontermsummary}

Let us summarize several characteristic properties of 
the phase diagram of QCD which are originated from the
explicit chiral symmetry violation
of the Wilson term.
They are in sharp contrast with those of staggered quarks where 
at least a part of chiral symmetry is preserved. 

(i) In the coupling parameter space, 
the location of the point where $m_\pi=0$ 
in the confining phase 
is not protected by chiral symmetry off the continuum limit. 
Therefore, the chiral limit $K_c$, 
defined by $m_q=0$ or $m_\pi=0$ at zero temperature, 
is different from the bare massless limit $K=1/8$ except
at $\beta = \infty$. 

(ii) As a statistical system on the lattice, 
QCD with Wilson quarks is well-defined 
also in the region above the $K_c$ line. 
At zero temperature, the $K_c$ line is a second order transition 
line between the conventional parity conserving phase at $K<K_c$ 
and a parity violating phase at $K>K_c$ \cite{aoki}.

(iii) At finite temperatures, the critical line 
$K_c(m_\pi^2; T\neq 0)$ 
where the screening pion mass vanishes
is not a line from $K \simeq 1/4$ at $\beta=0$ to 
an end at some finite $\beta$, 
but it sharply turns back toward larger $K$ region 
at the finite $\beta$ \cite{aokiukawaumemura}. 

(iv) Although the major part of the effects from the Wilson term 
can be absorbed by the shift of $K_c$ from $K=1/8$, 
there still exist 
additional small $O(a)$ effects 
which are related to the chiral symmetry violation.
In particular, the location of the point where $m_\pi=0$ in the 
confining phase slightly depends on $N_t$ \cite{aokiukawaumemura}. 
The continuum limit is not affected by these $O(a)$ effects.

\section{Simulation Parameters}
\label{sect:simulations}

In this article we mainly perform simulations on lattices with
the temporal direction extension $N_t=4$.
The spatial sizes are $8^2 \times 10$ and $12^3$. 
To study the $N_t$ dependence for the $N_F=2$ case, 
we also make simulations on $N_t=6$ and 8 lattices.
Simulations on an $N_t=8$ lattice 
are performed also for the case of $N_F=2+1$. 
When the hadron spectrum is calculated, 
the lattice is duplicated 
in a direction of lattice size 10 or 12.
We use an anti-periodic boundary condition for quarks in 
the $t$ direction 
and periodic boundary conditions otherwise.

We generate gauge configurations for $N_F=2$ 
by the Hybrid Monte Calro (HMC) algorithm \cite{HMC} 
with a molecular dynamics time step $\Delta\tau$ 
chosen in such a way that the acceptance rate is
about 80 --- 90\%.
For $N_F \ge 3$ and $N_F=2+1$ 
we use the hybrid R algorithm \cite{Ralgo} with 
$\Delta \tau =0.01$, unless otherwise stated.
We fix the time length of each molecular dynamics evolution 
to $\tau=1$.
The R algorithm introduces errors of $O(\Delta \tau^2)$,
while the HMC algorithm is exact.
As reported recently also for staggered quarks 
\cite{BlumKTG95}, 
we note that step size errors with the R algorithm 
are large in the confining phase near the chiral limit. 
In the immediate vicinity of the chiral transition, 
we observe step size errors also in the deconfining phase 
where a large $\Delta\tau$ can even push the phase into the 
confining phase, as reported previously with staggered quarks 
\cite{ColumbiaNf8}. 
In these cases, we apply a sufficiently small $\Delta\tau$ 
so that the results for physical quantities become stable
for a change of $\Delta\tau$.

The inversion of the quark matrix 
is done by the minimal conjugate residual (CR) method 
with the ILU preconditioning \cite{ILU} or 
the conjugate gradient (CG) method without preconditioning. 
We find that the CR method is efficient 
in the confining phase when it is not very close to the chiral limit 
and also in the deconfining phase at large $\beta$ 
and small $K$. 
In other cases we use the CG method.
The convergence condition for the norm of the residual 
$r$ is $\sqrt{\|r\|^2/(12V)} \leq 4.5\times10^{-7}$ 
($1.0\times10^{-8}$) for configuration generations 
(hadron measurements), where $V$ is the lattice volume. 
We also check that the relative changes of the quark propagator 
at several test points on the $x$ and $t$ axes 
are smaller than $10^{-3}$ 
for the last iteration of the matrix inversion steps: 
$|(G_n-G_{n-1})/G_n| \le 10^{-3}$ where $n$ denotes 
the last iteration. 
In the HMC calculations, we check that the difference 
of the action after molecular dynamic evolutions 
is sufficiently small with this convergence condition. 

The statistics is in general totally $\tau=$ several hundreds.
The initial configuration is taken from a thermalized one
at similar simulation parameters when such a configuration is 
available. 
In most cases, 
the plaquette and the Polyakov loop are measured every 
simulation time unit and
hadron spectrum is calculated every $\delta\tau=10$ (or less 
depending on the total statistics). 
When the value of $\beta$ is small the fluctuation of
physical quantities are small \cite{ourStrong}, 
and therefore we think the lattice sizes and the statistics
are sufficient for our purpose to determine 
the global phase structure of QCD at finite temperature.
Errors are estimated by the single-elimination jackknife method. 

Simulation parameters are summarized 
in Tables~\ref{tab:paramF2Nt4} --- \ref{tab:paramF21Nt8}.

\section{Numerical results for $K_c$}
\label{sect:numericalKc}

As discussed in Sec.~\ref{subsect:chirallim}, 
the chiral limit $K_c$ is defined by the vanishing
point of $m_q$ at zero temperature.
One straightforward way to determine
numerically the chiral limit at a fixed value of $\beta$
is to calculate the the quark mass through Eq.(\ref{eq:mq}) 
at several hopping parameters and extrapolate them 
to its vanishing point
in terms of a linear function of $1/K$. 
We denote the $K_c$ thus determined by $K_c(m_q)$.
Because we expect the PCAC relation (\ref{eq:pcac}) to hold 
also at finite $\beta$, we may alternatively calculate $K_c$ by
the vanishing point of $m_\pi$ using a linear extrapolation of 
$m_\pi^2$ in $1/K$.
We denote this $K_c$ by $K_c(m_\pi^2)$.

On finite temperature lattices, it was previously shown that 
the value of the quark mass at given $(\beta,K)$
does not depend on whether the system is in 
the deconfining or confining phase
at $\beta=5.85$ in the quenched QCD \cite{ChQ} and
at $\beta=5.5$ for the $N_F=2$ case \cite{ChNf2}.
This enables us to determine the chiral limit, 
for these values of $\beta$, alternatively by the vanishing 
point of $m_q$ at finite temperatures. 
Strictly speaking there are systematic errors which come from 
finite $N_t$, as mentioned earlier. 
On the other hand, in the deconfining phase,
one is able to perform simulations around the $K_c$ line 
as discussed later, i.e.\ 
we can determine $K_c$ without an extrapolation 
which usually leads to a considerable amount of systematic errors.
Therefore, the determination of $K_c$ from $m_q$ 
in the deconfining phase is useful in particular at large $\beta$.

\begin{figure}
\centerline{ \epsfxsize=8.5cm \epsfbox{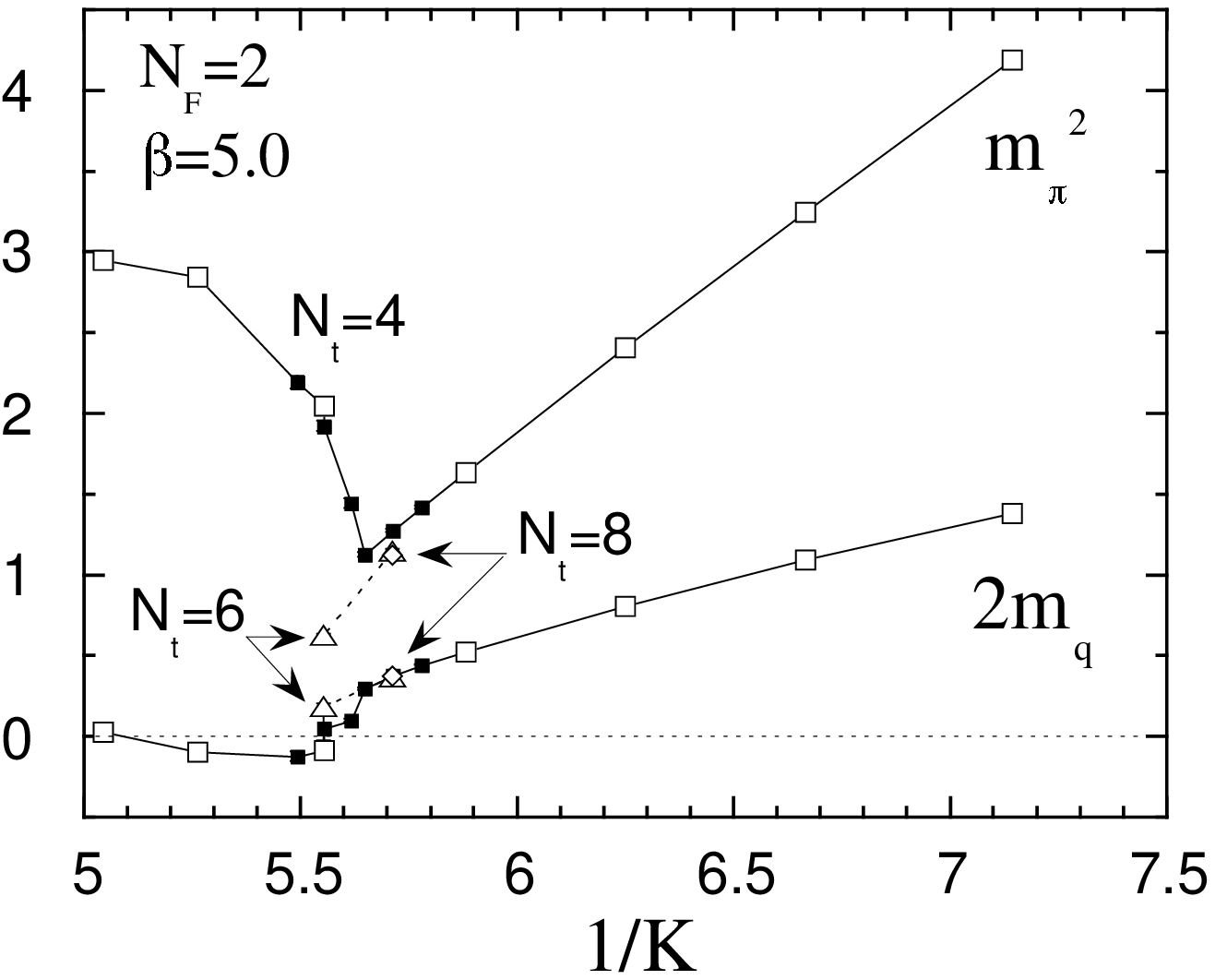} }
\centerline{ \epsfxsize=8.5cm \epsfbox{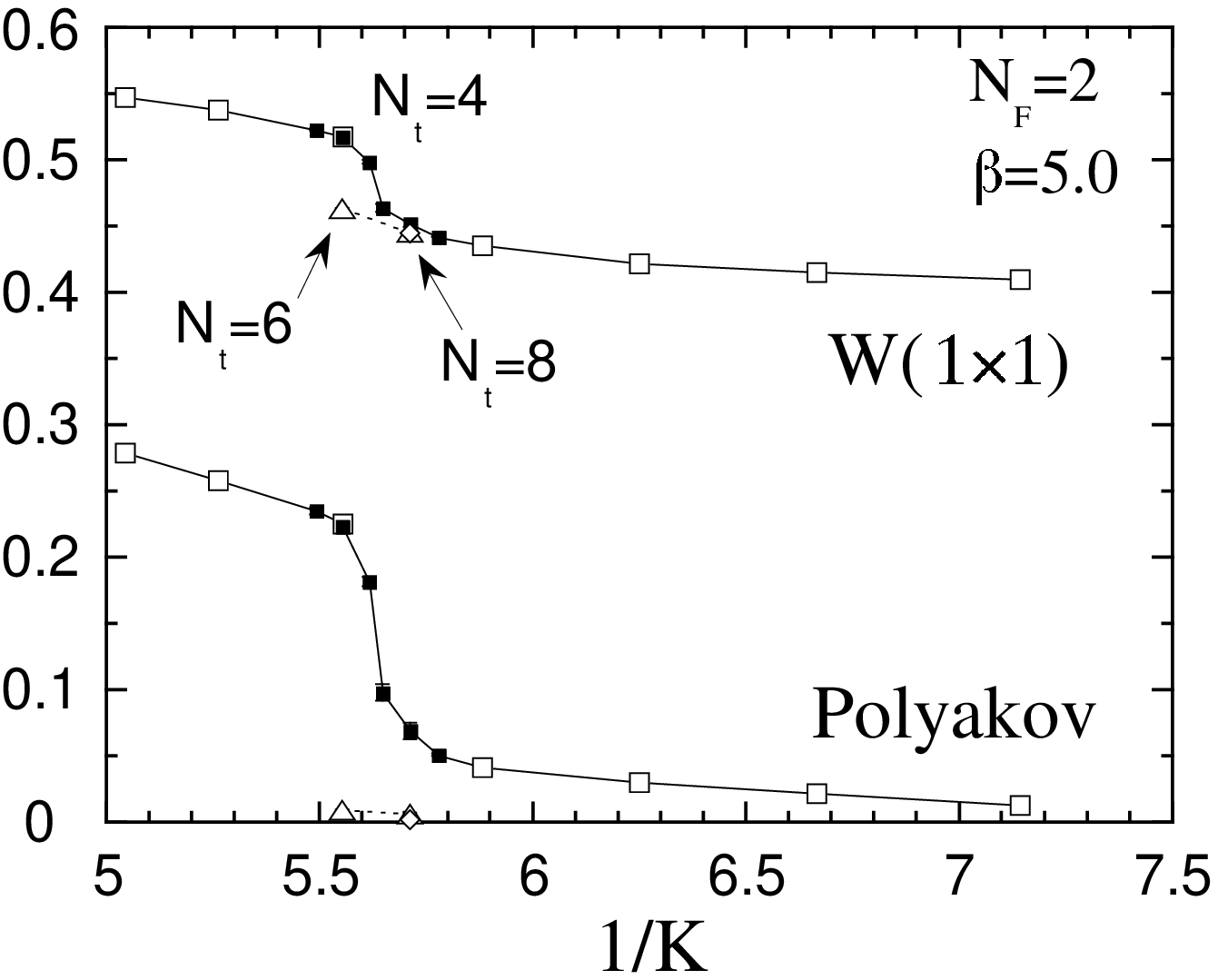} }
\caption{\baselineskip=.8cm
Physical quantities for $N_F=2$ 
at $\beta=5.0$ on an $8^2\times10\times 4$ lattice
(open squares): 
(a) $m_\pi^2 a^2$ and $2m_q a$, 
(b) the plaquette and the Polyakov loop. 
Plotted together are the data by the MILC collaboration 
on an $8^2\times20\times N_t$ lattice 
with $N_t=4$ (filled squares), 6 (triangles), and 8 (diamonds) 
\protect\cite{MILC4}. 
The finite temperature transition $K_t$ obtained by the 
MILC data locates at $K=0.177$ --- 0.178 
($1/K=5.62$ --- 5.65) for $N_t=4$. 
}
\label{fig:F2B5.0}
\end{figure}

\begin{figure}
\centerline{ \epsfxsize=10cm \epsfbox{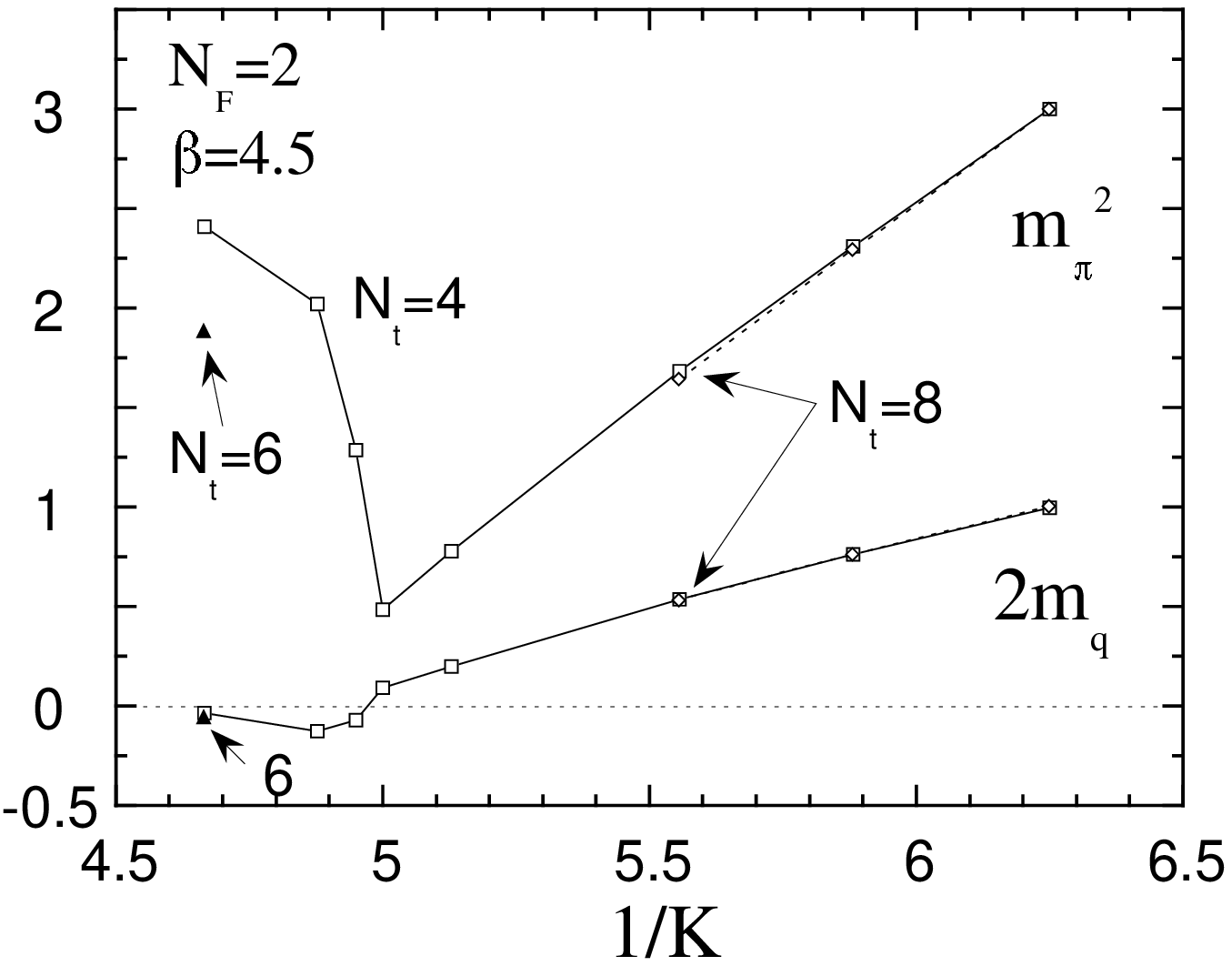} }
\centerline{ \epsfxsize=10cm \epsfbox{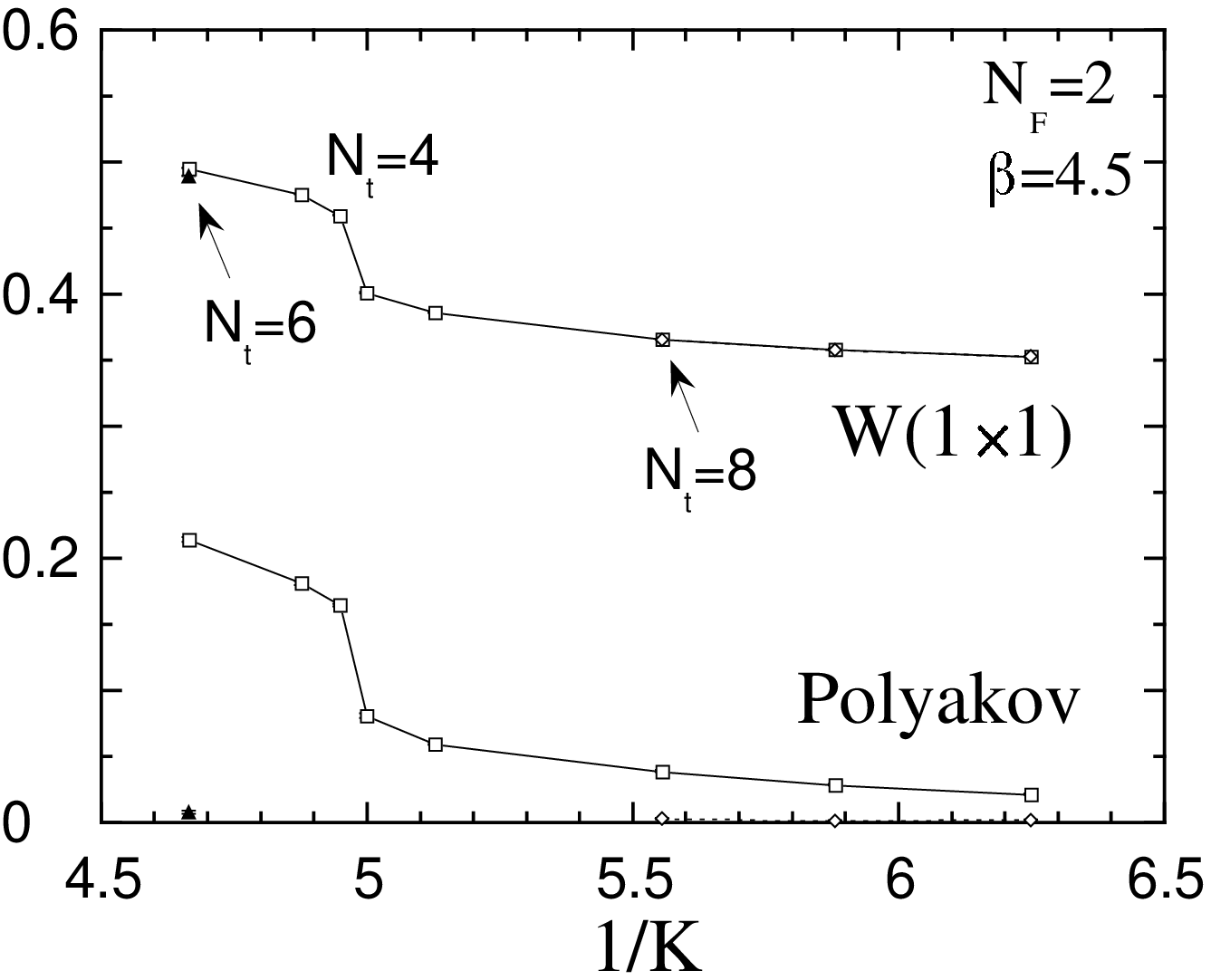} }
\caption{\baselineskip=.8cm
The same as Fig.~\protect\ref{fig:F2B5.0} 
at $\beta=4.5$ on $8^2\times10\times N_t$ lattices, 
where $N_t=4$ (squares), 6 (triangles), and 8 (diamonds). 
The finite temperature transition $K_t$ locates 
at $K=0.200$ --- 0.202 ($1/K=4.95$ --- 5.0) for $N_t=4$. 
}
\label{fig:F2B4.5}
\end{figure}

At small $\beta$ region ($\beta$
\raise0.3ex\hbox{$<$\kern-0.75em\raise-1.1ex\hbox{$\sim$}} 5.3)
where we mainly perform simulations in this work, 
$m_q$ in the deconfining phase
does not agree with that in the confining phase.
Therefore, the proportionality between
$m_q$ in the deconfining phase
and $m_\pi^2$ in the confining phase is lost,
contrary to the case $\beta$
\raise0.3ex\hbox{$>$\kern-0.75em\raise-1.1ex\hbox{$\sim$}} 5.5
discussed above.
This behavior is seen in Figs.~\ref{fig:F2B5.0} 
and \ref{fig:F2B4.5},
where physical quantities for $N_F=2$ at $\beta=5.0$ and 4.5,
respectively, are shown.
As we discuss in Sec.~\ref{sect:problems},
we interpret this unexpected phenomenon at $\beta$
\raise0.3ex\hbox{$<$\kern-0.75em\raise-1.1ex\hbox{$\sim$}} 5.3
in the deconfining phase as a lattice artifact.

In the confining phase, on the other hand,
the proportionality between $m_q$
and $m_\pi^2$ 
is well satisfied for all values of $\beta$ 
\cite{ourOldLatC,ourStrong,ChNf2,ChQ,DanielGKPS92,MILC4}. 
We also find that $m_q$ and $m_\pi$ are almost 
independent of $N_t$ in the confining phase.
See Fig.~\ref{fig:F2B4.5} for $N_F=2$ at $\beta=4.5$.
Therefore we can calculate $K_c$ approximately also by
the vanishing point of $m_q$, $K_c(m_q)$, or that of $m_\pi^2$,
$K_c(m_\pi^2)$, in the confining phase at $T>0$.

\begin{figure}
\centerline{ \epsfxsize=14cm \epsfbox{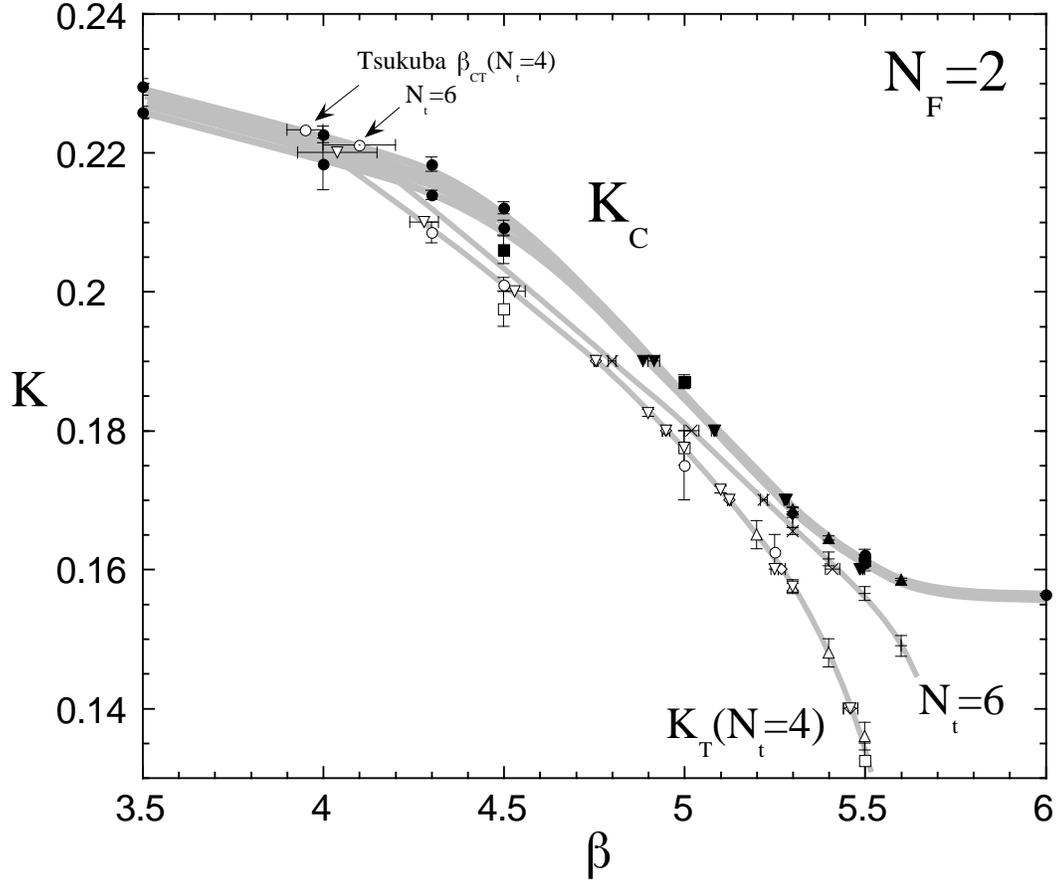} }
\caption{\baselineskip=.8cm
Phase diagram for $N_F=2$.
Filled symbols are for $K_c$ determined by $m_\pi=0$ 
and $m_q=0$.
Open symbols are for $K_t(N_t=4)$ 
and other symbols such as
crosses except filled ones are for $K_t(N_t=6)$. 
Circles are our data. 
Lines are to guide the eyes. 
}
\label{fig:F2KcKt}
\end{figure}

The numerical results for $K_c$ for $N_F=2$
obtained by various groups
\cite{ourOldLatC,ChNf2,kc_data,UkawaLat88,kc_data2,kc_data3,MILC6}
are plotted in Fig.~\ref{fig:F2KcKt}
together with finite temperature transition lines discussed
in the following sections.
The values of $K_c$ show a slight dependence
(at most of the order of 0.01)
on the choice of $K_c(m_q)$ or $K_c(m_\pi^2)$,
which can be probably attributed to
the systematic errors in the extrapolation of 
$m_\pi^2$ and $m_q$ in $1/K$,\footnote{
The range of the quark mass value we use in this article
for the extrapolation to determine the $K_c$ is 
mainly about 0.2 --- 0.5
in lattice units in the confining phase. 
As seen from Fig.~\protect\ref{fig:F2KcKt}, $m_\pi^2$ and $m_q$ 
sometimes show slightly convex curves in $1/K$. 
In such cases, a choice of the fit range at smaller $m_q$ 
will lead to slightly smaller values for $K_c$. 
}
because, as discussed above,
we expect that $K_c(m_q)$ and $K_c(m_\pi^2)$ are identical.
The values of $K_c(m_\pi^2)$ for $N_F=2$ for various $\beta$'s
are listed in Table~\ref{tab:1}.
We estimate the systematic errors due to the extrapolation are 
of the same order as the differences 
between $K_c(m_q)$ and $K_c(m_\pi^2)$.

The $N_t$ dependence of $K_c$
at $\beta=4.5$ are listed in Table~\ref{tab:2}.
The $N_F$ dependence is also given.
We find that the differences due to $N_F$ and $N_t$ are
of the same order of
magnitude as the difference between $K_c(m_q)$ 
and $K_c(m_\pi^2)$.

To summarize this section, we note that
although the chiral limit is defined by the vanishing point 
of $m_q$ at zero temperature, 
there are several practically useful
ways to determine $K_c$: 
$K_c(m_q)$ and $K_c(m_\pi^2)$ at $T=0$ and
in the confining phase, 
and $K_c(m_q)$ in the deconfining phase.
They all gives the same results within present numerical 
errors.

\section{Finite Temperature Transition and Problems with 
Wilson Quarks}
\label{sect:problems}

The location of the finite temperature phase transition 
$K_t$ is identified by a sudden change of
physical observables such as the plaquette, the Polyakov 
line and screening hadron masses.
(A more precise determination of the location will be given by
the maximum point of the susceptibility of a physical quantity 
such as the Polyakov loop. However, our statistics is 
not high enough for it.)
See Figs.~\ref{fig:F2B5.0} and \ref{fig:F2B4.5} 
for the case of $N_F=2$ at $\beta=5.0$ and 4.5.
Our numerical results of $K_t$ are summarized 
in Table~\ref{tab:Kt}.
Results of $K_t$ for $N_F=2$ at $N_t=4$ and 6 
obtained by us and other groups 
\cite{MILC4,UkawaLat88,MILC6,kt_data,kt_data2}
are compiled in Fig.~\ref{fig:F2KcKt}. 
(Results for $N_F=3$ will be discussed 
in Sec.~\ref{sect:strange}.)

We expect, at least near the continuum limit, 
that as the quark mass increases from the chiral limit,
the transition becomes weaker with the quark mass 
and it becomes strong again when the quark mass is heavy enough 
to recover the first order transition of the SU(3) gauge theory.
The MILC collaboration performed a systematic study 
of the transition at various $K$ and $\beta$ 
and found that, 
contrary to the expectation, 
when we decrease $K$ from the chiral limit $K_c$ 
on an $N_t=4$ lattice, 
the $K_t$ transition becomes once very strong 
at $K \simeq 0.18$ and becomes weaker again at smaller $K$
\cite{MILC4}. 
On a lattice with $N_t=6$ they even found a first order transition 
at $K=0.17$ --- 0.19 \cite{MILC6}. 

Looking at the phase diagram shown in Fig.~\ref{fig:F2KcKt} closely, 
we note that the $K_t$ lines initially deviate from the $K_c$ line 
and then approach the $K_c$ line at $\beta \sim 4.8$ and 
$K \sim 0.18$ for $N_t=4$ and at $\beta \sim 4.8$ --- 5.2 
and $K \sim 0.17$ --- 0.19
for $N_t=6$, contrary to the naive expectation that they 
monotonously deviate from the $K_c$ line.
The points where strong transitions occur are just in the region
where the $K_t$ lines approach the $K_c$ line.
Therefore, it is plausible that the strong transition 
at intermediate values of $K$ is a result of lattice artifacts 
caused by this unusual relation of the $K_t$ and $K_c$ lines 
\cite{IwasakiLat94}.
This unusual relation is probably due to
the sharp bend of the $K_c$ line at $\beta \simeq 5.0$
which is caused by the cross-over phenomenon between weak and 
strong coupling regions of QCD. 
Our recent study indeed shows that, 
with an improved lattice action, 
the distance between the the $K_c$ and $K_t$ lines
becomes monotonically large
when we decrease $K$ and, correspondingly, 
the $K_t$ transition becomes rapidly weak as we 
decrease $K$ from the chiral limit \cite{ourImprove}.
Also the unexpected $N_t$ dependence of $m_q$ 
in the deconfining phase 
at small $\beta$ discussed in the previous section, 
is removed with the same improved lattice action. 

The appearance of the lattice artifacts implies that
we have to be cautious 
when we try to derive the conclusions in the continuum limit
from the numerical results at finite $\beta$.
We also note that $N_t=4$ is far from the continuum limit
and therefore we should take with reservation,
in particular, 
quantitative values in physical units 
which are quoted in the following.
We, however, note that the PCAC relation $m_\pi^2 \propto m_q$
expected from chiral symmetry in the confining phase is well
satisfied even in the strong coupling region and therefore
we expect that qualitative feature of the chiral transition
such as the order of the transition does not affected by lattice
artifacts. We certainly have to check in future that 
the conclusions in this
article are also satisfied when an improved action is adopted.

\section{Numerical Results for Chiral Transitions}
\label{sect:chiral}

As discussed in Sec.~\ref{subsect:finitetemp},
the chiral transition can be studied along the $K_c$ line
at the crossing point of the $K_t$ and $K_c$ lines,
which we denote as the chiral 
transition point $\beta_{ct}$.
We first address ourselves to the problem of
whether the chiral limit of the finite temperature 
transition exists at all.
We then study the order of the chiral transition. 

In a previous paper \cite{ourStrong} 
we showed that, when $N_F \ge 7$, there is a bulk 
first order phase transition at $\beta=0$ 
which separates the confining phase at small $K$ from 
a deconfining phase near the chiral limit at $K = 1/4$.
This implies that the $K_t$ line 
does not cross the $K_c$ line at finite $\beta$ for any $N_t$. 
On the other hand, when $N_F \le 6$, the chiral limit 
belongs to the confining phase at $\beta=0$,
which implies that there is
a crossing point somewhere at finite $\beta$ for the case 
$N_F \le 6$. 

\subsection{On-$K_c$ method}

In order to identify the crossing point $\beta_{ct}$ 
and study the order of the chiral transition there, 
we take the strategy of 
performing simulations on the $K_c$ line starting from
a value of $\beta$ in the deconfining phase
and reducing $\beta$.
We call this method ``on-$K_c$'' simulation method.
The number of iterations $N_{\rm inv}$ needed for the quark 
matrix inversion, in general, provides a good indicator
to discriminate the deconfining phase from the confining 
phase \cite{ChQ,guptainv}.
The use of $N_{\rm inv}$ as an indicator
is extremely useful on the $K_c$ line,
because $N_{\rm inv}$ is enormously large on the $K_c$ line 
in the confining phase,
while it is of order several hundreds in the deconfining phase.
Therefore there is a sudden drastic 
change of $N_{\rm inv}$ across the boundary of the two
phases.
This difference is due to the fact that
there are zero modes around $K_c$ in the confining phase, 
while none exists in the deconfining phase
\cite{ourStrong,guptainv,zero_mode}:
We have checked this difference for the existence of zero modes
in various cases discussed below and conclude that the 
difference of $N_{\rm inv}$ is not a numerical artifact. 

In the deconfining phase on the $K_c$ line, 
we measure physical observables such as 
the Polyakov loop, the plaquette and hadron screening 
masses, as usual, after thermalization.
From the behavior of physical quantities toward $\beta_{ct}$,
we are able to study the nature of the chiral transition.
In the confining phase, on the other hand, 
it is hard to make the system on the $K_c$ line
thermalized due to the enormously large $N_{\rm inv}$ 
we encounter in the configuration generation. 
In this case, we only obtain at most bounds for 
several physical quantities 
by measuring the molecular dynamic time evolution of them 
starting a hot state or a mix state. 
Although it is unsatisfactory that we cannot obtain 
expectation values for
physical quantities in the confining phase,
the on-$K_c$ method is very powerful to identify the 
critical point 
because the difference between the two phases is clear 
already with short time-histories. 
We also check that the crossing point thus determined is 
consistent with
a linear extrapolation of the line $K_t$ 
toward the chiral limit.

\subsection{Chiral transition for $N_F=2$}

For the case of QCD with two flavors, 
studies of an effective $\sigma$ model\cite{Pisarski,Wilczek}
imply that
the order of the chiral transition depends 
on the the strength of the U$\!_A$(1) anomaly term at the 
transition temperature. 
When the strength is zero, it is of first order.
However, if the strength of the anomaly term 
in the effective $\sigma$ model is 
non-zero at the starting point of renormalization transformation,
it is likely that
the effective action is attracted to a $O(4)$ symmetric 
fixed point under renormalization group transformation 
\cite{Ukawa95}.
Therefore, it is plausible that 
the chiral transition is of second order.

Our main results of the measurements for $N_F=2$ are summarized 
in Tables~\ref{tab:resultsF2Nt4} --- \ref{tab:resultsF2Nt8}. 

Let us first discuss the results at $N_t = 4$. 
In order to confirm the existence of the crossing point, 
we take the largest (farthest) values of $K_c$ 
for on-$K_c$ simulations, 
that is, $K_c(m_\pi^2)$ for $N_F=2$ in Table~\ref{tab:1}
and interpolated ones. 
As discussed previously, $K_c(m_\pi^2)$ in general 
depend on the value of $N_t$. 
However, 
the differences between those on the $N_t=4$ and 8 lattices
are within numerical uncertainties as shown Table~\ref{tab:2}.
Therefore, 
we take the stringent condition to verify the existence of
the crossing point, taking the farthest values of $K_c$.

When we take into account the structure of $K_c(m_\pi^2; T\neq 0)$
that it sharply turns back at finite $\beta$, we may hit the upper 
part of it by taking the largest values of $K_c$
for the ``on-$K_c$'' method.  
This, however, does not affect the
conclusion that the $K_t$ line crosses the $K_c$ line.
Our estimates for the value of $\beta_{ct}$ in this case will be
slightly underestimated (cf.\ Fig.~\ref{fig:PDfiniteT}). 
This comment applies also for $N_F=3$ and 6.

\begin{figure}
\centerline{ \epsfxsize=14cm \epsfbox{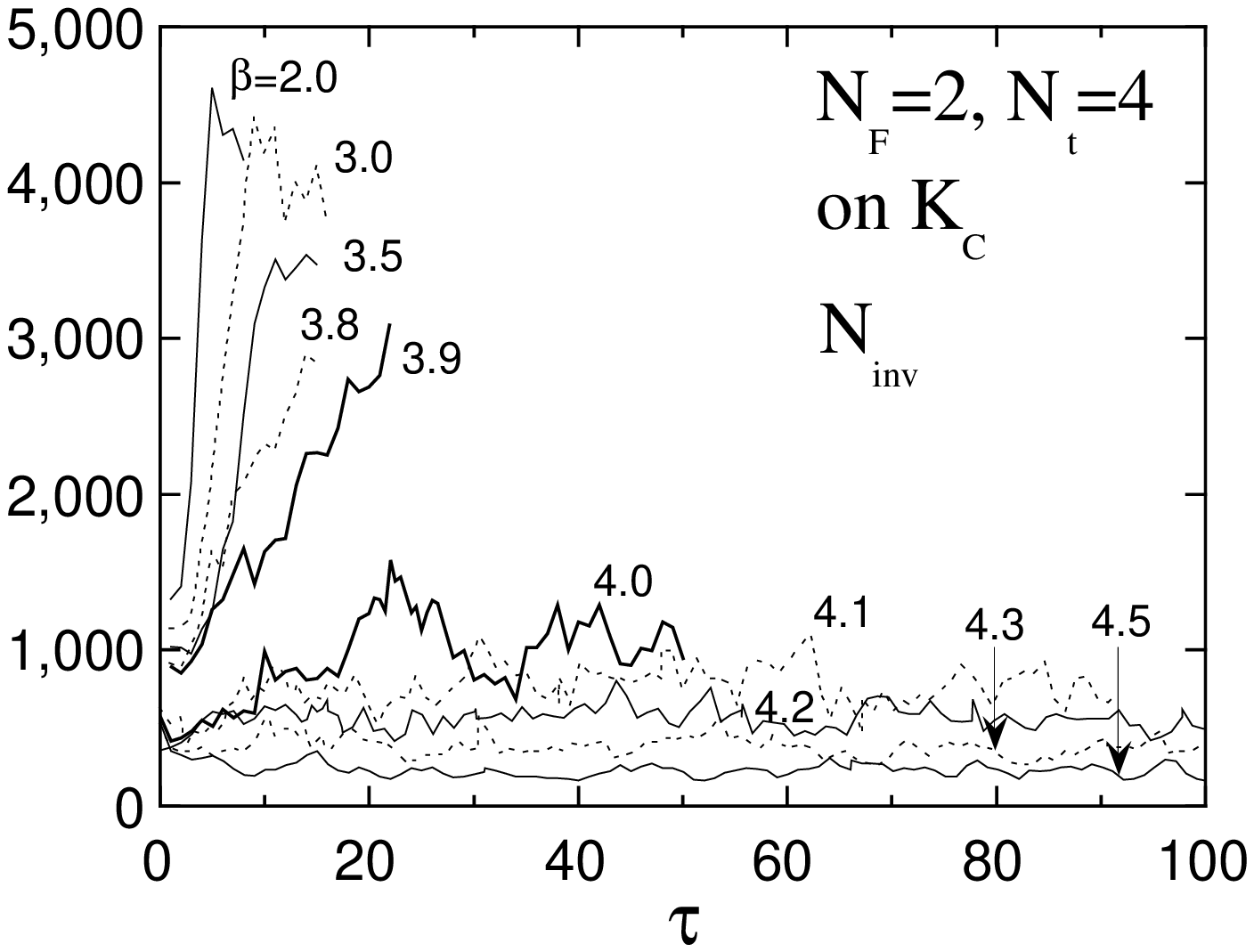}}
\caption{\baselineskip=.8cm
Molecular-dynamics time history of $N_{\rm inv}$ for 
$N_F=2$ on the $K_c$ line 
obtained on an $8^2\times10\times 4$ lattice. 
}
\label{fig:H2Ninv}
\end{figure}

We first perform on-$K_c$ simulations 
by the R algorithm to identify the crossing point, because 
it is very time consuming 
to perform simulations with the HMC algorithm 
due to a low acceptance rate 
on the $K_c$ line in the confining phase.
We find that when $\beta \ge 4.0$, $N_{\rm inv}$ stays around 
several hundreds,
while for $\beta \le 3.9$ it increases with $\tau$ and 
exceeds several thousands (see Fig.~\ref{fig:H2Ninv})
and in accord with this behavior the plaquette, the Polyakov 
loop and
$m_\pi$ decrease rapidly toward those in the confining phase.
Therefore we identify the crossing point
at $\beta_{ct}\sim 3.9$ --- 4.0.
This $\beta_{ct}$ is
consistent with a linear extrapolation of the $K_t$ line as
is shown in Fig.~\ref{fig:F2KcKt}.

\begin{figure}
\centerline{ \epsfxsize=14cm \epsfbox{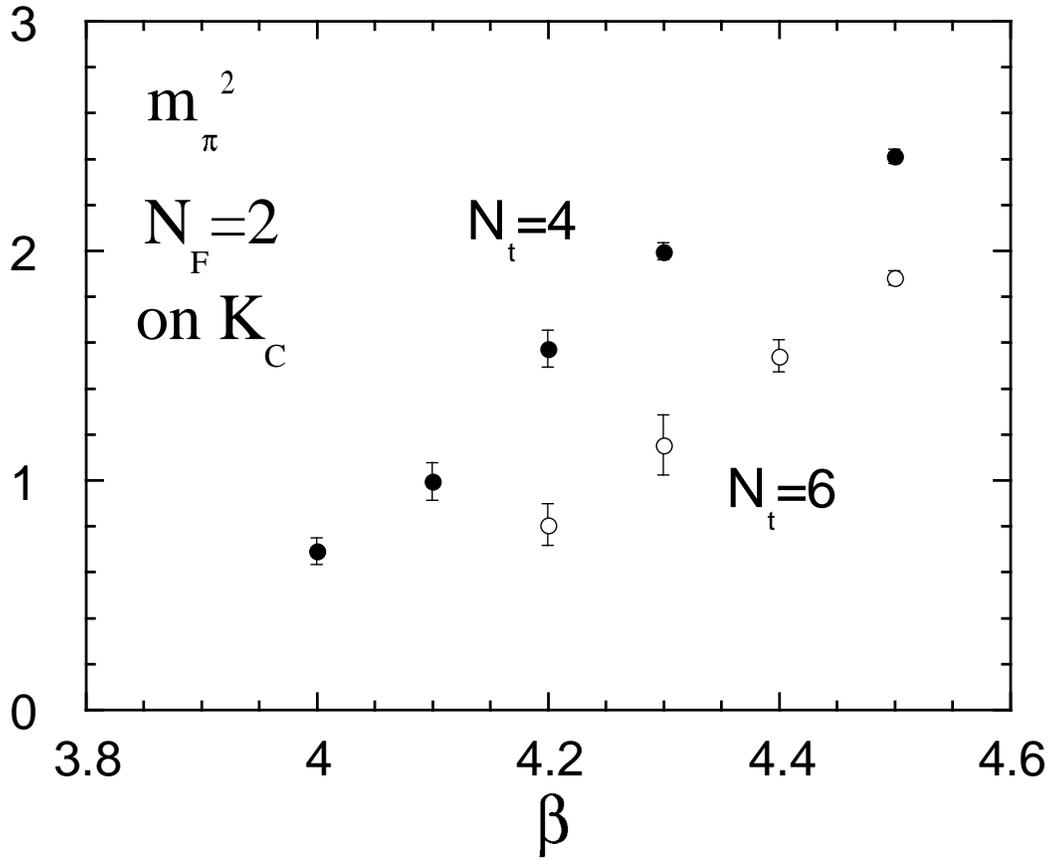}}
\caption{\baselineskip=.8cm
The pion screening mass squared $m_\pi^2 a^2$ for $N_F=2$ 
on the $K_c$ line
obtained on $8^2\times10\times 4$ and $12^3\times6$ lattices. 
}
\label{fig:F2Mpi}
\end{figure}

Then we repeat on-$K_c$ simulations by the HMC algorithm
for $\beta \ge 4.0$
in order to measure physical observables. 
The time histories for $N_{\rm inv}$ at $\beta \ge 4.0$ plotted
in Fig.~\ref{fig:H2Ninv} are obtained with the HMC algorithm, 
which are similar to those with the R algorithm.
The $\Delta\tau$ should be taken small near $\beta_{ct}$ 
in order to keep the acceptance rate reasonably high 
(for $\beta=4.0$, 4.1 and 4.2 we use $\Delta\tau=0.002$, 0.005
and 0.005 to get acceptance rates 0.91, 0.79 and 0.93, 
respectively).
The value of $m_\pi^2$ thus obtained
decreases smoothly toward zero 
as the chiral transition is approached 
and is consistent with zero at the estimated $\beta_{ct}$ 
(see Fig.~\ref{fig:F2Mpi}).

We find no two-state signals around $\beta_{ct}$.
This is in sharp contrast with the $N_F=3$ and 6 cases
where we find clear two-state signals at $\beta_{ct}$,
as discussed below. 
This, together with the vanishing $m_\pi^2$ toward $\beta_{ct}$,
indicates that the chiral transition is continuous 
(second order or crossover) for $N_F=2$.

The results from on-$K_c$ simulations on the $N_t=6$ lattice
are similar to those on the $N_t=4$ lattice. 
The estimated transition point is $\beta_{ct}\sim $ 4.0 --- 4.2.
The value of $m_{\pi}^2$ listed in Table \ref{tab:resultsF2Nt6} 
and plotted in Fig.~\ref{fig:F2Mpi}, 
again decreases toward zero as $\beta$ approaches $\beta_{ct}$.
For $N_t=18$ with the spatial size $18^2 \times 24$,
we previously found that the transition is at
$\beta_{ct} \sim 4.5$ --- 5.0 \cite{ourOldLatC}. 
Although the spatial size is not large enough, 
this result suggests
that the shift of $\beta_{ct}$ with $N_t$ is very slow.

\subsection{Chiral transition for $N_F=3$}

\begin{figure}
\centerline{ \epsfxsize=14cm \epsfbox{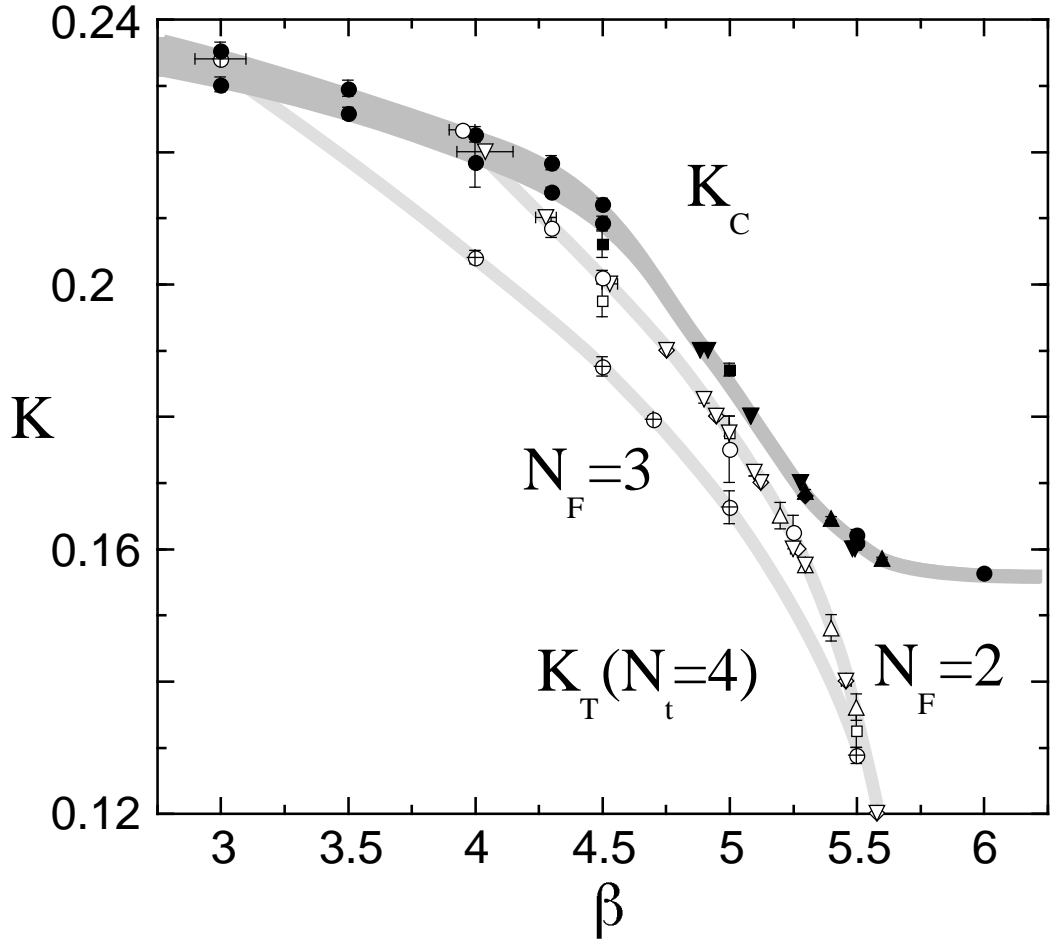} }
\caption{\baselineskip=.8cm
Phase diagram for $N_F=2$ and 3.
Filled symbols are for $K_c(m_\pi^2)$ and $K_c(m_q)$.
Open symbols are for $K_t(N_t=4)$ for $N_F=2$ and 
open circles with cross for $N_F=3$. 
Circles are our data. 
On the $K_t$ line for $N_F=3$, 
clear two-state signals are observed at $\beta \leq 4.7$ 
both on $8^2\times10\times4$ and $12^3\times4$ lattices.
Lines are to guide the eyes. 
}
\label{fig:F2F3KcKt}
\end{figure}

Main results of measurements for $N_F=3$ are summarized in 
Tables \ref{tab:resultsF3Nt4Ns8} and \ref{tab:resultsF3Nt4Ns12}. 
The phase diagram for $N_F=3$ obtained from our simulations at 
$\beta$=4.0, 4.5, 4.7, 5.0 and 5.5 
is shown in Fig.~\ref{fig:F2F3KcKt}. 
We find that the $K_t$ line linearly approaches 
to the $K_c$ line.
In order to confirm the existence of the crossing point 
by on-$K_c$ simulations, 
we take the largest (farthest)
$K_c$, that is $K_c(m_\pi^2)$ for $N_F=2$
at $\beta$'s we have studied,
since this is the most stringent condition for the existence 
of $\beta_{ct}$.
We use them and interpolated values for on-$K_c$ simulations here. 
For $N_F=6$ discussed in the next subsection, 
we interpolate these values of $K_c$ with $K_c=0.25$ at $\beta=0$.
Note that the differences of $K_c$'s for $N_F=2$, 3 and 6 are of the
same magnitude of numerical uncertainties of $K_c$.

\begin{figure}
\centerline{ \epsfxsize=14cm \epsfbox{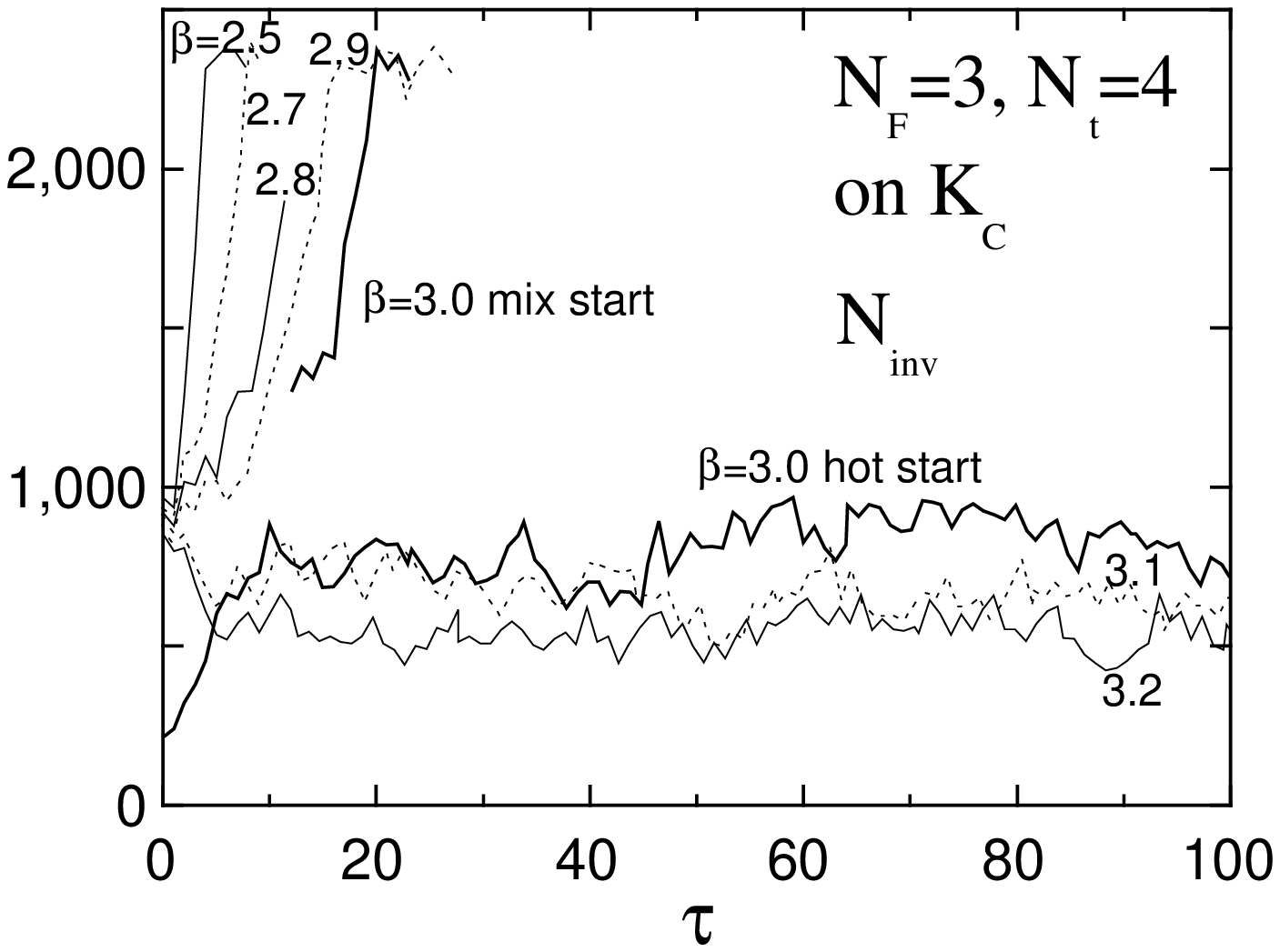}}
\caption{\baselineskip=.8cm
Time history of $N_{\rm inv}$ for 
$N_F=3$ on the $K_c$ line
obtained on an $8^2\times10\times 4$ lattice. 
}
\label{fig:H3Ninv}
\end{figure}

Fig.~\ref{fig:H3Ninv} shows $N_{\rm inv}$ as a function of 
the molecular-dynamics time $\tau$
for several values of $\beta$'s.
When $\beta \ge 3.1$, $N_{\rm inv}$ is of order of several 
hundreds, while
when $\beta \le 2.9$, $N_{\rm inv}$
shows a rapid increase with $\tau$.
At $\beta = 3.0$ we see a clear two-state signal
depending on the initial condition:
For a hot start, $N_{\rm inv}$
is quite stable around $\sim 800$ and $m_\pi^2$ is large 
($\sim 1.0$).
On the other hand, for a mix start, $N_{\rm inv}$ 
shows a rapid increase with $\tau$ and exceeds
2,000 in $\tau \sim 20$, and in accordance with this, $m_\pi^2$ 
decreases 
with $\tau$. 

\begin{figure}
\centerline{ \epsfxsize=14cm \epsfbox{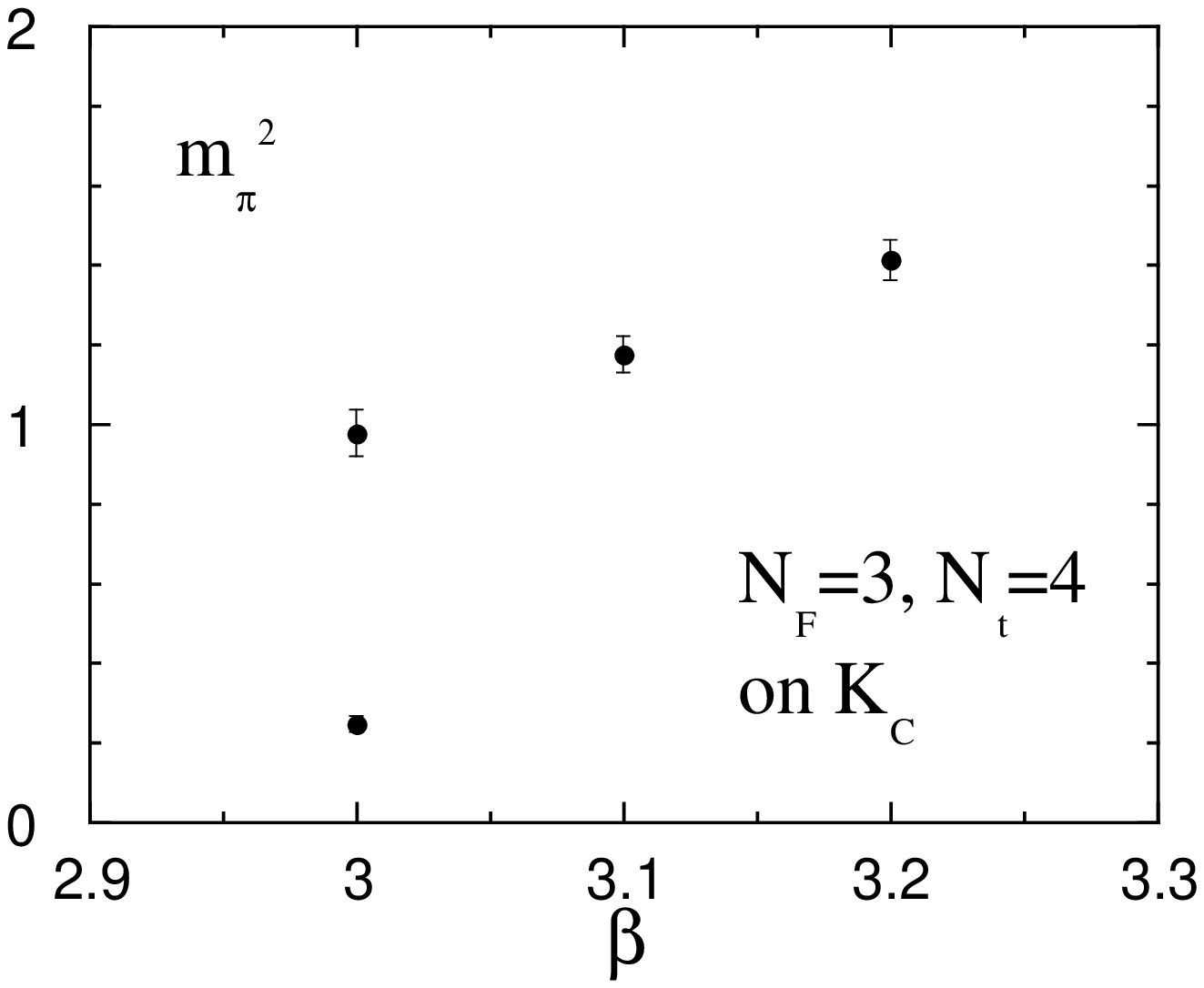}}
\caption{\baselineskip=.8cm
$m_\pi^2 a^2$ for $N_F=3$ on the $K_c$ line
obtained on an $8^2\times10\times 4$ lattice. 
}
\label{fig:F3Mpi}
\end{figure}

The value of $m_\pi^2$ is plotted in Fig.~\ref{fig:F3Mpi}.
At $\beta =3.0$ we have two values for $m_\pi^2$ depending on the
initial configuration. 
The larger one obtained for the hot start
is of order 1.0, which is a smooth extrapolation of the
values at $\beta \sim 3.1$ - 3.2.
The smaller one is an upper bound for $m_\pi^2$ for the mix start.

We note that the result of
$\beta_{ct} \sim 3.0$ is consistent with an extrapolation of 
$K_t$ points listed in Table~\ref{tab:Kt} 
as is shown in Fig.~\ref{fig:F2F3KcKt}.
(The nature of the transition $K_t$ off the chiral limit 
is discussed in Sec.~\ref{sect:strange}.)
Thus we identify the crossing point at $\beta_{ct} \sim 3.0(1)$.
With the clear two-state signal we 
conclude that the chiral transition is of first order for $N_F=3$.

\subsection{Chiral transition for $N_F$ = 6}

\begin{figure}
\centerline{ \epsfxsize=14cm \epsfbox{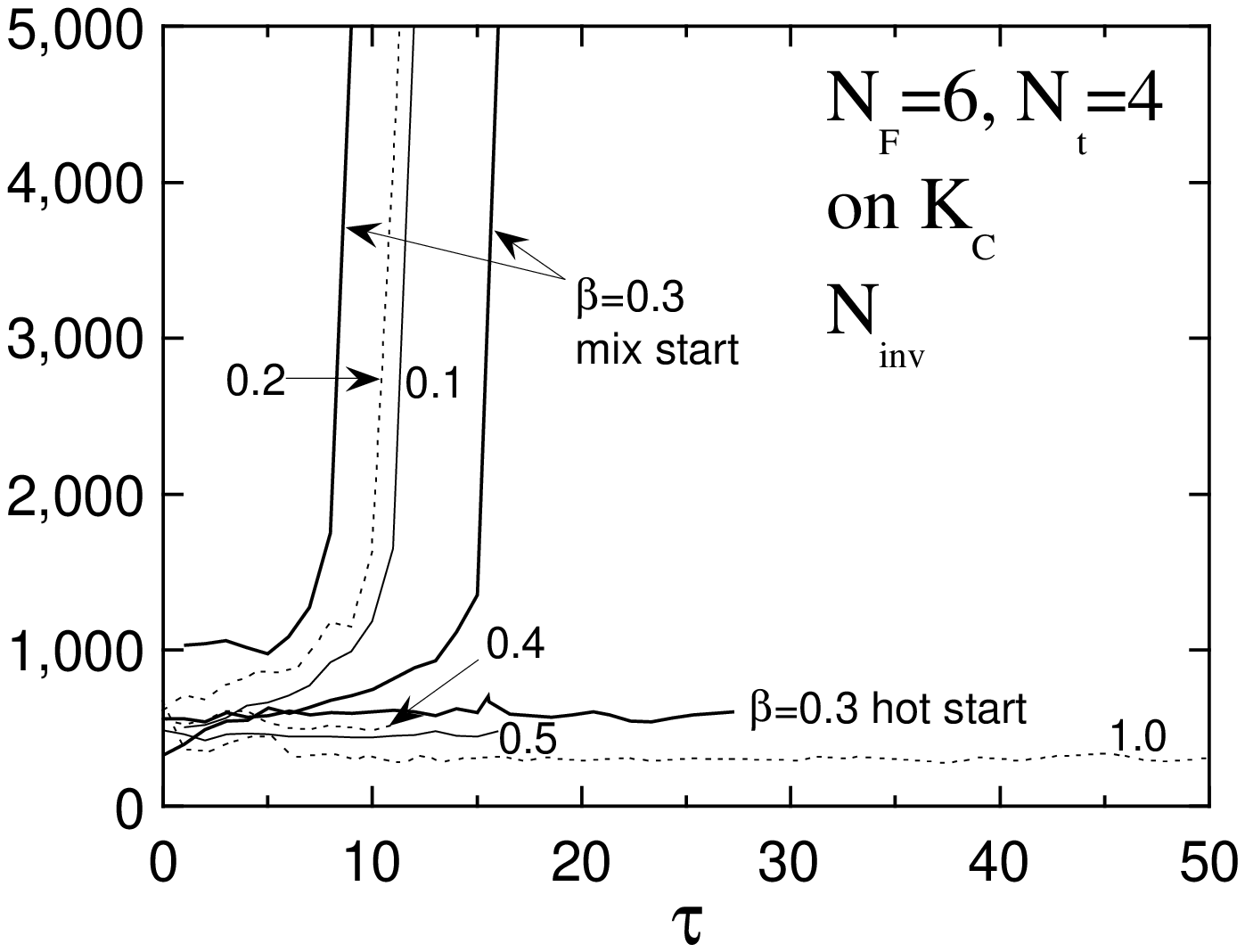}}
\caption{\baselineskip=.8cm
Time history of $N_{\rm inv}$ for $N_F=6$ on the $K_c$ line
obtained on an $8^2\times10\times 4$ lattice. 
}
\label{fig:H6Ninv}
\end{figure}

\begin{figure}
\centerline{ \epsfxsize=14cm \epsfbox{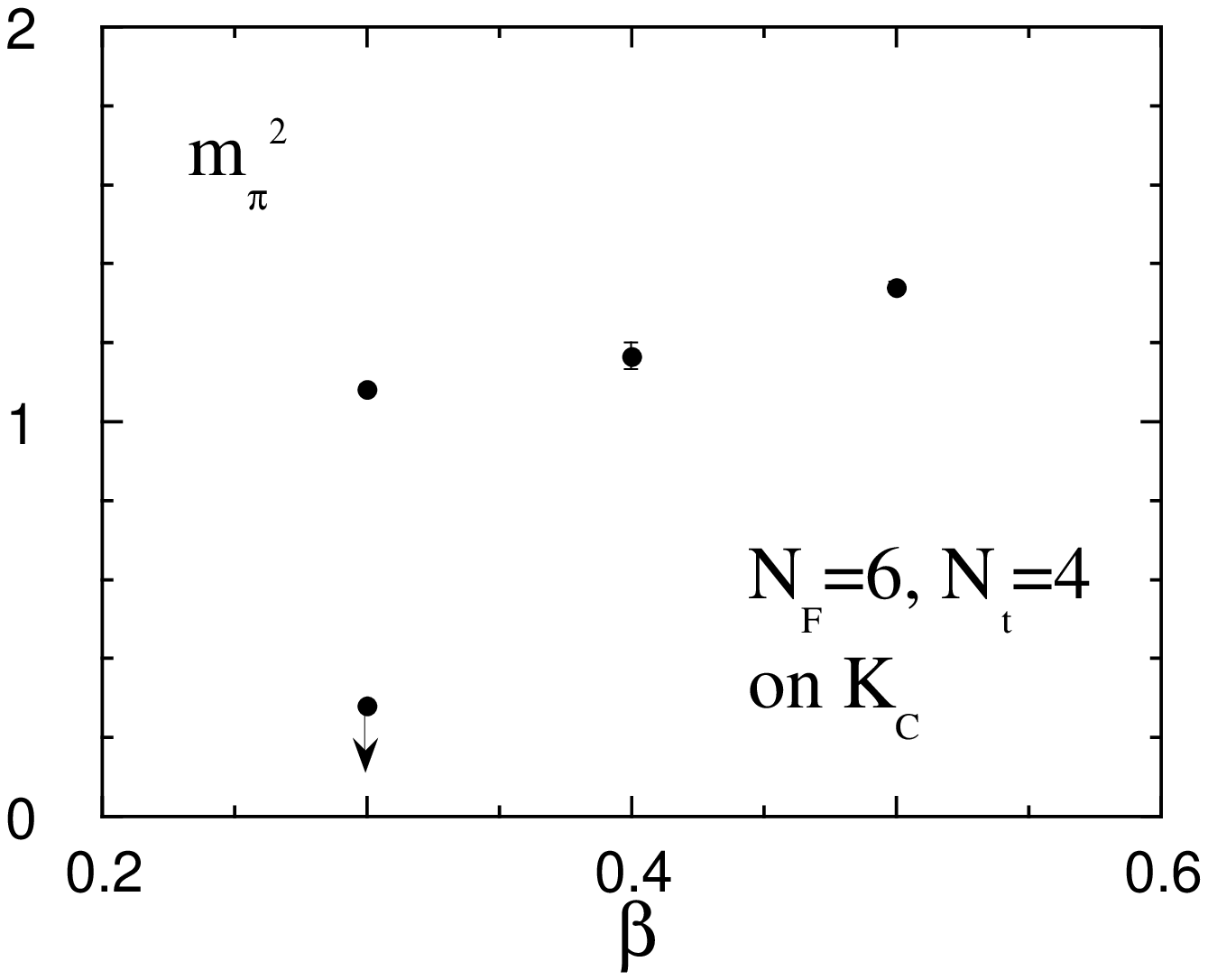}}
\caption{{\baselineskip=.8cm
$m_\pi^2 a^2$ for $N_F=6$ on the $K_c$ line
obtained on an $8^2\times10\times 4$ lattice. 
}
\label{fig:F6Mpi}}
\end{figure}

Our previous study at $\beta=0$ \cite{ourStrong} 
shows that for $N_F=7$ there is no crossing point of 
the $K_c$ and $K_t$ lines
and that $N_F=6$ is the largest number of flavors for which
a crossing point exists.
Main results of measurements for $N_F=6$ are summarized in 
Table \ref{tab:resultsF6Nt4}. 
Overall features of the transition obtained from 
numerical simulations for $N_F=6$
are very similar to those for $N_F=3$ except for the
location of $\beta_{ct}$, which moves to a smaller $\beta$ as 
expected.
Fig.~\ref{fig:H6Ninv} shows that $N_{\rm inv}$ on the $K_c$ line 
stays at several hundreds for $\beta \ge 0.4$
and for a hot start at $\beta=0.3$.
On the other hand,
$N_{\rm inv}$ grows rapidly with $\tau$ and
exceeds 5,000 for $\beta \le 0.2$ and for a mix start at 
$\beta=0.3$.
In accord with this, we have two values of $m_\pi^2$ at 
$\beta=0.3$ 
(cf.\ Fig.~\ref{fig:F6Mpi}).
Therefore we identify the crossing point at 
$\beta_{ct} \sim 0.3(1)$
and conclude that the chiral transition is of first order
for $N_F=6$.
This $\beta_{ct}$ is consistent with a linear extrapolation 
of the $K_t$ line (cf.\ Table~\ref{tab:Kt}).

For QCD with $N_F \geq 3$, 
Pisarski and Wilczek predicted a first order chiral transition 
from a renormalization group study of 
an effective $\sigma$ model \cite{Pisarski}. 
Our results for $N_F=3$ and 6 
are consistent with their prediction.

\section{Influence of the Strange Quark}
\label{sect:strange}

In the previous section, we have seen that the chiral transition 
is consistent with a second order transition for $N_F=2$, 
while it is of first order for $N_F \geq 3$, 
both in accordance with theoretical expectations. 
Off the chiral limit, we expect that 
the first order transition for $N_F \geq 3$ smoothens 
into a crossover at sufficiently large $m_q$. 
In this way the nature of the transition sensitively 
depends on $N_F$ and $m_q$. 
Therefore,
in order to study the nature of the transition in the real world, 
we should include the strange quark properly 
whose mass $m_s$ is of the same order of magnitude 
as the transition temperature $T_c \simeq 100$ --- 200 MeV. 

In a numerical study we are able to vary the mass of the 
strange quark.
When the mass of the strange quark is reduced from infinity 
to zero with up and down quarks fixed to the chiral limit, 
the nature of the transition must change from continuous 
to first order at some quark mass $m_s^*$. 
Assuming that the chiral transition is of second order 
for $N_F=2$ (i.e.\ $m_s=\infty$), 
this point at $m_s^*$ is a tricritical point \cite{Wilczek}. 
The crucial question is whether
the physical strange quark mass is larger or smaller 
than $m_s^*$.
Studies with an effective linear $\sigma$ model 
suggest a crossover for the case of realistic quark masses 
in meanfield approximation 
and in a large $1/N_F$ approximation 
\cite{SigmaModel,SigmaModel1}, 
while the possibility of a weakly first order transition 
is not excluded 
when numerical errors in the calculation of basic parameters 
are taken into account
\cite{SigmaModel1}.

\subsection{$N_F=3$}

\begin{figure}
\centerline{ \epsfxsize=10cm \epsfbox{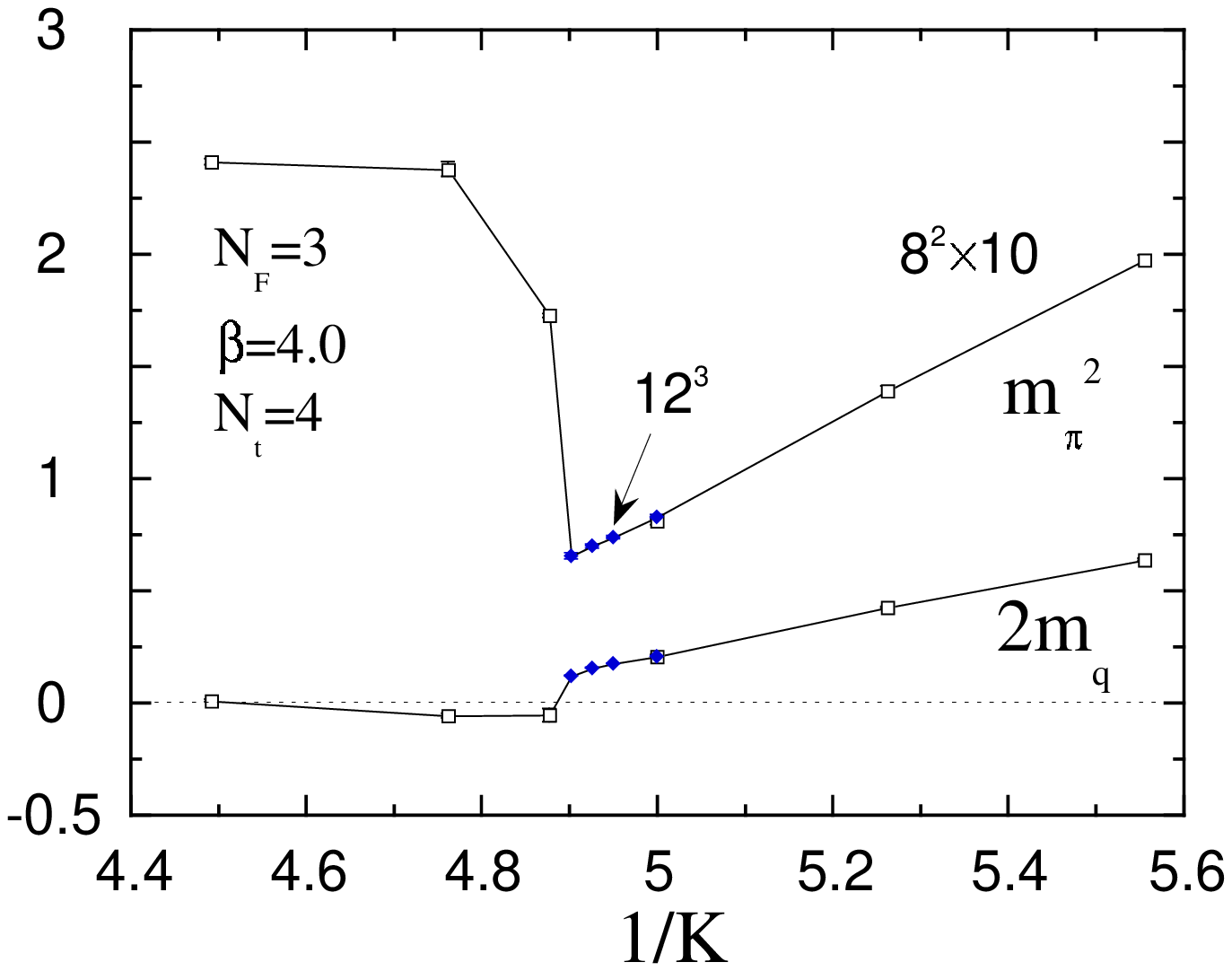} }
\centerline{ \epsfxsize=9.7cm \epsfbox{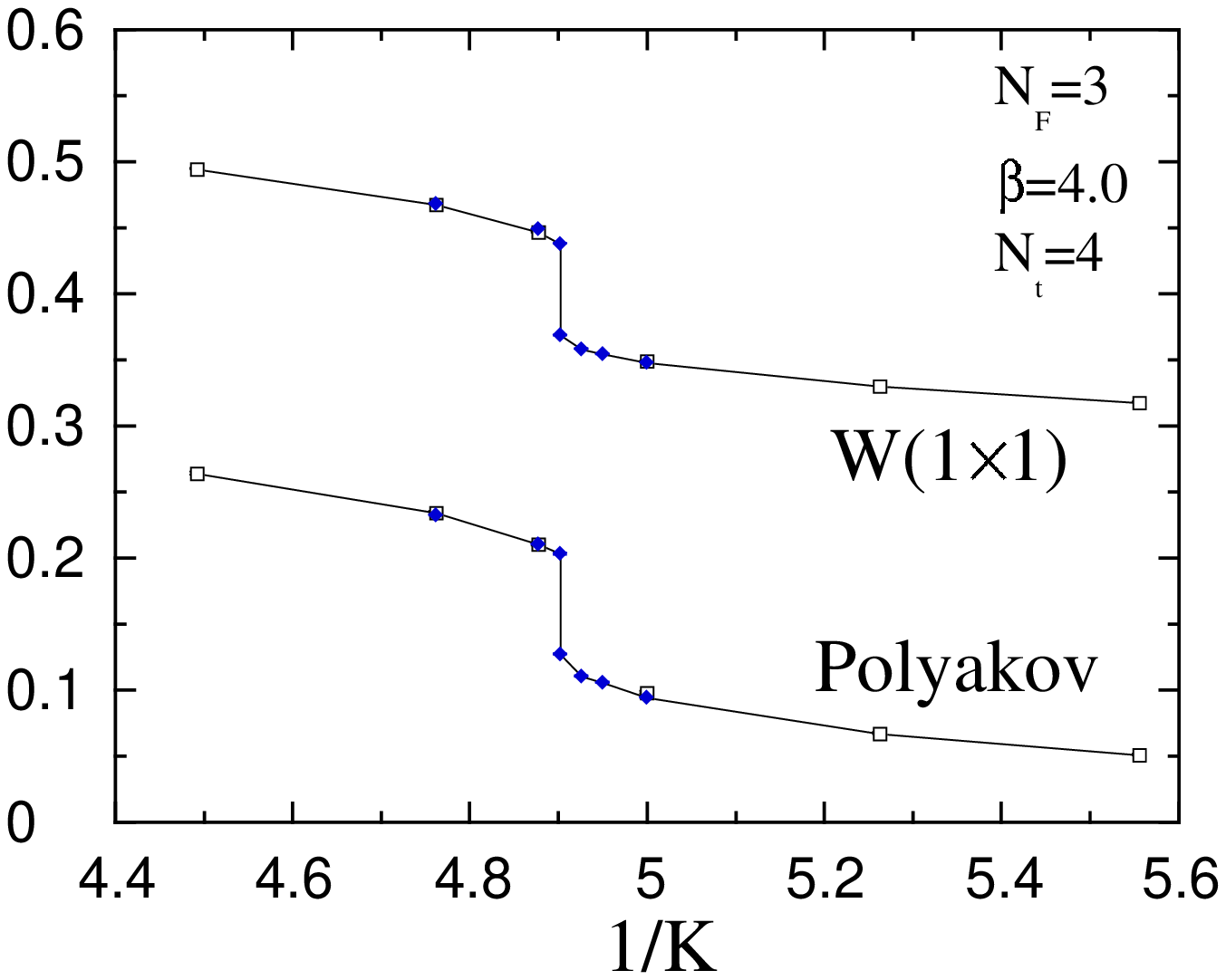} }
\caption{\baselineskip=.8cm
Physical quantities for $N_F=3$ 
at $\beta=4.0$ on $8^2\times10\times 4$ (open squares)
and $12^3\times4$ (filled diamonds) lattices: 
(a) $m_\pi^2 a^2$ and $2m_q a$, 
(b) the plaquette and the Polyakov loop. 
The finite temperature transition $K_t$ locates 
at $K \simeq 0.204$ ($1/K \simeq 4.90$). 
}
\label{fig:F3B4.0}
\end{figure}

\begin{figure}
\centerline{ \epsfxsize=10cm \epsfbox{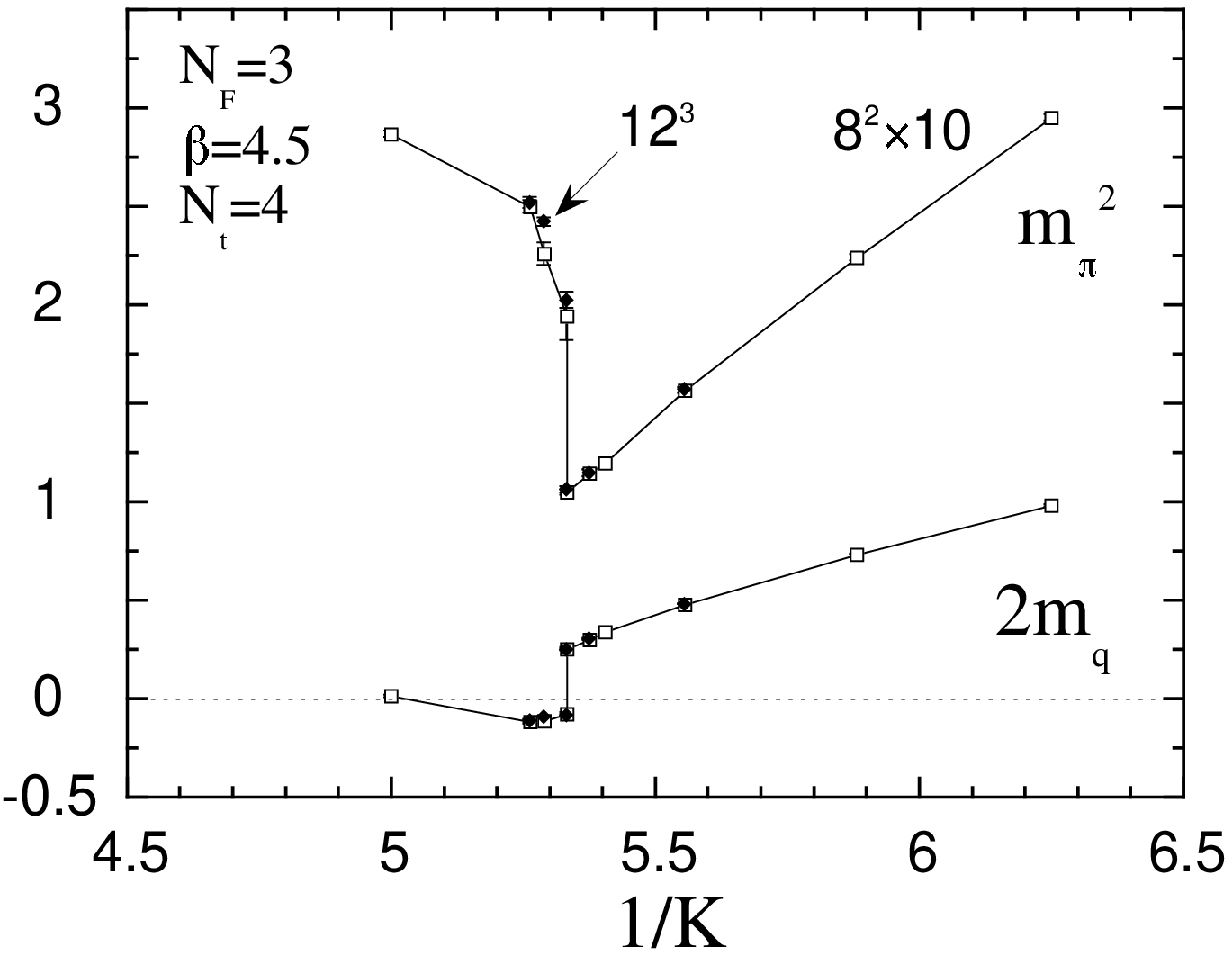} }
\centerline{ \epsfxsize=9.7cm \epsfbox{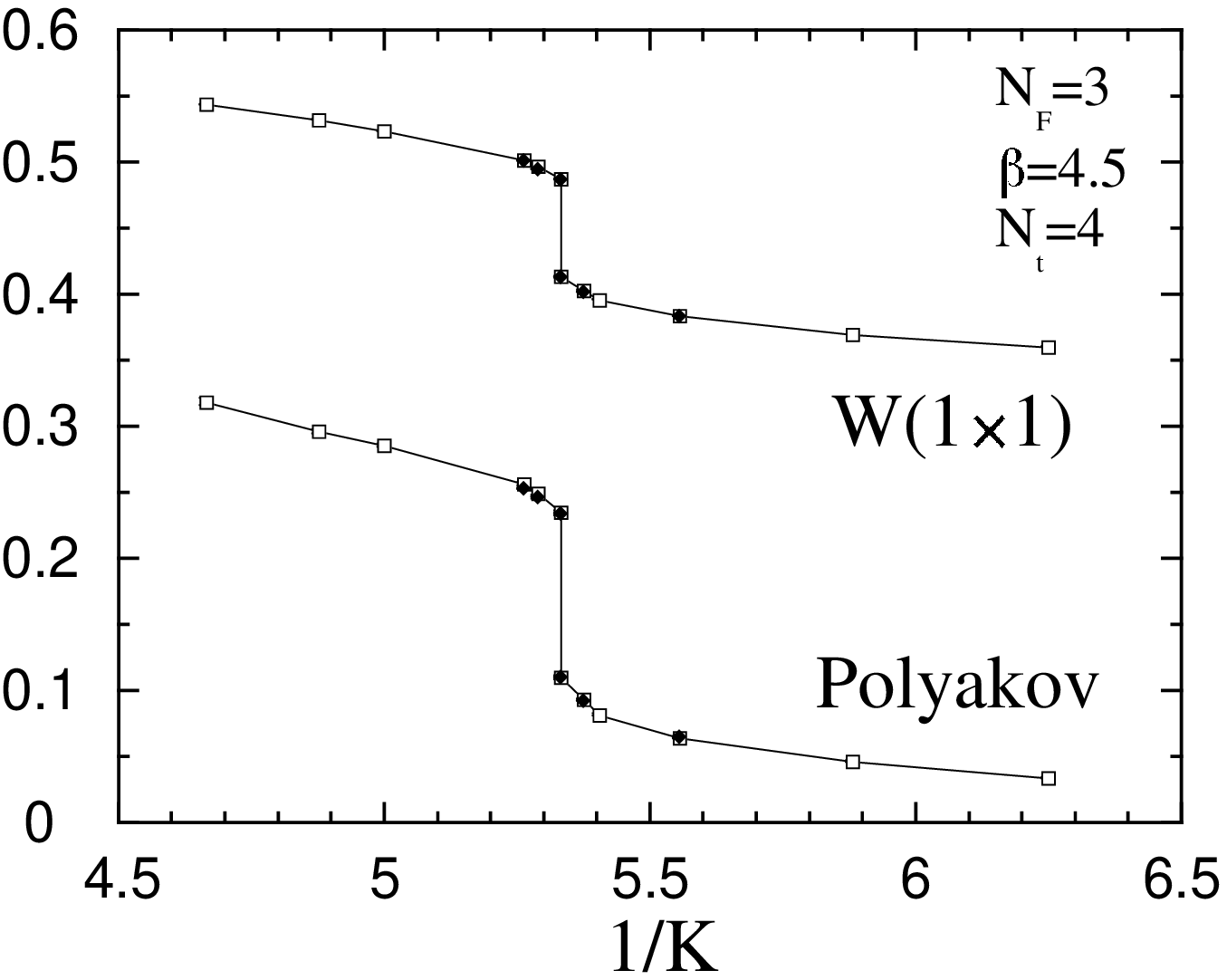} }
\caption{\baselineskip=.8cm
The same as Fig.~\protect\ref{fig:F3B4.0} at $\beta=4.5$. 
The finite temperature transition $K_t$ locates 
at $K \simeq 0.1875$ ($1/K \simeq 5.33$). 
}
\label{fig:F3B4.5}
\end{figure}

\begin{figure}
\centerline{ \epsfxsize=10cm \epsfbox{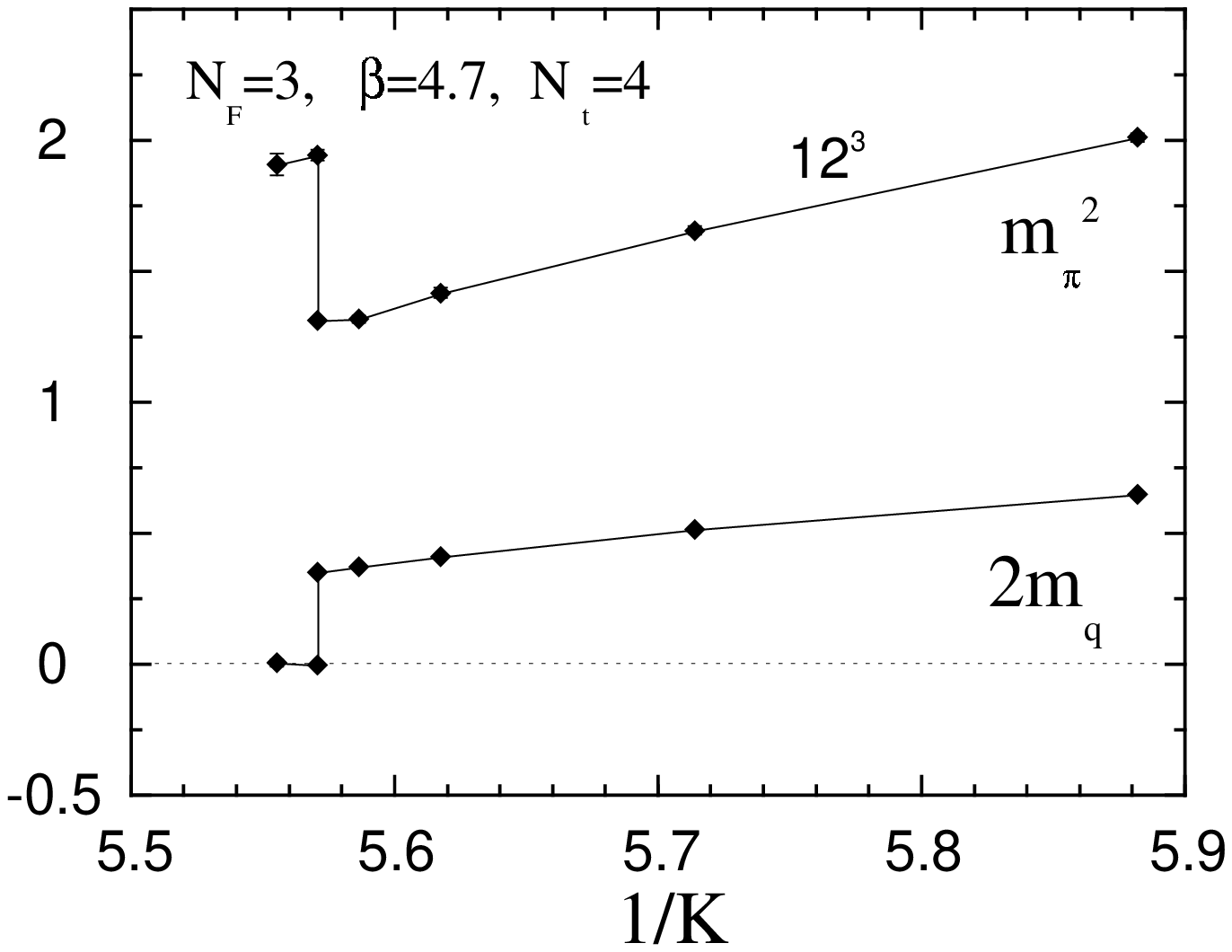} }
\centerline{ \epsfxsize=9.7cm \epsfbox{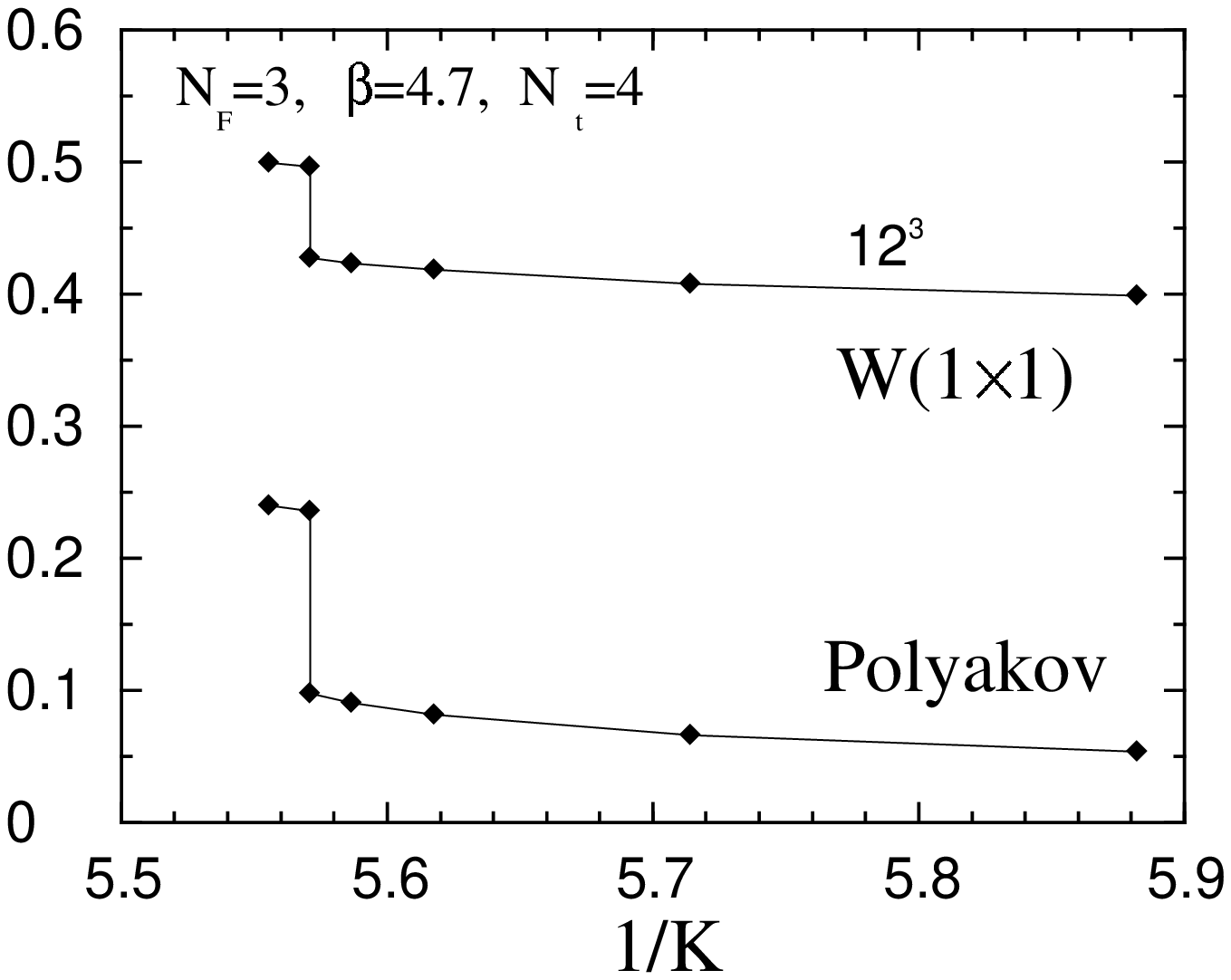} }
\caption{\baselineskip=.8cm
The same as Fig.~\protect\ref{fig:F3B4.0} at $\beta=4.7$ 
obtained on a $12^3\times 4$ lattice.
The finite temperature transition $K_t$ locates 
at $K \simeq 0.1795$ ($1/K \simeq 5.57$). 
}
\label{fig:F3B4.7}
\end{figure}

\begin{figure}
\centerline{ \epsfxsize=10cm \epsfbox{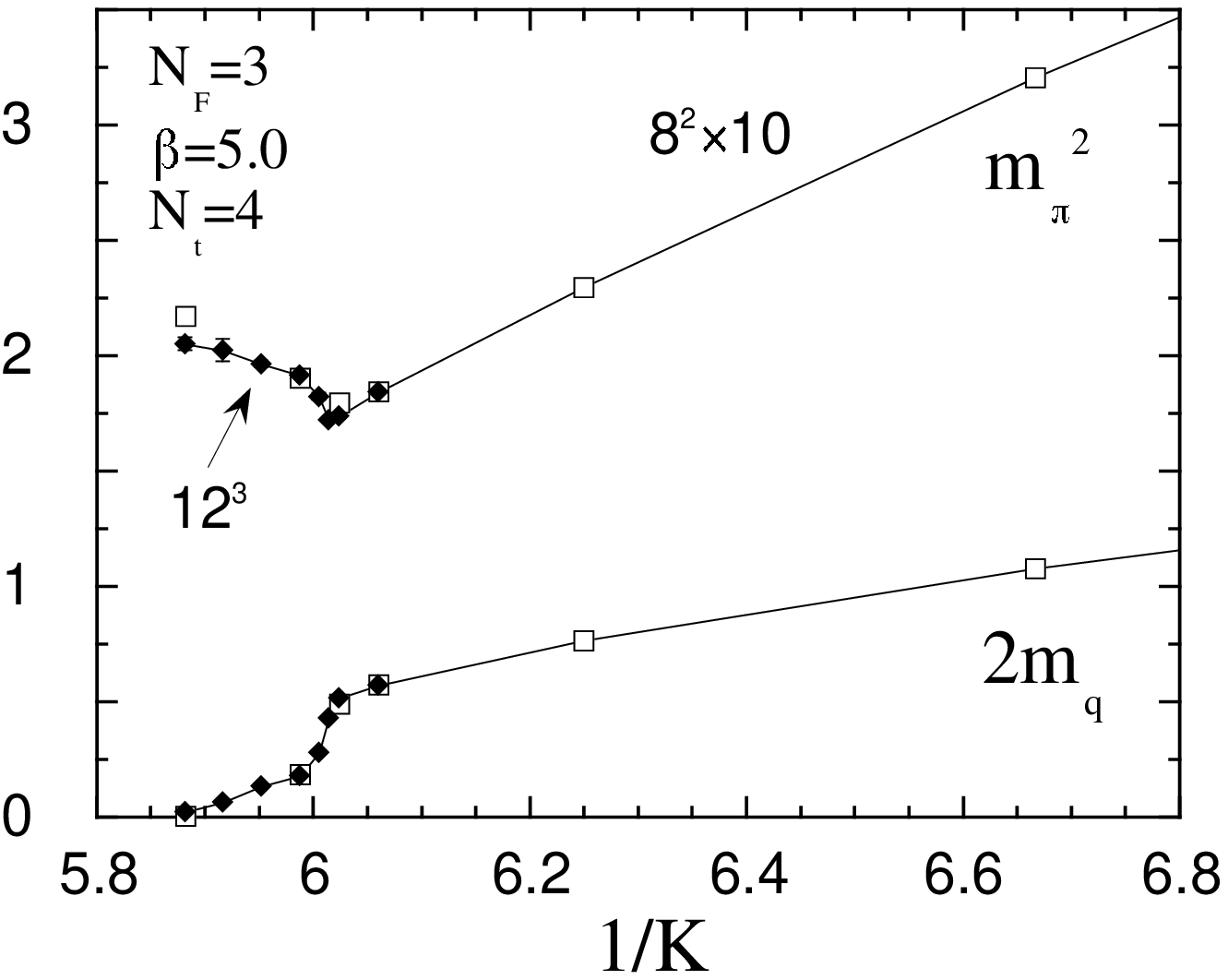} }
\centerline{ \epsfxsize=10cm \epsfbox{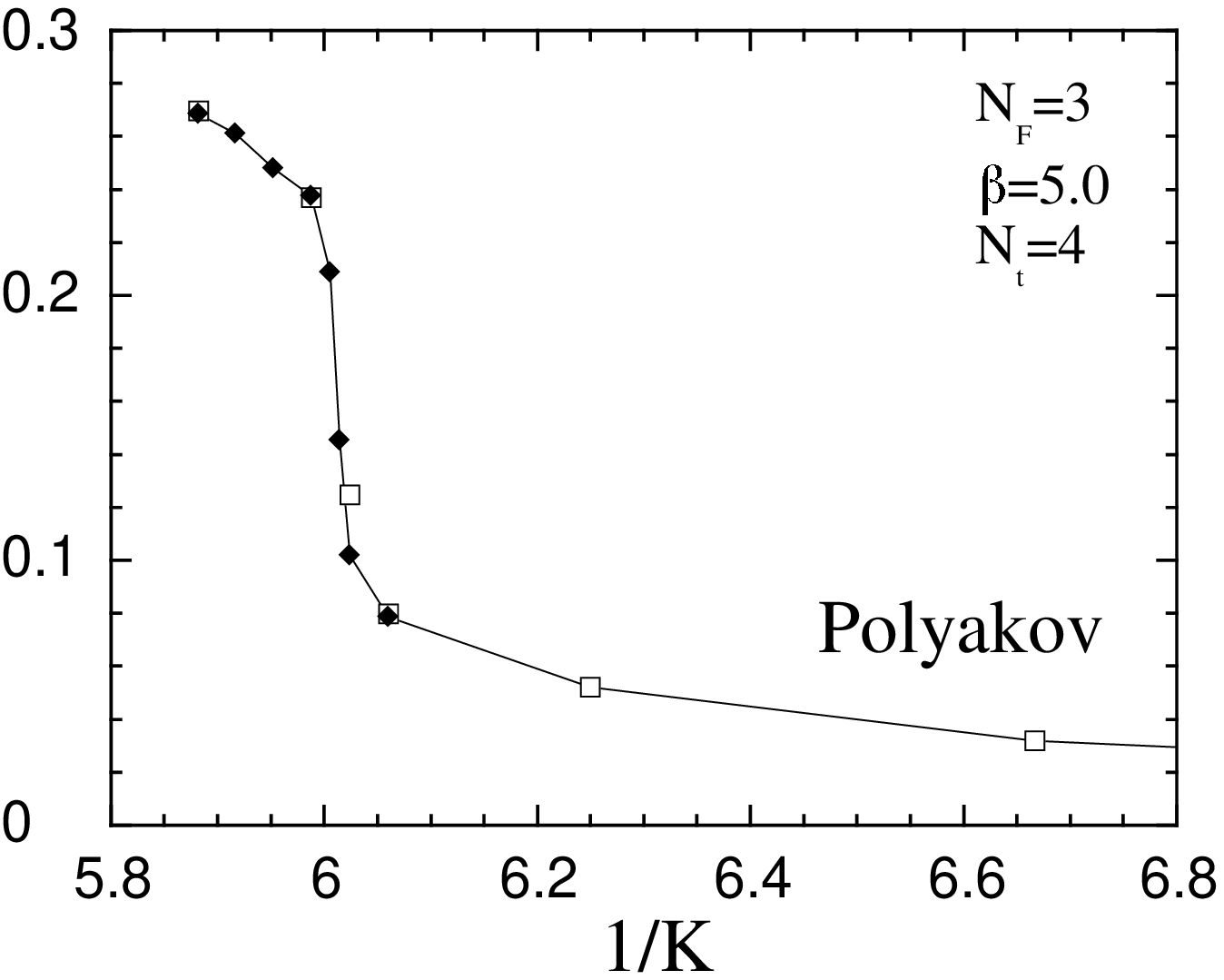} }
\caption{\baselineskip=.8cm
The same as Fig.~\protect\ref{fig:F3B4.0} at $\beta=5.0$:
(a) $m_\pi^2 a^2$ and $2m_q a$, 
(b) the Polyakov loop. 
The finite temperature crossover $K_t$ locates 
at $K \simeq 0.166$ --- 0.1665 ($1/K \simeq 6.01$ --- 6.02). 
}
\label{fig:F3B5.0}
\end{figure}

\begin{figure}
\centerline{ \epsfxsize=10cm \epsfbox{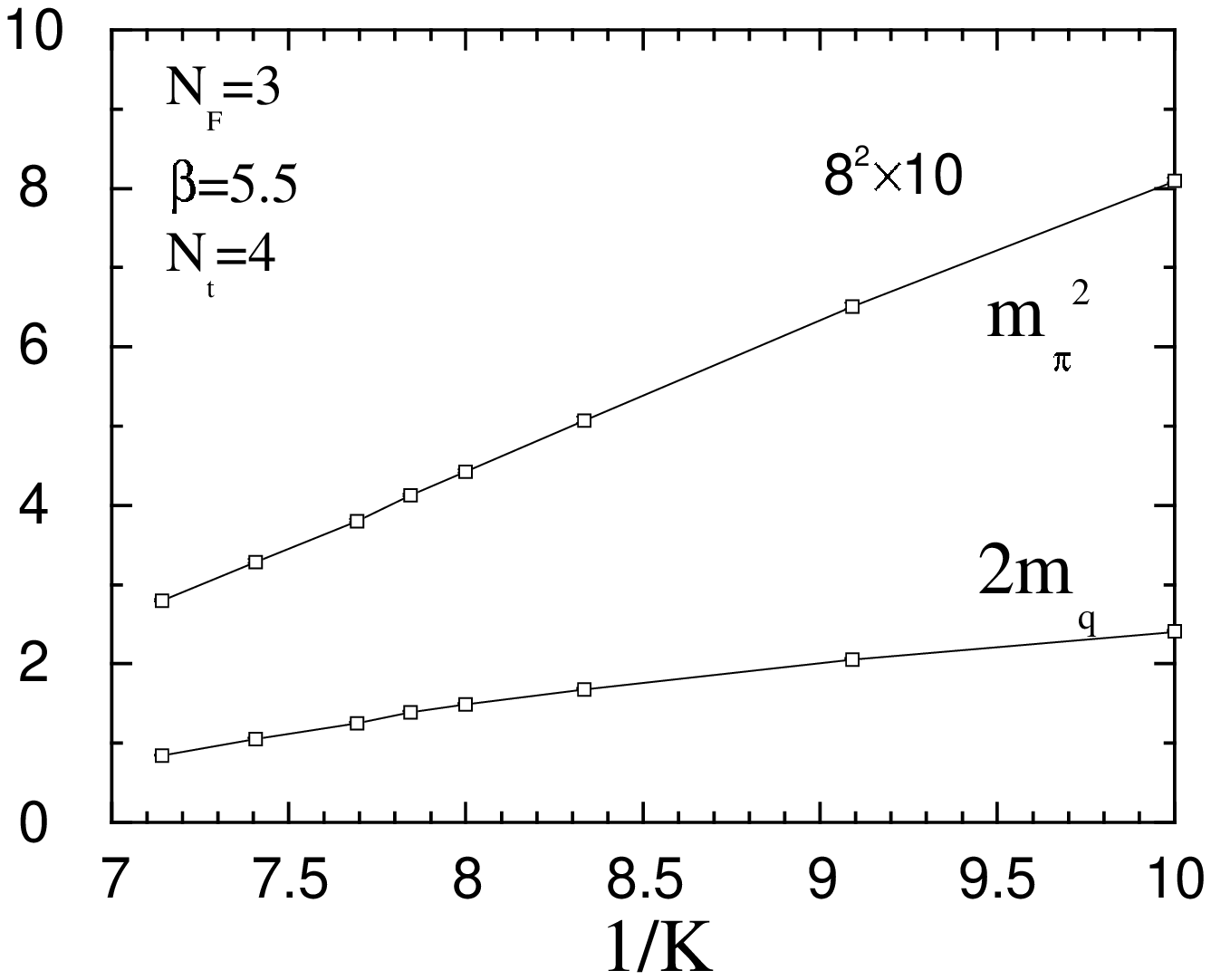} }
\centerline{ \epsfxsize=10cm \epsfbox{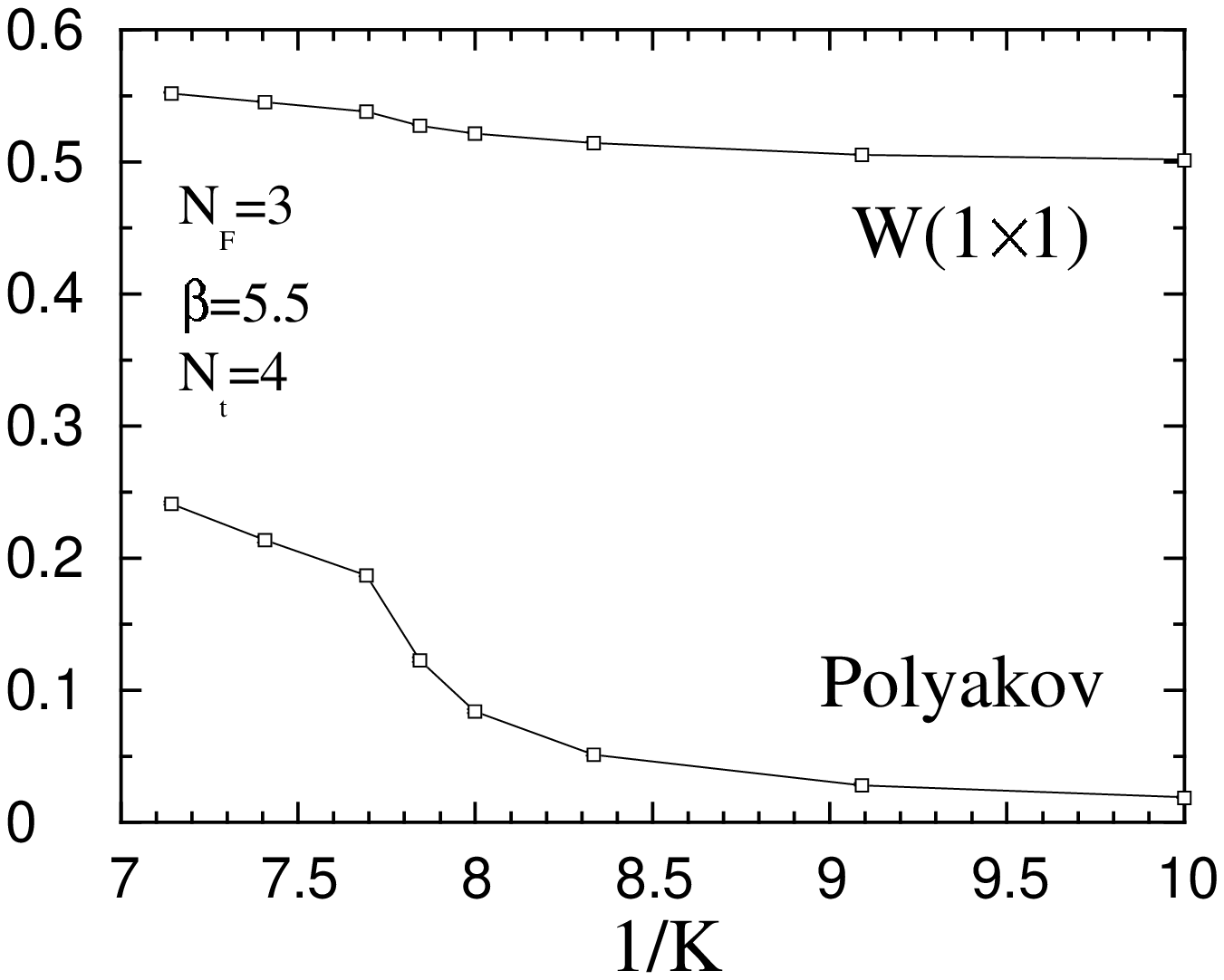} }
\caption{\baselineskip=.8cm
The same as Fig.~\protect\ref{fig:F3B4.0} at $\beta=5.5$ 
obtained on an $8^2\times10\times 4$ lattice.
The finite temperature crossover $K_t$ locates 
at $K \simeq 0.125$ --- 0.130 ($1/K \simeq 7.7$ --- 8.0). 
}
\label{fig:F3B5.5}
\end{figure}

Let us first discuss the case of the degenerate $N_F=3$:
$K_{u}=K_{d}=K_{s} \equiv K$.
As we have already discussed the chiral transition previously,
we are mainly interested in the transition for the massive quarks.
In order to find the transition points 
we perform simulations at $\beta$=4.0, 4.5, 4.7, 5.0 and 5.5.
The results for physical quantities are plotted in 
Figs.~\ref{fig:F3B4.0} --- \ref{fig:F3B5.5}.
The transition points identified by a sudden change of
physical observables are given in Table~\ref{tab:Kt} 
and plotted in Fig.~\ref{fig:F2F3KcKt}. 
We note that the $K_t$ line for $N_F=3$ at $N_t=4$ 
locates sufficiently far from the points where the $K_c$ line 
bends rapidly. 
This situation is quite different from the $N_F=2$ case
where the unusual relation between the $K_t$ line and $K_c$ line
causes the lattice artifacts.
Therefore, 
we expect that these lattice artifacts are small 
in the $N_F=3$ case.

In the previous section we have seen that the transition 
is of first order in the chiral limit 
$K_c=0.235$ at $\beta = 3.0$ for $N_F=3$.
For phenomenological applications, it is important to 
estimate the critical value
of the quark mass $m_q^{crit}$ up to which
the first order phase transition persists.

\begin{figure}
\centerline{ \epsfxsize=10.3cm \epsfbox{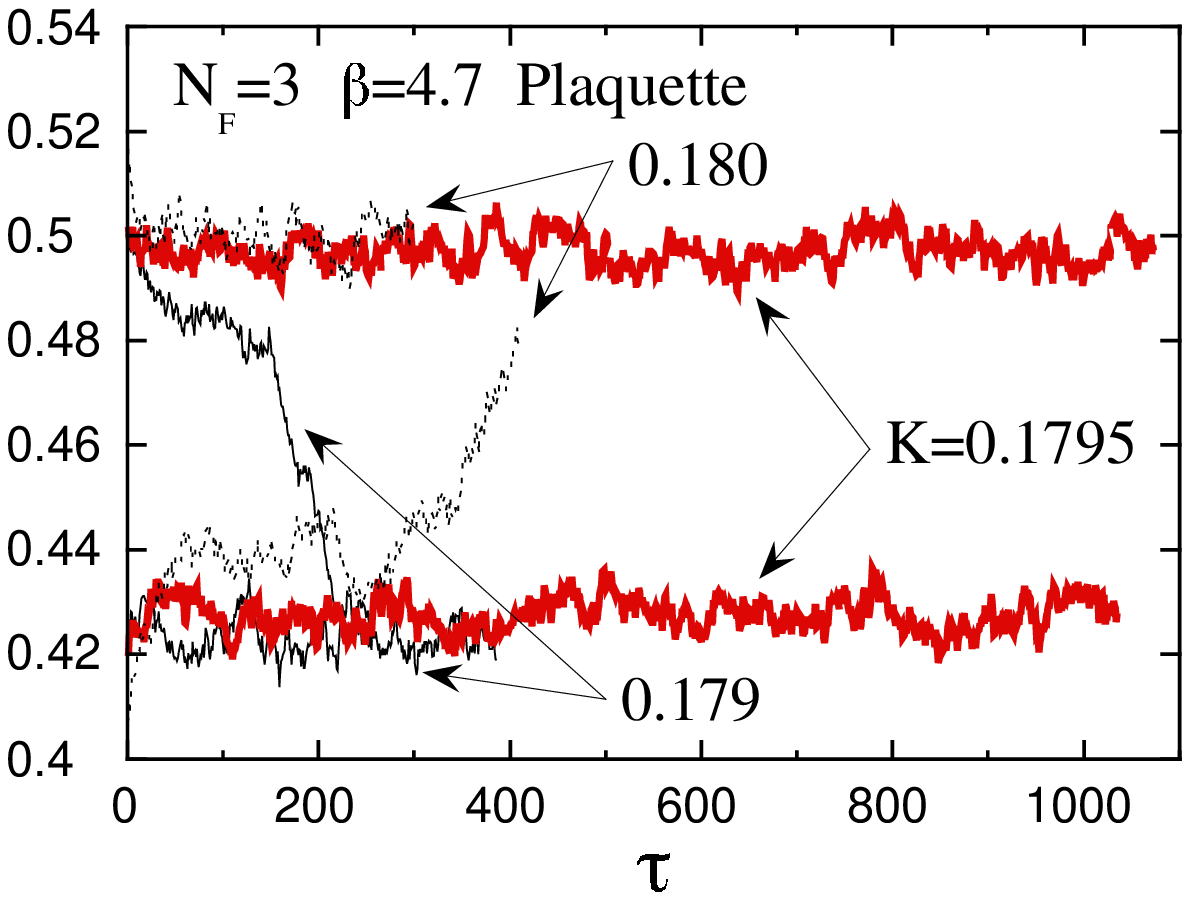} }
\centerline{\makebox[2mm]{} 
\epsfxsize=10.5cm \epsfbox{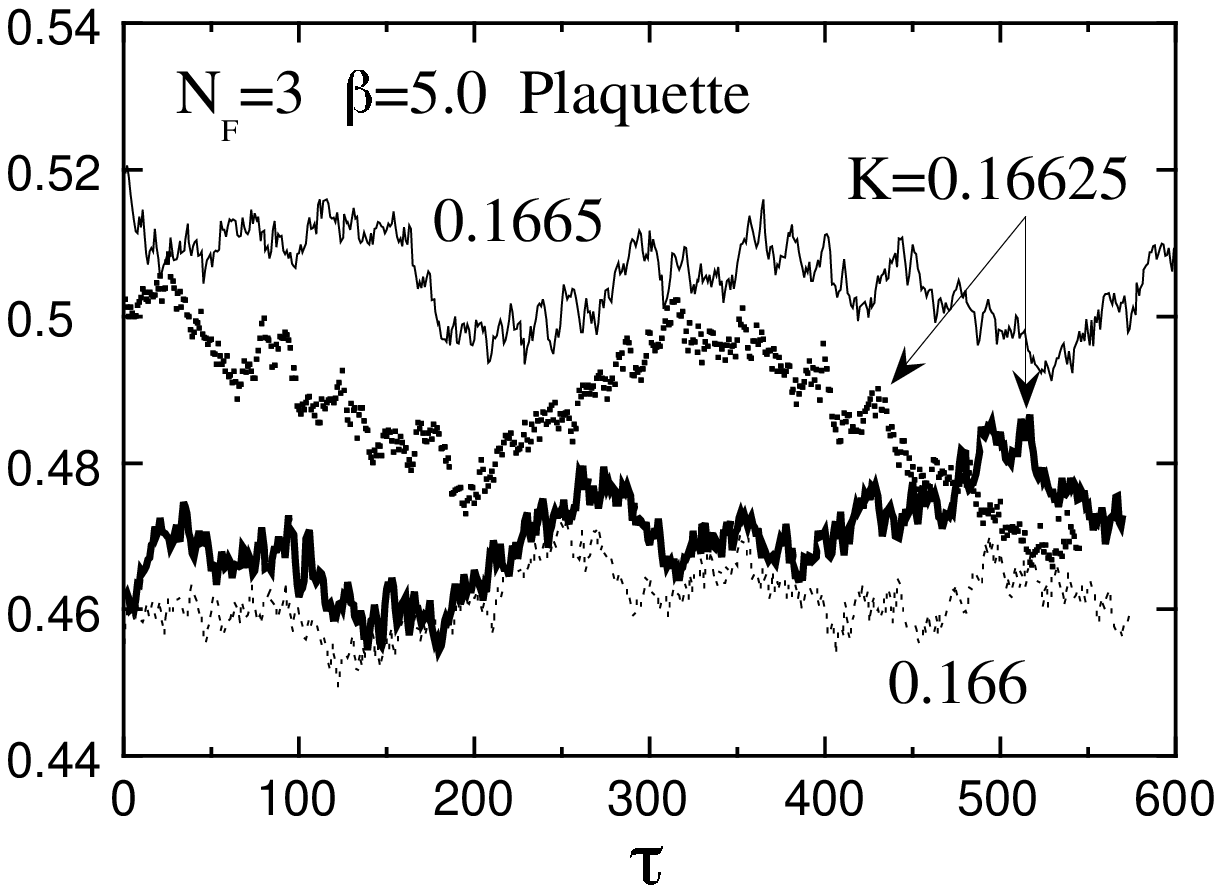}}
\caption{\baselineskip=.8cm
Time history of the plaquette for $N_F=3$ at (a) $\beta=4.7$ 
and (b) 5.0 on a $12^3\times4$ lattice.
}
\label{fig:H3W}
\end{figure}

We observe clear two state signals at 
$\beta$ = 4.0, 4.5 and 4.7, while 
for $\beta=5.0$ and $5.5$ no such signals have been seen:
The simulation time history of the plaquette 
at $\beta=4.7$ on a $12^3\times4$ lattice is plotted in 
Fig.~\ref{fig:H3W}(a). 
The confining and deconfining phases coexist over 1,000 
trajectories at $K=0.1795$ and, in accordance with this,
we find two-state signals also in other observables such as 
the plaquette and the pion screening mass $m_\pi$
(cf.\ Fig.~\ref{fig:F3B4.7}). 
From them we conclude that the transition at $K=0.1795(5)$ and
$\beta=4.7$ is first order. 
On the other hand, the time history of the plaquette at 
$\beta=5.0$ shown in Fig.~\ref{fig:H3W}(b) suggests that the 
transition is a crossover there.

\begin{figure}
\centerline{ \epsfxsize=14cm \epsfbox{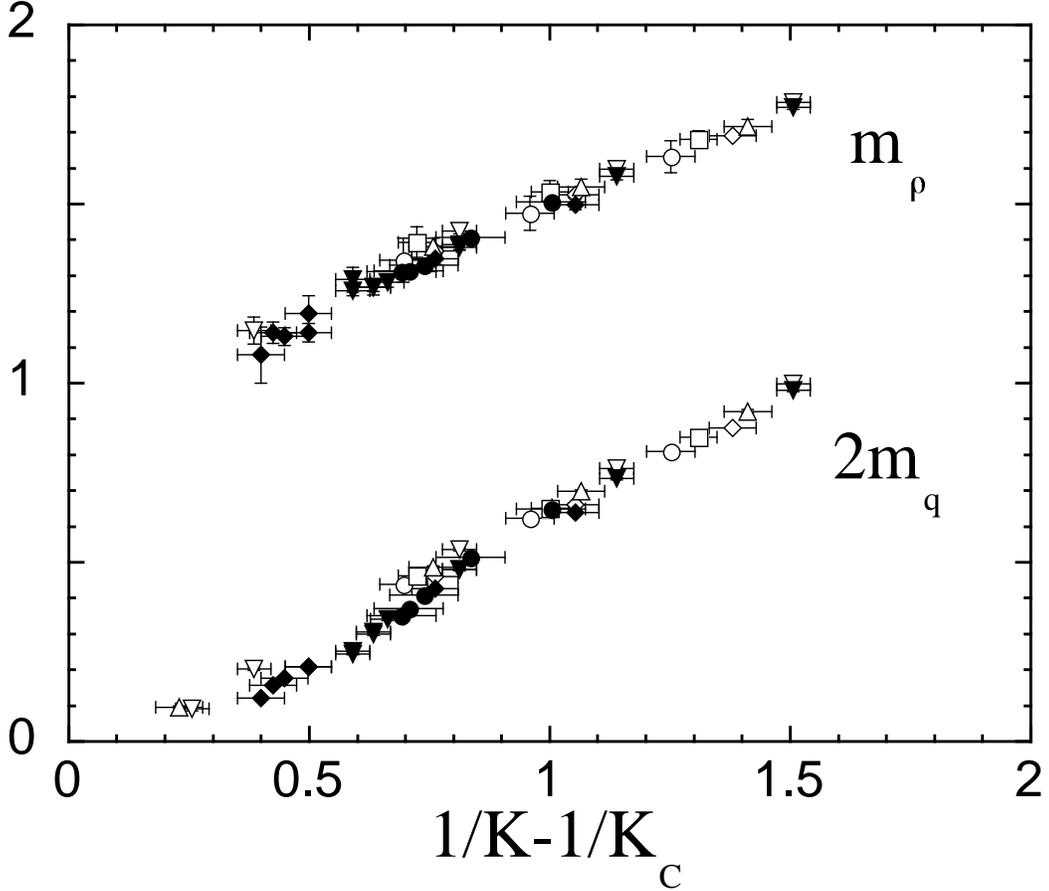} }
\caption{\baselineskip=.8cm
The rho meson screening mass $m_{\rho}a$ and 
twice the quark mass $2m_q a$ 
in the confining phase as a function of $1/K-1/K_c$. 
Open symbols are for $N_F=2$, $\beta =  3.0$, 
3.5, 4.0, 4.3, and 4.5 on an $8^2\times10\times4$ lattice. 
Filled symbols are for $N_F=3$, $\beta=4.0$, 4.5 and 4.7 on 
$8^2\times10\times4$ and $12^3\times4$ lattices. 
The values of $K_c(\beta)$ for $N_F=2$ is used. 
Horizontal errors are from those for 
$K_c$ with taking into account the difference due to 
definitions, either the vanishing point
of $m_{\pi}^2$ or $m_q$.
}
\label{fig:F2F3Rho}
\end{figure}

At the transition point (in the confining phase) 
of $\beta=4.7$
the value of $m_q a$ is 0.175(2) and $m_\pi /m_\rho=0.873(6)$.
The results of the hadron spectrum in the range of $\beta = 3.0$ 
-- 4.7 for $N_F=2$ and 3 (cf.\ Fig.~\ref{fig:F2F3Rho}) indicate 
that the inverse lattice spacing $a^{-1}$
estimated from the rho meson mass is almost independent on 
$\beta$ in this range and $a^{-1} \sim 0.8$ GeV.
(Hereafter we use $a^{-1}$ determined from $m_\rho$ in the 
chiral limit.)
Therefore we obtain a bound on the critical quark mass 
$m_q^{crit} 
\mathop{\vtop{\ialign{#\crcr
$\hfil\displaystyle{>}\hfil$\crcr
\noalign{\kern0.5pt\nointerlineskip}
$\sim$\crcr\noalign{\kern0.5pt}}}}\limits 
140$ MeV,
or equivalently $(m_{\pi}/m_{\rho})^{crit} \ge 0.873(6)$.
It should be noted that the physical strange quark mass 
determined from $m_\phi =1020$ MeV, 
using the data shown in Fig.~\ref{fig:F2F3Rho},
turns out to be
$m_s \sim 150$ MeV in this $\beta$ range
with our definition of the quark mass.

We note that these values for the critical quark mass 
are much larger than those with staggered quarks 
where $m_q^{crit} a=0.025$ --- 0.075 
\cite{Gavai,Columbia} 
($m_q^{crit} \sim 10$ --- 40 MeV 
using $a^{-1} \sim 0.5$ GeV at $\beta=5.2$ for 
$N_F=2$ \cite{Ukawa93}) 
which means that 
$(m_{\pi}/m_{\rho})^{crit} \simeq 0.42$ --- 0.58
(using the results of meson masses for $N_F=4$ at 
$\beta=5.2$ \cite{Born89}, 
because the data for $N_F=3$ are not available). 

\subsection{$N_F=2+1$}

Now let us discuss a more realistic case of 
massless up and down quarks and a light strange quark 
($N_F=2+1$).
Main results of measurements are summarized in 
Tables \ref{tab:resultsF21Nt4} --- \ref{tab:resultsF21Nt8}. 
Our strategy to study the phase structure
is similar to that applied in Sec.~\ref{sect:chiral} 
for the investigation of the chiral
transition in the degenerate quark mass cases,
which we called the on-$K_c$ method.
We set the value of masses for the up and down quarks 
$m_{ud}$ to zero ($K_{ud}= K_c$)
and fix the strange quark mass $m_s$ to some value,
and make simulations starting 
from a value of $\beta$ in the deconfining phase and
reducing the value of $\beta$.
When u and d quarks are massless, the number of iteration 
$N_{\rm inv}$
needed for the quark matrix inversion (for u and d quarks) 
is enormously large in the confining phase, 
while it is of order of several hundreds 
in the deconfining phase. 
The values which we take for $K_c$ are given in Table~\ref{tab:Ks}.
They are the vanishing point of extrapolated $m_\pi^2$ for 
$N_F=2$ and interpolated ones.
We have used those for $N_F=2$, because we have the data most 
in this case,
and the difference between that for $N_F=2$ and 3 is
of the same order of magnitude as the difference
due to the definition of $K_c$ (cf.\ discussions in 
Sec.~\ref{sect:chiral}).

We study two cases of $m_s \sim 150$ MeV and 400 MeV. 
From the value of $a^{-1} \sim 0.8$ GeV and 
an empirical rule $m_q a \simeq (2/3)(1/K-1/K_c)$ 
satisfied for $N_F=2$ and 3 in the $\beta$ region we have studied
(cf.\ Fig.~\ref{fig:F2F3Rho}),
we get the values for $K_s$ shown in Table~\ref{tab:Ks}.

In order to confirm that our choice of parameters for 
the case $m_s \sim 150$ MeV is really close to the physical values, 
we have also made a zero-temperature
spectroscopy calculation for the $N_F=2+1$ case
at $\beta=3.5$ on an $8^3 \times 10$ lattice. 
Keeping $K_s=0.2017$ ($m_s \sim 150$ MeV), 
we vary $K_{ud}$ from 0.195 to 0.210 in steps of 0.005. 
Taking the chiral limit of $K_{ud}$, we obtain
$a^{-1}=903(38)$ MeV from the rho meson mass 
($m_\rho a=0.853(36)$ at $K_c=0.2227$, 
where $K_c$ is determined by a linear extrapolation of 
$m_\pi^2 a^2$ in terms of $1/K$).  
The mass of the $\phi$ meson at the simulation point 
turns out to be 1.03(5) GeV which should be compared 
with the physical value 1.02 GeV.
Thus the hopping parameter chosen for $m_s \sim 150$ MeV 
corresponds to the physical strange quark mass, in this sense. 
As far as we consider the meson sector the numerical results 
for mass ratio do not differ so much from the physical values.
However, we emphasize one caveat here. 
The nucleon-rho mass ratio $m_N/m_\rho$ turns out to be 2.0(1) 
which is the same as the result 2.0 in the strong coupling limit 
and is much larger than the physical value 1.22. 
This implies that $\beta$=3.5 is far from the continuum limit. 

\begin{figure}
\centerline{ \epsfxsize=10.5cm \epsfbox{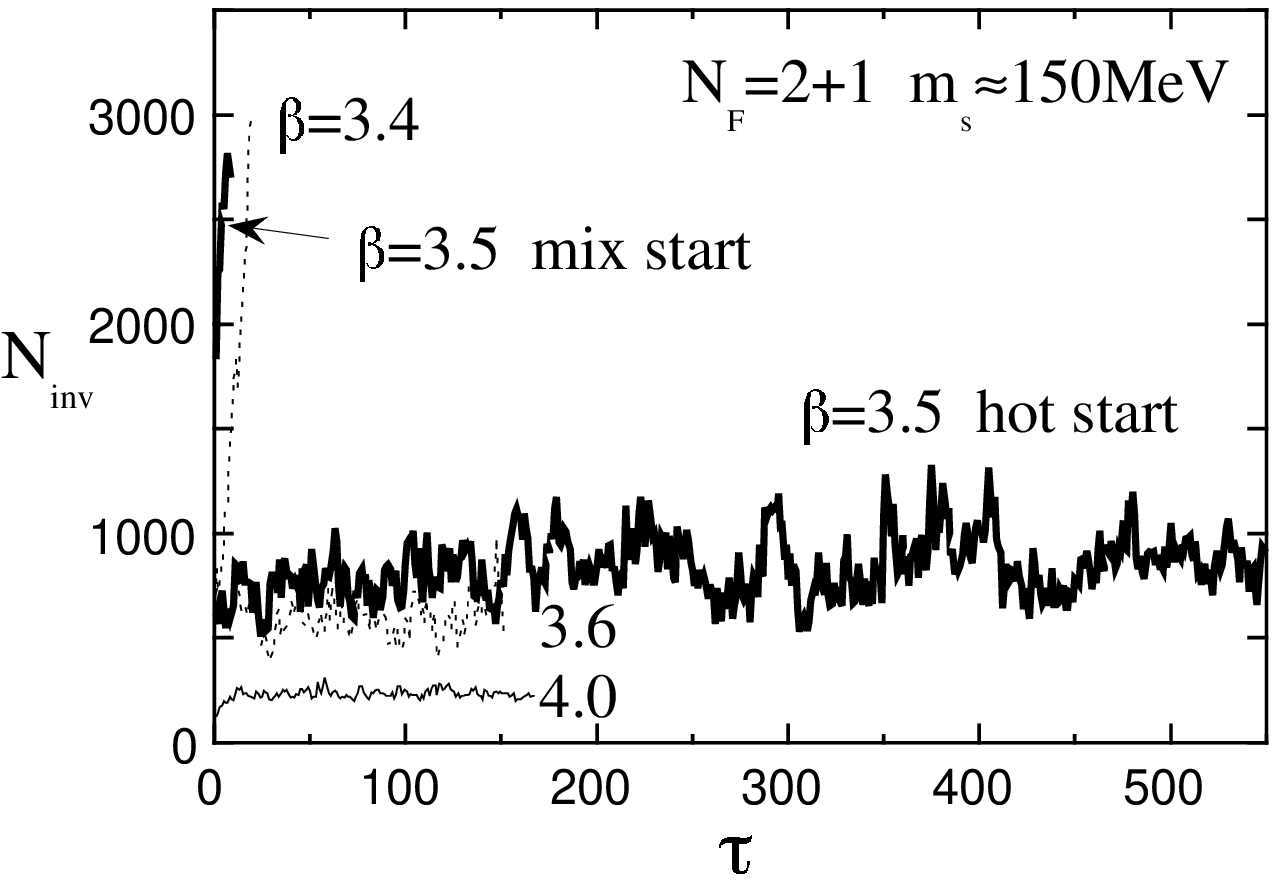} }
\centerline{\makebox[2mm]{} 
\epsfxsize=10.3cm \epsfbox{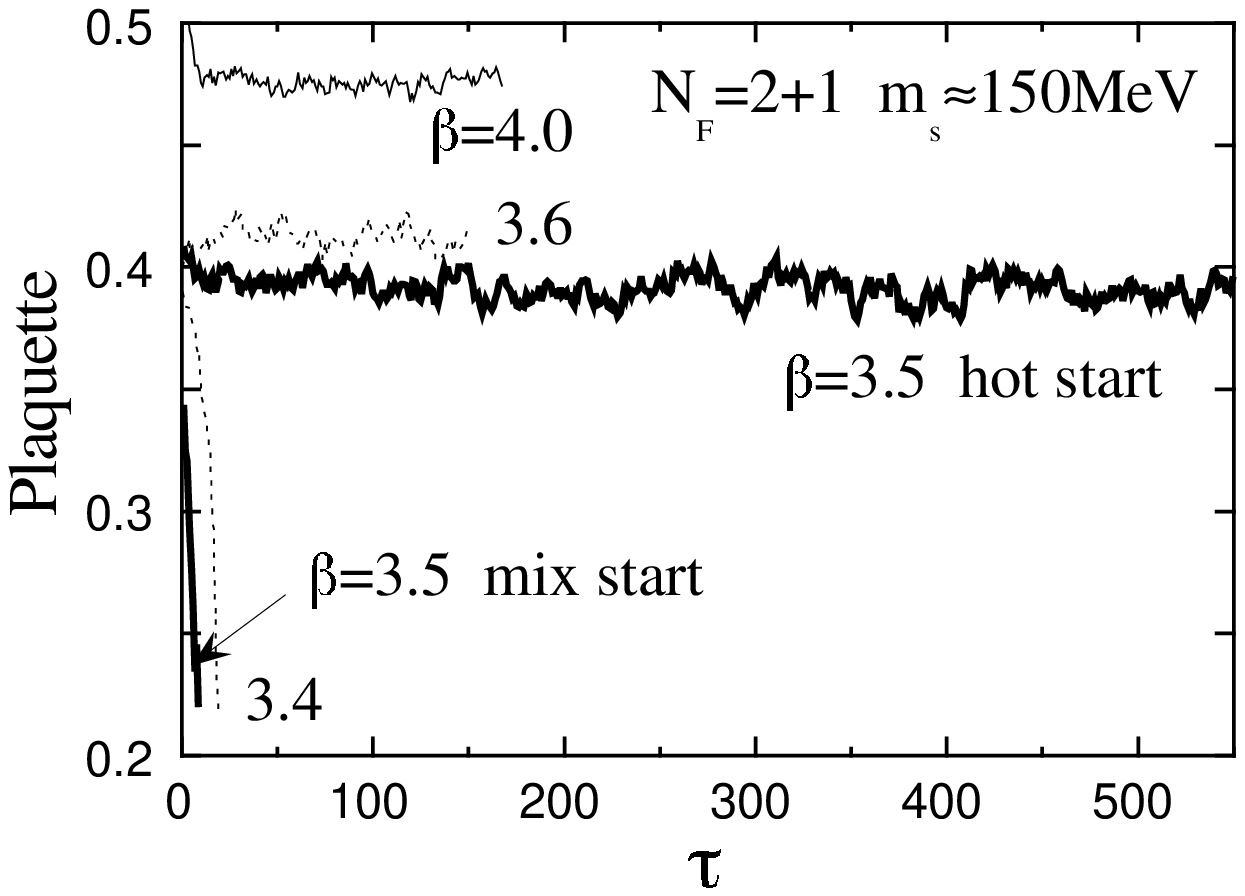}}
\caption{\baselineskip=.8cm
Time history of (a) $N_{\rm inv}$ and (b) the plaquette for 
$m_s \sim 150$ MeV on an $8^2\times10\times4$ lattice.
}
\label{fig:H21S150}
\end{figure}

\begin{figure}
\centerline{ \epsfxsize=14cm \epsfbox{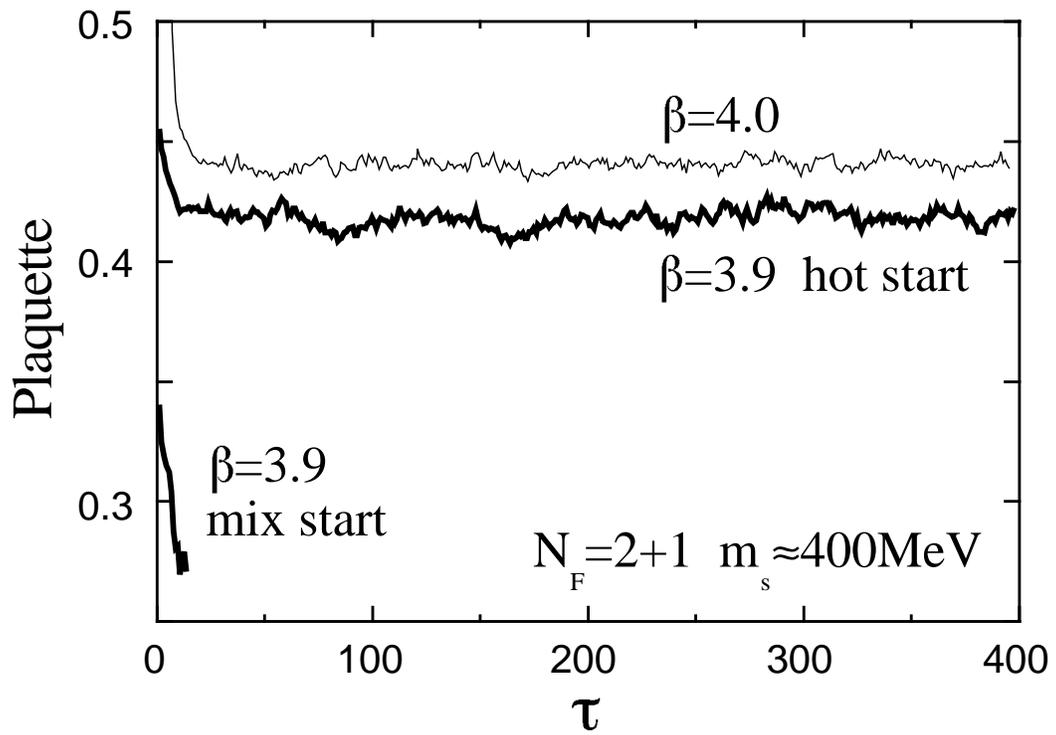} }
\caption{\baselineskip=.8cm
Time history of the plaquette for 
$m_s \sim 400$ MeV on a $12^3\times4$ lattice.
}
\label{fig:H21S400}
\end{figure}

\begin{figure}
\centerline{ \epsfxsize=14cm \epsfbox{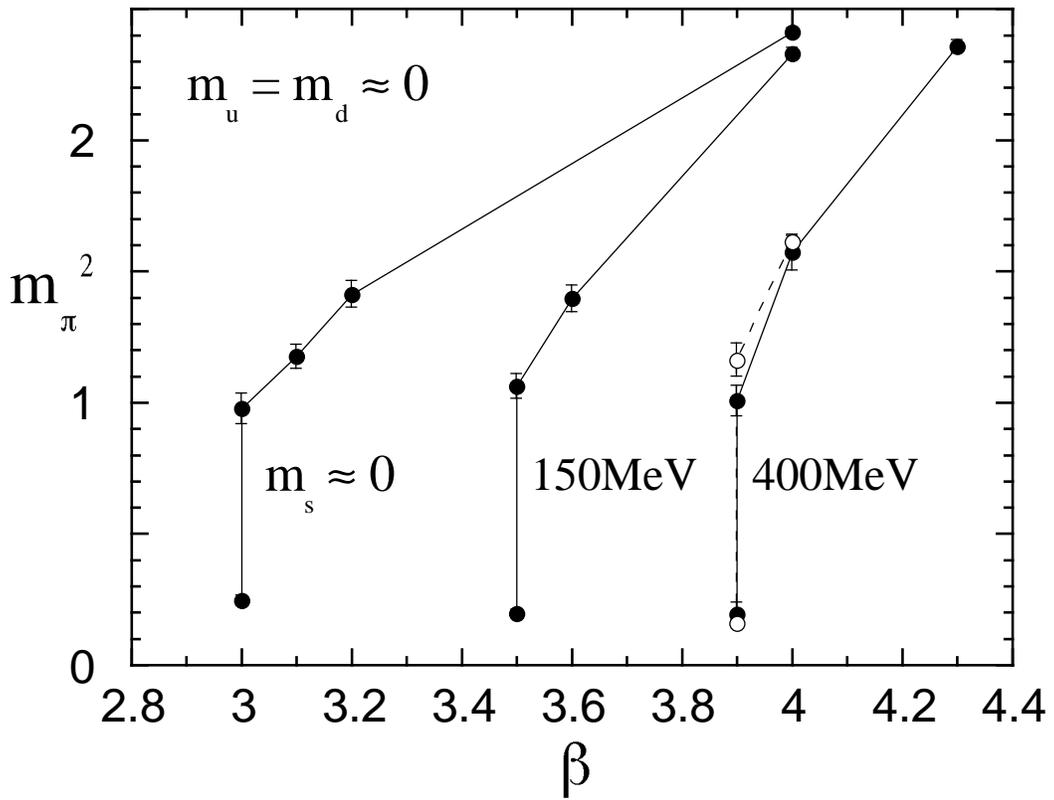} }
\caption{\baselineskip=.8cm
$m_\pi^2 a^2$ versus $\beta$ for $m_s \simeq 0$, 150 and 400 
MeV with $m_{ud} \simeq 0$. 
Filled and open symbols are for $8^2 \times 10 \times 4$ 
and $12^3 \times 4$ lattices, respectively. 
}
\label{fig:F321Pi}
\end{figure}

The simulation time history of $N_{\rm inv}$ on the 
$8^2 \times 10$ spatial lattice is plotted in 
Fig.~\ref{fig:H21S150}(a) for the case of 150 MeV.
When $\beta \ge 3.6$, $N_{\rm inv}$ is of order of several
hundreds, while when $\beta \le 3.4$, $N_{\rm inv}$
shows a rapid increase with $\tau$.
At $\beta = 3.5$ we see a clear two-state signal
depending on the initial condition:
For a hot start, $N_{\rm inv}$
is quite stable around 900 and $m_\pi^2$ is large
($\sim 1.0$ in lattice units).
On the other hand, for a mix start, $N_{\rm inv}$
shows a rapid increase with $\tau$ and exceeds
2,500 in $\tau \sim 10$, and in accord with this, 
the plaquette and $m_\pi^2$
decreases with $\tau$ as shown in Fig.~\ref{fig:H21S150}(b) 
for the plaquette. 
For the case of 400 MeV a similar clear two-state signal 
is observed at $\beta=3.9$
both on the $8^2 \times 10$ and $12^3$ spatial lattices 
(cf.\ Fig.~\ref{fig:H21S400}). 
The values of $m_\pi^2$ versus $\beta$ are plotted in 
Fig.~\ref{fig:F321Pi} together with those in 
the case of degenerate $N_F=3$ on the $K_c$ line.
At $\beta =3.5$ for the case of 150 MeV and at $\beta=3.9$ for 
the case of 400 MeV,
we have two values for $m_\pi^2$ depending
on the initial configuration.
The larger ones of order 1.0 are for hot starts, 
while the smaller ones are upper bounds for 
mix starts.
These results imply that $m_s^* 
\mathop{\vtop{\ialign{#\crcr
$\hfil\displaystyle{>}\hfil$\crcr
\noalign{\kern0.5pt\nointerlineskip}
$\sim$\crcr\noalign{\kern0.5pt}}}}\limits
400\, {\rm MeV}
$ in our normalization for quark masses.

\begin{figure}
\centerline{ \epsfxsize=12cm \epsfbox{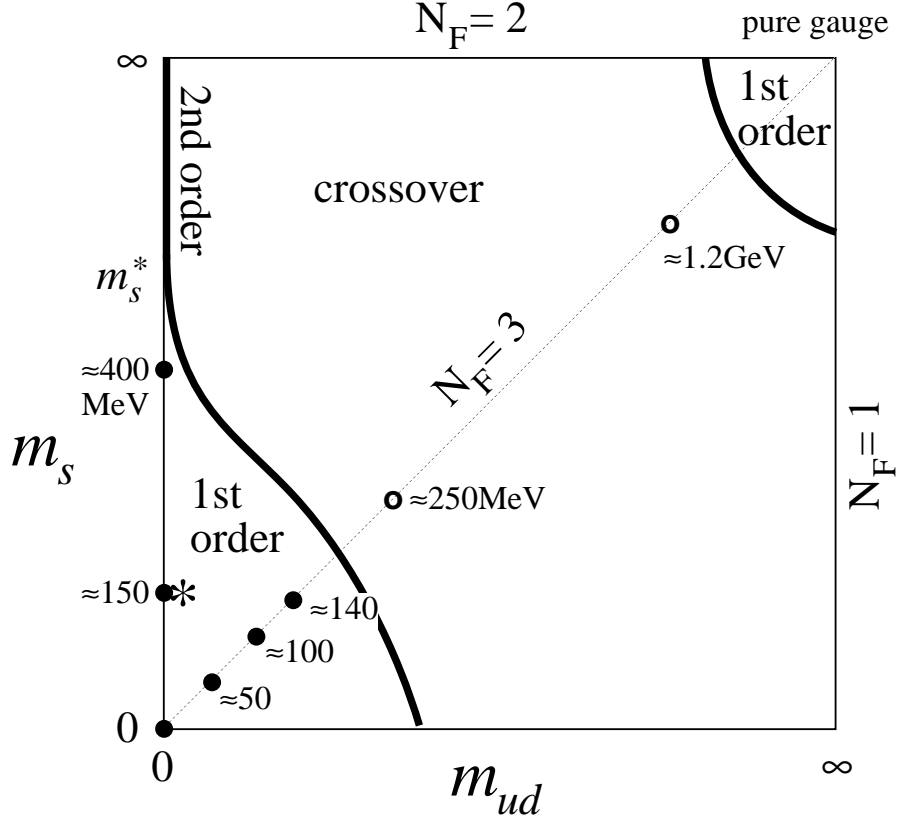} }
\caption{\baselineskip=.8cm
Order of the finite temperature QCD transition in the 
$(m_{ud},m_s)$ plane.
First order signals are observed at the points marked with 
filled circle, while no clear two state signals are found 
at the points with open circle. 
The second order transition line is 
suggested \protect\cite{Rajagopal95}
to deviate from 
the vertical axis as $m_{ud} \propto (m_s^{*} - m_s)^{5/2}$
below $m_s^{*}$.
The values of quark mass in physical units are computed 
using $a^{-1}$ determined from $m_\rho$:
$a^{-1} \sim 0.8$ GeV for $\beta \leq 4.7$ and 
$\sim 1.0 (1.8)$ GeV for $\beta = 5.0 (5.5)$.
See Sec.~\protect\ref{sect:strange} for more detailed discussion
on the values of the quark mass in physical units.
The real world determined by the value of 
$m_\phi/m_\rho$ and $m_\pi/m_\rho$
corresponds to the point marked with star. 
}
\label{fig:MsMud}
\end{figure}

Following the Columbia group \cite{Columbia}, 
we summarize our results about the nature of the 
QCD transition at $N_t=4$ as a function of $m_{ud}$ and $m_s$ 
in Fig.~\ref{fig:MsMud}, 
together with theoretical expectations 
\cite{Pisarski,Wilczek,Rajagopal95} 
assuming that the chiral transition is of second order for 
$N_F=2$. 
Clearly the point which corresponds to the physical values of 
the up, down and strange quark masses measured by 
$m_\phi/m_\rho$ and $m_\pi/m_\rho$
exists in the range of the first order transition.
If this situation persists in the continuum limit,
the transition for the physical quark masses is of 
first order.

The Columbia group studied the influence of the strange quark 
for the case of staggered quarks \cite{Columbia}. 
Their result shows that no transition occurs
at $m_u a = m_d a = 0.025$, $m_s a = 0.1$
($m_u=m_d \sim 12$ MeV, $m_s \sim 50$ MeV using 
$a^{-1} \sim 0.5$ GeV). 
Their zero-temperature values for $m_K/m_\rho$ and $m_\pi/m_\rho$ 
obtained at this simulation point suggest that 
this value for $m_s$ is smaller than its physical value 
and those for $m_u$ and $m_d$ are larger than their physical 
values.
This implies that the transition in the real world is also 
a crossover, 
unless the second order transition line, 
which has a sharp $m_{ud}$ dependence 
near $m_s^*$ as shown in Fig.~\ref{fig:MsMud} \cite{Rajagopal95}, 
crosses between the physical point and the simulation point. 

Although both staggered and Wilson simulations give phase 
structures qualitatively consistent with theoretical 
expectations \cite{Pisarski,Wilczek,Rajagopal95},
we note that 
Wilson quarks tend to give larger values for critical 
quark masses (measured by $m_\phi / m_\rho$ etc.) 
than those with staggered quarks. 
This leads to the difference in the conclusions 
about the nature of the physical transition. 
However, 
since the deviation from the continuum limit is 
large in both of the studies at $N_t=4$, 
we certainly should make a calculation with larger $N_t$ 
\cite{KogutSW91} 
or using an improved action \cite{ourImprove} 
to get closer to the continuum limit 
and to obtain a definite conclusion about 
the nature of the QCD transition. 
With Wilson quarks using the standard gauge action, however, 
$N_t$ should be enormously large ($\ge 18$) \cite{ourOldLatC} 
in order to avoid the lattice 
artifacts discussed in Sec.~\ref{sect:problems}. 
Improvement of the lattice action will be essential 
especially for Wilson quarks. 

\section{Conclusions}
\label{sect:conclusions}

We have studied the nature of finite temperature
transitions near the chiral limit 
for various numbers of flavors ($N_F=2$, 3, and 6)
and also for the case of massless up and down quarks and a 
light strange quark ($N_F=2+1$),
mainly on lattices with $N_t=4$,
using the Wilson formalism of quarks on the lattice.

We have found that the chiral transition is continuous
(second order or crossover) for $N_F=2$, 
while it is of first order for $N_F=3$ and 6. 
These results are in accordance with theoretical predictions
based on universality \cite{Pisarski,Wilczek}. 
Our results with Wilson quarks are also consistent with 
those with staggered quarks \cite{KSfl}. 

Our results 
for QCD with a strange quark as well as up and down quarks 
obtained on $N_t=4$ lattices 
are summarized in Fig.~\ref{fig:MsMud}. 
Clearly, the point which corresponds to the physical values of 
the up, down and strange quark masses measured by 
$m_\phi/m_\rho$ and $m_\pi/m_\rho$,
marked with star in Fig.~\ref{fig:MsMud},
exists in the range of first order transition.
If this situation persists in the continuum limit,
the transition for the physical quark masses is of 
first order.

We have found that 
Wilson quarks tend to give larger values for critical 
quark masses 
(measured, for example, by $m_\phi/m_\rho$ and $m_\pi/m_\rho$ ) 
than those with staggered quarks. 
This leads to the difference in the conclusions 
about the nature of the physical transition. 
Because the deviation from the continuum limit is 
large on the $N_t=4$ lattices, 
we certainly should make a calculation with larger $N_t$ 
or with an improved action \cite{ourImprove} 
in order to get closer to the continuum limit 
and to obtain a definite conclusion about the nature 
of the physical QCD transition, by resolving the discrepancy
between Wilson and staggered quarks for the conclusions. 
Studies with an improved gauge action and the Wilson quark 
action are in progress.

\section*{Acknowledgements}

The simulations have been performed with
HITAC S820/80 at the National Laboratory for High Energy Physics 
(KEK), 
Fujitsu VPP500/30 at the Science Information Processing Center of 
the University of Tsukuba, and HITAC H6080-FP12 at the 
Center for Computational Physics of the University of Tsukuba.
We would like to thank members of KEK
for their hospitality and strong support
and we also thank Sinya Aoki and Akira Ukawa 
for valuable discussions.
This project is in part supported by the Grants-in-Aid
of Ministry of Education, Science and Culture
(Nos.07NP0401, 07640375 and 07640376).


\clearpage

\begin{table}
\begin{center}
\begin{tabular}{lllrrcrc}
\hline
$\beta$ & $K$ & $\Delta\tau$ & $\tau_{\rm tot}$ & 
$\tau_{\rm therm}$ & algo.\ & $N_{\rm inv}$ & phase \\
\hline
0 & 0.2 & 0.02 & 1132 & 500 & H-CR & 37 & c \\
0 & 0.21 & 0.01 & 1005 & 500 & H-CR & 48 & c \\
0 & 0.22 & 0.01 & 1041 & 500 & H-CR & 45 & c \\
0 & 0.23 & 0.01 & 700 & 500 & H-CR & 95 & c \\
3 & 0.18 & 0.01 & 250 & 100 & H-CR & 37 & c \\
3 & 0.19 & 0.01 & 150 & 100 & H-CR & 35 & c \\
3 & 0.2 & 0.01 & 160 & 100 & H-CR & 48 & c \\
3.5 & 0.175 & 0.01 & 160 & 100 & H-CR & 27 & c \\
3.5 & 0.185 & 0.01 & 160 & 100 & H-CR & 34 & c \\
3.5 & 0.195 & 0.01 & 160 & 100 & H-CR & 46 & c \\
4 & 0.17 & 0.02 & 1650 & 500 & H-CR & 15 & c \\
4 & 0.18 & 0.02 & 2188 & 1000 & H-CR & 18 & c \\
4 & 0.19 & 0.02 & 1550 & 500 & H-CR & 23 & c \\
4 & 0.2226 & 0.002 & 50 & 24 & H-CG & 1054 & d \\
4.1 & 0.2211 & 0.005 & 92 & 50 & H-CG & 781 & d \\
4.2 & 0.2195 & 0.005 & 206 & 100 & H-CG & 430 & d \\
\hline
\end{tabular}
\caption{\baselineskip=.8cm
Table of job parameters for $N_F=2$ simulations performed 
on an $8^2\times10\times4$ lattice.
Data marked with \dag \ are taken from our previous simulation 
\protect\cite{ChNf2} performed on an $8^2\times20\times4$ lattice. 
The column ``algo.'' is for the algorithm used for update 
(HMC or R) and for quark matrix inversion (CR or CG).
$N_{\rm inv}$ is an average number of iterations 
needed for the quark matrix inversion. 
Errors for $N_{\rm inv}$ are in most cases about 1\%. 
The last column is for the initial and final phases
(c: the low temperature confining phase, d: the high temperature 
deconfining phase, and m: mix state), 
where parentheses mean that the system is not completely 
thermalized. 
}
\protect\label{tab:paramF2Nt4}
\end{center}
\end{table}
\clearpage
\addtocounter{table}{-1}
\begin{table}
\begin{center}
\begin{tabular}{lllrrcrc}
\hline
$\beta$ & $K$ & $\Delta\tau$ & $\tau_{\rm tot}$ & 
$\tau_{\rm therm}$ & algo.\ & $N_{\rm inv}$ & phase \\
\hline
4.3 & 0.165 & 0.02 & 520 & 320 & H-CR & 23 & c \\
4.3 & 0.175 & 0.01 & 490 & 290 & H-CR & 28 & c \\
4.3 & 0.185 & 0.01 & 400 & 200 & H-CR & 39 & c \\
4.3 & 0.205 & 0.008 & 460 & 250 & H-CR & 250 & d$\rightarrow$c \\
4.3 & 0.207 & 0.005 & 16 &  & H-CG &  & c \\
4.3 & 0.207 & 0.005 & 30 &  & H-CG &  & d$\rightarrow$(c) \\
4.3 & 0.208 & 0.005 & 38 &  & H-CG &  & c \\
4.3 & 0.208 & 0.005 & 45 &  & H-CG &  & d$\rightarrow$(c) \\
4.3 & 0.21 & 0.005 & 150 & 50 & H-CG & 820 & d \\
4.3 & 0.218 & 0.01 & 196 & 100 & H-CG & 338 & d \\
4.5 & 0.16 & 0.02 & 500 & 300 & H-CR & 25 & c \\
4.5 & 0.17 & 0.01 & 580 & 300 & H-CR & 34 & c \\
4.5 & 0.18 & 0.01 & 530 & 300 & H-CR & 42 & c \\
4.5 & 0.195 & 0.01 & 310 & 100 & H-CR & 92 & c \\
4.5 & 0.2 & 0.005 & 175 & 135 & H-CR & 280 & c \\
4.5 & 0.202 & 0.008 & 700 & 300 & H-CG & 473 & d \\
4.5 & 0.205 & 0.01 & 190 & 100 & H-CG & 314 & d \\
4.5 & 0.2143 & 0.01 & 197 & 100 & H-CG & 209 & d \\
5 & 0.14 & 0.02 & 500 & 300 & H-CR & 17 & c \\
5 & 0.15 & 0.02 & 520 & 300 & H-CR & 20 & c \\
5 & 0.16 & 0.02 & 600 & 300 & H-CR & 24 & c \\
5 & 0.17 & 0.01 & 540 & 300 & H-CR & 41 & d$\rightarrow$c \\
5 & 0.18 & 0.01 & 640 & 200 & H-CG & 169 & c$\rightarrow$d \\
5 & 0.19 & 0.01 & 720 & 300 & H-CG & 132 & d \\
5 & 0.1982 & 0.01 & 761 & 300 & H-CG & 118 & c$\rightarrow$d \\
\hline
\end{tabular}
\caption{\baselineskip=.8cm
{\it Continued.}
}
\end{center}
\end{table}
\clearpage
\addtocounter{table}{-1}
\begin{table}
\begin{center}
\begin{tabular}{lllrrcrc}
\hline
$\beta$ & $K$ & $\Delta\tau$ & $\tau_{\rm tot}$ & 
$\tau_{\rm therm}$ & algo.\ & $N_{\rm inv}$ & phase \\
\hline
5.25 & 0.1 & 0.01 & 520 & 300 & H-CR & 12 & c \\
5.25 & 0.11 & 0.01 & 600 & 300 & H-CR & 13 & c \\
5.25 & 0.12 & 0.01 & 600 & 300 & H-CR & 15 & c \\
5.25 & 0.13 & 0.01 & 560 & 300 & H-CR & 17 & c \\
5.25 & 0.14 & 0.01 & 580 & 300 & H-CR & 20 & c \\
5.25 & 0.15 & 0.01 & 520 & 300 & H-CR & 25 & c \\
5.25 & 0.155 & 0.01 & 520 & 300 & H-CR & 31 & d$\rightarrow$c \\
5.25 & 0.16 & 0.01 & 540 & 300 & H-CR & 39 & d$\rightarrow$c \\
5.25 & 0.165 & 0.01 & 600 & 300 & H-CG & 121 & d \\
5.25 & 0.175 & 0.01 & 610 & 300 & H-CG & 118 & d \\
5.25 & 0.18 & 0.01 & 640 & 300 & H-CG & 111 & d \\
5.5\dag & 0.15 & 0.025 & 2500 & 800 & H-CR & 17 & d \\
5.5\dag & 0.16 & 0.025 & 1572 & 500 & H-CR & 37 & d \\
5.5\dag & 0.1615 & 0.025 & 1532 & 500 & H-CR & 43 & d \\
5.5\dag & 0.163 & 0.025 & 1458 & 500 & H-CR & 53 & d \\
6 & 0.15 & 0.01 & 427 & 200 & H-CG & 73 & d \\
6 & 0.1524 & 0.01 & 230 & 150 & H-CG & 78 & d \\
6 & 0.155 & 0.01 & 427 & 200 & H-CG & 80 & d \\
6 & 0.16 & 0.01 & 400 & 200 & H-CG & 83 & d \\
10 & 0.13 & 0.01 & 351 & 200 & H-CG & 48 & d \\
10 & 0.14 & 0.01 & 400 & 200 & H-CG & 78 & d \\
10 & 0.15 & 0.01 & 338 & 200 & H-CG & 60 & d \\
\hline
\end{tabular}
\caption{\baselineskip=.8cm
{\it Continued.}
}
\end{center}
\end{table}

\clearpage

\begin{table}
\begin{center}
\begin{tabular}{lllrrcrc}
\hline
$\beta$ & $K$ & $\Delta\tau$ & $\tau_{\rm tot}$ & 
$\tau_{\rm therm}$ & algo.\ & $N_{\rm inv}$ & phase \\
\hline
4.2 & 0.2195 & 0.00125 & 56 & 30 & H-CG & 1119 & d  \\
4.3 & 0.2183 & 0.002 & 138 & 40 & H-CG & 863 & d \\
4.4 & 0.2163 & 0.005 & 160 & 30 & H-CG & 678 & d \\
4.5 & 0.2143 & 0.008 & 130 & 80 & H-CG & 505 & d \\
5 & 0.1982 & 0.01 & 224 & 100 & H-CG & 160 & d \\
5.02 & 0.16 & 0.01 & 560 & 300 & H-CR & 27 & c \\
5.02 & 0.17 & 0.01 & 560 & 300 & H-CR & 36 & c \\
5.02 & 0.18 & 0.01 & 180 & 100 & H-CG & 143 & c \\
5.02 & 0.18 & 0.01 & 210 & 100 & H-CG & 529 & d \\
\hline
\end{tabular}
\caption{\baselineskip=.8cm
The same as Table~\protect\ref{tab:paramF2Nt4} 
for $N_F=2$ simulations performed on a $12^3\times6$ lattice.
}
\protect\label{tab:paramF2Nt6}
\end{center}
\end{table}


\begin{table}
\begin{center}
\begin{tabular}{lllrrcrc}
\hline
$\beta$ & $K$ & $\Delta\tau$ & $\tau_{\rm tot}$ & 
$\tau_{\rm therm}$ & algo.\ & $N_{\rm inv}$ & phase \\
\hline
4.5 & 0.16 & 0.02 & 500 & 300 & H-CR & 23 & c \\
4.5 & 0.17 & 0.01 & 540 & 300 & H-CR & 29 & c \\
4.5 & 0.18 & 0.01 & 540 & 300 & H-CR & 35 & c \\
5.5\dag & 0.15 & 0.025 & 2050 & 1000 & H-CR & 8 & c \\
5.5\dag & 0.155 & 0.02 & 1600 & 500 & H-CR & 23 & c \\
6 & 0.1524 & 0.01 & 230 & 150 & H-CG & 78 & d \\
\hline
\end{tabular}
\caption{\baselineskip=.8cm
The same as Table~\protect\ref{tab:paramF2Nt4} 
for $N_F=2$ on an $8^3\times10$ lattice.
Data marked with \dag \ are taken from 
Ref.~\protect\cite{ChNf2} 
obtained on an $8^3\times20$ lattice. 
}
\protect\label{tab:paramF2Nt8}
\end{center}
\end{table}

\clearpage

\begin{table}
\begin{center}
\begin{tabular}{lllrrcrc}
\hline
$\beta$ & $K$ & $\Delta\tau$ & $\tau_{\rm tot}$ & 
$\tau_{\rm therm}$ & algo.\ & $N_{\rm inv}$ & phase \\
\hline
2.5 & 0.2381 & 0.01 & 8 &  & R-CG & $\sim$2300 & d$\rightarrow$(c) \\
2.7 & 0.2369 & 0.01 & 10 &  & R-CG & $\sim$2300 & d$\rightarrow$(c) \\
2.8 & 0.2364 & 0.01 & 12 &  & R-CG & $>$1900 & d$\rightarrow$(c) \\
2.9 & 0.2358 & 0.01 & 28 &  & R-CG & $\sim$2300 & d$\rightarrow$(c) \\
3 & 0.205 & 0.01 & 280 & 170 & R-CR & 64 & c \\
3 & 0.205 & 0.01 & 202 & 100 & R-CR & 64 & c \\
3 & 0.215 & 0.01 & 190 & 100 & R-CR & 117 & c \\
3 & 0.225 & 0.005 & 75 &  & R-CR & $\sim$563 & d$\rightarrow$(c) \\
3 & 0.23 & 0.0025 & 18 &  & R-CG &  & d$\rightarrow$(c) \\
3 & 0.2352 & 0.01 & 23 &  & R-CG & $\sim$2300 & m$\rightarrow$(c) \\
3 & 0.2352 & 0.01 & 68 &  & R-CG & $\sim$2300 & c$\rightarrow$(c) \\
3 & 0.2352 & 0.01 & 159 & 100 & R-CG & 851 & d \\
3.1 & 0.2341 & 0.01 & 160 & 50 & R-CG & 650 & d \\
3.2 & 0.2329 & 0.01 & 114 & 50 & R-CG & 556 & d \\
3.2 & 0.2329 & 0.01 & 169 & 100 & R-CG & 504 & d \\
4 & 0.18 & 0.01 & 520 & 300 & R-CR & 35 & c \\
4 & 0.19 & 0.01 & 520 & 300 & R-CR & 47 & c \\
4 & 0.2 & 0.01 & 391 & 200 & R-CR & 84 & d$\rightarrow$c \\
4 & 0.205 & 0.01 & 320 & 200 & R-CG & 351 & d \\
4 & 0.21 & 0.01 & 308 & 200 & R-CG & 247 & d \\
4 & 0.2226 & 0.01 & 320 & 200 & R-CG & 188 & d \\
4.5 & 0.16 & 0.01 & 500 & 300 & R-CR & 25 & c \\
4.5 & 0.17 & 0.01 & 542 & 300 & R-CR & 30 & c \\
4.5 & 0.18 & 0.01 & 545 & 300 & R-CR & 40 & d$\rightarrow$c \\
4.5 & 0.185 & 0.01 & 534 & 300 & R-CR & 51 & d$\rightarrow$c \\
4.5 & 0.186 & 0.01 & 301 & 150 & R-CR & 56 & c \\
4.5 & 0.1875 & 0.01 & 191 & 100 & R-CR & 82 & c \\
4.5 & 0.1875 & 0.01 & 181 & 100 & R-CG & 248 & d \\
4.5 & 0.189 & 0.01 & 207 & 100 & R-CG & 214 & d \\
\hline
\end{tabular}
\caption{\baselineskip=.8cm
The same as Table~\protect\ref{tab:paramF2Nt4} 
for $N_F=3$ on an $8^2\times10\times4$ lattice.
}
\protect\label{tab:paramF3Nt4Ns8}
\end{center}
\end{table}
\clearpage
\addtocounter{table}{-1}
\begin{table}
\begin{center}
\begin{tabular}{lllrrcrc}
\hline
$\beta$ & $K$ & $\Delta\tau$ & $\tau_{\rm tot}$ & 
$\tau_{\rm therm}$ & algo.\ & $N_{\rm inv}$ & phase \\
\hline
4.5 & 0.19 & 0.01 & 336 & 200 & R-CG & 200 & d \\
4.5 & 0.2 & 0.01 & 394 & 200 & R-CG & 158 & d \\
4.5 & 0.205 & 0.01 & 190 &  & R-CG & 142 & d \\
4.5 & 0.2143 & 0.01 & 101 &  & R-CG & 132 & d \\
5 & 0.13 & 0.01 & 313 & 150 & R-CR & 49 & c \\
5 & 0.14 & 0.01 & 275 & 150 & R-CR & 20 & c \\
5 & 0.15 & 0.01 & 310 & 150 & R-CR & 22 & c \\
5 & 0.16 & 0.01 & 324 & 150 & R-CR & 27 & c \\
5 & 0.165 & 0.01 & 373 & 150 & R-CR & 65 & c \\
5 & 0.165 & 0.01 & 202 & 150 & R-CG & 48 & d$\rightarrow$c \\
5 & 0.166 & 0.01 & 120 &  & R-CG &  & d$\rightarrow$(c) \\
5 & 0.166 & 0.01 & 264 & 150 & R-CR & 35 & c \\
5 & 0.167 & 0.01 & 145 &  & R-CR &  & c$\rightarrow$(d) \\
5 & 0.167 & 0.01 & 187 & 100 & R-CG & 155 & d \\
5 & 0.17 & 0.01 & 291 & 150 & R-CG & 119 & d \\
5.5 & 0.1 & 0.01 & 652 & 150 & R-CR & 13 & c \\
5.5 & 0.11 & 0.01 & 505 & 150 & R-CR & 15 & c \\
5.5 & 0.12 & 0.01 & 571 & 250 & R-CR & 16 & c \\
5.5 & 0.125 & 0.01 & 695 & 250 & R-CR & 17 & c$\rightarrow$m \\
5.5 & 0.1275 & 0.01 & 676 & 100 & R-CR & 18 & d \\
5.5 & 0.13 & 0.01 & 364 & 150 & R-CR & 18 & c$\rightarrow$d \\
5.5 & 0.135 & 0.01 & 174 & 100 & R-CR & 20 & d$\rightarrow$d \\
5.5 & 0.14 & 0.01 & 296 & 100 & R-CR & 23 & d \\
6 & 0.08 & 0.01 & 355 & 100 & R-CG & 23 & d \\
6 & 0.09 & 0.01 & 194 & 100 & R-CG & 27 & d \\
6 & 0.1 & 0.01 & 320 & 100 & R-CG & 33 & d \\
6 & 0.11 & 0.01 & 270 & 100 & R-CG & 41 & d \\
6 & 0.12 & 0.01 & 244 & 100 & R-CG & 51 & d \\
6 & 0.135 & 0.01 & 180 & 100 & R-CG & 72 & d \\
\hline
\end{tabular}
\caption{\baselineskip=.8cm
{\it Continued.}
}
\end{center}
\end{table}

\clearpage

\begin{table}
\begin{center}
\begin{tabular}{lllrrcrc}
\hline
$\beta$ & $K$ & $\Delta\tau$ & $\tau_{\rm tot}$ & 
$\tau_{\rm therm}$ & algo.\ & $N_{\rm inv}$ & phase \\
\hline
4 & 0.2 & 0.01 & 198 & 100 & R-CR & 82 & c \\
4 & 0.202 & 0.01 & 273 & 100 & R-CR & 101 & c \\
4 & 0.203 & 0.01 & 229 & 100 & R-CR & 117 & c \\
4 & 0.203 & 0.01 & 63 &  & R-CG &  & d$\rightarrow$(c) \\
4 & 0.204 & 0.01 & 219 & 100 & R-CG & 152 & c \\
4 & 0.204 & 0.01 & 169 & 100 & R-CG & 449 & d \\
4 & 0.205 & 0.01 & 93 &  & R-CG &  & c$\rightarrow$(d) \\
4 & 0.205 & 0.01 & 192 & 100 & R-CG & 380 & d \\
4 & 0.21 & 0.01 & 203 & 100 & R-CG & 272 & d \\
4.5 & 0.18 & 0.01 & 282 & 100 & R-CR & 40 & c \\
4.5 & 0.186 & 0.01 & 230 & 100 & R-CR & 56 & c \\
4.5 & 0.1875 & 0.01 & 1040 & 369 & R-CR & 74 & c \\
4.5 & 0.1875 & 0.01 & 1072 & 100 & R-CG & 264 & d \\
4.5 & 0.189 & 0.01 & 183 & 100 & R-CG & 230 & d \\
4.5 & 0.19 & 0.01 & 196 & 100 & R-CG & 219 & d \\
4.7 & 0.17 & 0.01 & 307 & 100 & R-CR & 32 & c \\
4.7 & 0.175 & 0.01 & 225 & 100 & R-CR & 38 & c \\
4.7 & 0.178 & 0.01 & 232 & 100 & R-CG & 117 & d$\rightarrow$c \\
4.7 & 0.179 & 0.01 & 335 & 100 & R-CR & 48 & c \\
4.7 & 0.179 & 0.01 & 253 &  & R-CG &  & d$\rightarrow$(c) \\
4.7 & 0.1795 & 0.01 & 1035 & 100 & R-CR & 50 & c \\
4.7 & 0.1795 & 0.01 & 1073 & 100 & R-CG & 236 & d \\
4.7 & 0.18 & 0.01 & 299 & 100 & R-CG & 228 & d \\
4.7 & 0.18 & 0.01 & 410 &  & R-CG & & c$\rightarrow$(d) \\
\hline
\end{tabular}
\caption{\baselineskip=.8cm
The same as Table~\protect\ref{tab:paramF2Nt4} 
for $N_F=3$ on a $12^3\times4$ lattice.
}
\protect\label{tab:paramF3Nt4Ns12}
\end{center}
\end{table}
\clearpage
\addtocounter{table}{-1}
\begin{table}
\begin{center}
\begin{tabular}{lllrrcrc}
\hline
$\beta$ & $K$ & $\Delta\tau$ & $\tau_{\rm tot}$ & 
$\tau_{\rm therm}$ & algo.\ & $N_{\rm inv}$ & phase \\
\hline
5 & 0.165 & 0.01 & 203 & 100 & R-CR & 33 & c \\
5 & 0.166 & 0.01 & 574 & 200 & R-CR & 35 & c \\
5 & 0.166 & 0.01 & 405 &  & R-CG/CR &  & d$\rightarrow$(c) \\
5 & 0.16625 & 0.01 & 570 & 200 & R-CR & 37 & c$\rightarrow$m \\
5 & 0.16625 & 0.01 & 545 & 200 & R-CR & 47 & d$\rightarrow$m \\
5 & 0.1665 & 0.01 & 502 &  & R-CR &  & c$\rightarrow$(d) \\
5 & 0.1665 & 0.01 & 611 & 200 & R-CR & 75 & d \\
5 & 0.167 & 0.01 & 475 & 250 & R-CR & 53 & d \\
5 & 0.168 & 0.01 & 419 & 100 & R-CR & 104 & d \\
5 & 0.169 & 0.01 & 164 & 100 & R-CG & 163 & d \\
5 & 0.17 & 0.01 & 231 & 100 & R-CG & 166 & d \\
\hline
\end{tabular}
\caption{\baselineskip=.8cm
{\it Continued.}
}
\end{center}
\end{table}

\clearpage

\begin{table}
\begin{center}
\begin{tabular}{lllrrcrc}
\hline
$\beta$ & $K$ & $\Delta\tau$ & $\tau_{\rm tot}$ & 
$\tau_{\rm therm}$ & algo.\ & $N_{\rm inv}$ & phase \\
\hline
0 & 0.2 & 0.01 & 32 & 20 & R-CR & 38 & c \\
0 & 0.21 & 0.01 & 32 & 20 & R-CR & 49 & c \\
0 & 0.22 & 0.01 & 33 & 18 & R-CR & 67 & c \\
0 & 0.235 & 0.01 & 40 & 20 & R-CR & 155 & c \\
0.1 & 0.2495 & 0.01 & 11 &  & R-CG & $>$5000 & d$\rightarrow$(c) \\
0.2 & 0.249 & 0.01 & 11 &  & R-CG & $>$5000 & d$\rightarrow$(c) \\
0.2 & 0.24936 & 0.01 & 23 &  & R-CG & $>$10000 & d$\rightarrow$(c) \\
0.3 & 0.2485 & 0.01 & 16 &  & R-CG & $>$5000 & m$\rightarrow$(c) \\
0.3 & 0.2485 & 0.01 & 9 &  & R-CG & $>$5000 & m$\rightarrow$(c) \\
0.3 & 0.2485 & 0.01 & 27 &  & R-CG & 600 & d \\
0.3 & 0.249 & 0.01 & 16 &  &  & $>$5000 & m$\rightarrow$(c) \\
0.4 & 0.248 & 0.01 & 20 & 10 & R-CG & 500 & d \\
0.5 & 0.23 & 0.01 & 6 &  & R-CG &  & d$\rightarrow$(c) \\
0.5 & 0.235 & 0.01 & 6 &  & R-CG &  & d$\rightarrow$(c) \\
0.5 & 0.24 & 0.01 & 6 &  & R-CG &  & d$\rightarrow$(c) \\
0.5 & 0.245 & 0.01 & 53 &  & R-CG & $\sim$1400 & d$\rightarrow$c \\
0.5 & 0.2475 & 0.01 & 25 & 15 & R-CG & 445 & d \\
1 & 0.2 & 0.01 & 113 & 50 & R-CR & 42 & c \\
1 & 0.21 & 0.01 & 104 & 50 & R-CR & 60 & c \\
1 & 0.22 & 0.01 & 115 & 55 & R-CR & 80 & c \\
1 & 0.225 & 0.01 & 267 & 100 & R-CR & 126 & c \\
1 & 0.23 & 0.01 & 293 & 100 & R-CR & 192 & c \\
1 & 0.235 & 0.01 & 40 & & R-CG & & d$\rightarrow$c \\
1 & 0.235 & 0.005 & 112 & 60 & R-CG & 970 & c \\
1 & 0.235 & 0.005 & 19 &  & R-CG &  & d$\rightarrow$(c) \\
1 & 0.237 & 0.005 & 42 &  & R-CG &  & d \\
1 & 0.237 & 0.005 & 49 &  & R-CG &  & c$\rightarrow$(d) \\
1 & 0.238 & 0.005 & 28 &  & R-CG & 440 & d \\
1 & 0.24 & 0.005 & 108 & 40 & R-CG & 325 & d \\
1 & 0.245 & 0.01 & 114 & 60 & R-CG & 306 & d \\
\hline
\end{tabular}
\caption{\baselineskip=.8cm
The same as Table~\protect\ref{tab:paramF2Nt4} 
for $N_F=6$ on an $8^2\times10\times4$ lattice.
}
\protect\label{tab:paramF6Nt4}
\end{center}
\end{table}
\clearpage
\addtocounter{table}{-1}
\begin{table}
\begin{center}
\begin{tabular}{lllrrcrc}
\hline
$\beta$ & $K$ & $\Delta\tau$ & $\tau_{\rm tot}$ & 
$\tau_{\rm therm}$ & algo.\ & $N_{\rm inv}$ & phase \\
\hline
2 & 0.24 & 0.01 & 18 &  & R-CG & 162 & d \\
4 & 0.22 & 0.01 & 15 &  & R-CG & 88 & d \\
4.5 & 0.15 & 0.01 & 71 & 61 & R-CR & 21 & c \\
4.5 & 0.16 & 0.01 & 38 & 28 & R-CR & 27 & c \\
4.5 & 0.165 & 0.01 & 60 & 50 & R-CR & 32 & c \\
4.5 & 0.165 & 0.01 & 60 &  & R-CR &  & d$\rightarrow$(c) \\
4.5 & 0.166 & 0.01 & 277 & 267 & R-CR & 36 & d$\rightarrow$c \\
4.5 & 0.167 & 0.01 & 193 & 183 & R-CR & 36 & c \\
4.5 & 0.167 & 0.01 & 159 & 149 & R-CR & 105 & d \\
4.5 & 0.168 & 0.01 & 152 &  & R-CG &  & c$\rightarrow$(d) \\
4.5 & 0.17 & 0.01 & 73 &  & R-CG &  & c$\rightarrow$(d) \\
4.5 & 0.18 & 0.01 & 41 & 31 & R-CG & 115 & c$\rightarrow$d \\
4.5 & 0.19 & 0.01 & 38 & 28 & R-CG & 92 & c$\rightarrow$d \\
4.5 & 0.2143 & 0.01 & 181 & 150 & R-CG & 87 & d \\
\hline
\end{tabular}
\caption{\baselineskip=.8cm
{\it Continued.}
}
\end{center}
\end{table}

\clearpage

\begin{table}
\begin{center}
\begin{tabular}{lllrrrrc}
\hline
$\beta$ & $K_{ud}$ & $K_s$ & $\tau_{\rm tot}$ & 
$\tau_{\rm therm}$ & $N_{\rm inv}^{ud}$ & $N_{\rm inv}^s$ 
& phase \\
\hline
3.2 & 0.2329 & 0.2043 & 15 &  & $\sim$3000 & $\sim$250 
& d$\rightarrow$(c) \\
3.4 & 0.2306 & 0.2026 & 20 &  & $\sim$3000 & $\sim$290 
& d$\rightarrow$(c) \\
3.5 & 0.2295 & 0.2017 & 9 &  & $\sim$3000 & $\sim$260 
& m$\rightarrow$(c) \\
3.5 & 0.2295 & 0.2017 & 553 & 100 & 862 & 394 & d \\
3.6 & 0.2281 & 0.2006 & 153 & 100 & 622 & 344 & d \\
3.7 & 0.2267 & 0.1692 & 20 &  & $\sim$2500 & $\sim$100 
& d$\rightarrow$(c) \\
3.8 & 0.2254 & 0.1684 & 47 &  & $\sim$2500 & $\sim$93 
& d$\rightarrow$(c) \\
3.9 & 0.224 & 0.1677 & 12 &  & $\sim$2500 & $\sim$93 
& m$\rightarrow$(c) \\
3.9 & 0.224 & 0.1677 & 760 & 100 & 797 & 135 & d \\
4 & 0.2226 & 0.1669 & 159 & 100 & 521 & 137 & d \\
4 & 0.2226 & 0.1964 & 167 & 100 & 235 & 201 & d \\
4.3 & 0.218 & 0.1643 & 159 & 100 & 229 & 130 & d \\
5.5 & 0.163 & 0.15 & 376 & 208 & 119 & 97 & d \\
\hline
\end{tabular}
\caption{\baselineskip=.8cm
The same as Table~\protect\ref{tab:paramF2Nt4} 
for $N_F=2+1$ on an $8^2\times10\times4$ lattice. 
The molecular dynamics time step is $\Delta\tau=0.01$. 
Simulations are performed with the R algorithm 
for updating configurations 
and with the CG method for quark matrix inversions. 
}
\protect\label{tab:paramF21Nt4}
\end{center}
\end{table}


\begin{table}
\begin{center}
\begin{tabular}{lllrrrrc}
\hline
$\beta$ & $K_{ud}$ & $K_s$ & $\tau_{\rm tot}$ & 
$\tau_{\rm therm}$ & $N_{\rm inv}^{ud}$ & $N_{\rm inv}^s$ 
& phase \\
\hline
3.9 & 0.224 & 0.1677 & 14 &  & $\sim$3000 & $\sim$93 
& m$\rightarrow$(c) \\
3.9 & 0.224 & 0.1677 & 398 & 100 & 999 & 139 & d \\
4 & 0.2226 & 0.1669 & 396 & 100 & 636 & 141 & d \\
\hline
\end{tabular}
\caption{\baselineskip=.8cm
The same as Table~\protect\ref{tab:paramF21Nt4} 
for $N_F=2+1$ on an $8^2\times10\times4$ lattice. 
Simulations are performed with the R algorithm 
for updating configurations 
and with the CG method for quark matrix inversions. 
}
\protect\label{tab:paramF21Nt4Ns12}
\end{center}
\end{table}


\begin{table}
\begin{center}
\begin{tabular}{lllrrrrc}
\hline
$\beta$ & $K_{ud}$ & $K_s$ & $\tau_{\rm tot}$ & 
$\tau_{\rm therm}$ & $N_{\rm inv}^{ud}$ & $N_{\rm inv}^s$ 
& phase \\
\hline
3.5 & 0.195 & 0.2017 & 196 & 100 & 46 & 58 & c \\
3.5 & 0.2 & 0.2017 & 164 & 50 & 57 & 59 & c \\
3.5 & 0.205 & 0.2017 & 166 & 50 & 74 & 61 & c \\
3.5 & 0.21 & 0.2017 & 158 & 40 & 109 & 64 & c \\
\hline
\end{tabular}
\caption{\baselineskip=.8cm
The same as Table~\protect\ref{tab:paramF21Nt4} 
for $N_F=2+1$ on an $8^3\times10$ lattice. 
Simulations are performed with the R algorithm 
for updating configurations 
and with the CR method for quark matrix inversions. 
}
\protect\label{tab:paramF21Nt8}
\end{center}
\end{table}


\begin{table}
\begin{center}
\begin{tabular}{ccc}
\hline
$\beta$ & $K_c(m_{\pi}^2)$ & $K_c(m_q)$ \\
\hline
3.0     & 0.235(1) & 0.230(1) \\
3.5     & 0.230(1) & 0.226(1) \\
4.0     & 0.223(1) & 0.218(4) \\
4.3     & 0.218(1) & 0.214(1) \\
4.5     & 0.214(1) & 0.210(1) \\
6.0     &          & 0.1564(1)\\
10.0    &          & 0.1396(1)\\
\hline
\end{tabular}
\caption{\baselineskip=.8cm
The chiral limit $K_c$ for $N_F=2$ 
determined on an $8^2\times10\times4$ lattice. 
The results for $\beta=3.0$ --- 4.5 are determined 
by $m_\pi=0$ and $m_q=0$, where
values of $m_\pi^2$ and $m_q$ in the confining phase are 
linearly extrapolated in $1/K$. 
The results for $\beta=6.0$ and 10.0 are determined from 
an interpolation of $m_q$ in the deconfining phase.
\protect\label{tab:1}}
\end{center}
\end{table}


\begin{table}
\begin{center}
\begin{tabular}{c|cc|cc}
\hline
 & \multicolumn{2}{c|}{$N_t=4$} & \multicolumn{2}{c}{$N_t=8$}\\
$N_F$ & $K_c(m_{\pi}^2)$ & $K_c(m_q)$ & $K_c(m_{\pi}^2)$ 
& $K_c(m_q)$\\
\hline
2 & 0.214(1) & 0.210(1) & 0.212(1) & 0.209(1) \\
3 & 0.210(1) & 0.204(1) &          &          \\
6 & 0.205(2) & 0.200(1) &          &          \\
\hline
\end{tabular}
\caption{\baselineskip=.8cm
The chiral limit $K_c$ at $\beta=4.5$ 
determined by $m_\pi=0$ and $m_q=0$, where
values of $m_\pi^2$ and $m_q$ in the confining phase are 
linearly extrapolated in $1/K$ 
using data from $K=0.16$ --- 0.18 for $N_F=2$ and 3 and
$K=0.15$ --- 0.165 for $N_F=6$ 
(because $K_t = 0.167(1)$ for $N_F=6$ at $N_t=4$). 
The spatial lattice size is $8^2\times 10$.
\protect\label{tab:2}}
\end{center}
\end{table}


\begin{table}
\begin{center}
\begin{tabular}{cc|cc|cc}
\hline
\multicolumn{2}{c|}{$N_F=2$} & \multicolumn{2}{c|}{$N_F=3$} & 
\multicolumn{2}{c}{$N_F=6$} \\
$\beta$ & $K_t$ & $\beta$ & $K_t$ & $\beta$ & $K_t$\\
\hline
4.3  & 0.207--0.210 & 3.0  & $>$ 0.230      & 0.5 & 0.245--0.2475 \\
4.5  & 0.200--0.202 & 4.0  & 0.200--0.205 & 1.0 & 0.235--0.237 \\
5.0  & 0.170--0.180 & 4.5  & 0.186--0.189 & 4.5 & 0.166--0.168 \\
5.25 & 0.160--0.165 & 4.5* & 0.186--0.189 &     & \\
     &              & 4.7* & 0.179--0.180 &     & \\
     &              & 5.0  & 0.166--0.167 &     & \\
     &              & 5.0* & 0.166--0.1665 &     & \\
     &              & 5.5  & 0.125--0.130 &     & \\
\hline
\end{tabular}
\caption{\baselineskip=.8cm
Finite temperature transition $K_t$ for 
$N_F=2$, 3 and 6
obtained on an $8^2\times10\times4$ lattice
(data with * obtained on a $12^3\times4$ lattice). 
For $N_F=2$ at $\beta=5.0$, the data by the MILC collaboration 
\protect\cite{MILC4} give a more precise value 0.177 --- 0.178 
for $K_t$ (cf.\ Fig.~\protect\ref{fig:F2B5.0}).
\protect\label{tab:Kt}}
\end{center}
\end{table}

\clearpage

\begin{table}
\begin{center}
\begin{tabular}{lllllll}
\hline
$\beta$ & $K$ & plaquette & Polyakov & 
$m_\pi a$ & $2m_q a$ & $m_\rho a$\\
\hline
0 & 0.2 & 0.0088(1) & 0.0367(3) & 1.441(3) & 0.715(2) & 1.542(25)\\
0 & 0.21 & 0.0109(1) & 0.0449(3) & 1.272(4) & 0.552(3) & 1.405(34)\\
0 & 0.22 & 0.0134(1) & 0.0548(32) & 1.086(3) & 0.400(2) & 1.300(37)\\
0 & 0.23 & 0.0161(2) & 0.0681(10) & 0.871(5) & 0.253(2) & 1.074(65)\\
3 & 0.18 & 0.2174(3) & 0.0252(13) & 1.554(7) & 0.808(5) & 1.631(44)\\
3 & 0.19 & 0.2201(2) & 0.0361(7) & 1.376(7) & 0.621(5) & 1.473(48)\\
3 & 0.2 & 0.2247(3) & 0.0454(14) & 1.179(5) & 0.437(3) & 1.342(61)\\
3.5 & 0.175 & 0.2587(3) & 0.0268(12) & 1.606(5) & 0.848(5) & 1.679(24)\\
3.5 & 0.185 & 0.2624(3) & 0.0333(13) & 1.408(7) & 0.649(4) & 1.532(31)\\
3.5 & 0.195 & 0.267(4) & 0.0436(13) & 1.211(7) & 0.461(4) & 1.392(43)\\
4 & 0.17 & 0.3034(1) & 0.0249(2) & 1.623(3) & 0.874(2) & 1.688(5)\\
4 & 0.18 & 0.3079(1) & 0.0318(2) & 1.426(3) & 0.659(2) & 1.523(6)\\
4 & 0.19 & 0.3141(1) & 0.0408(2) & 1.207(4) & 0.458(2) & 1.367(14)\\
4 & 0.2226 & 0.4002(7) & 0.1328(24) & 0.831(35) & -.074(11) & \\
4.1 & 0.2211 & 0.4300(4) & 0.1345(20) & 0.997(41) & -.100(9) & 1.07(44)\\
4.2 & 0.2195 & 0.4445(6) & 0.1535(29) & 1.254(32) & -.081(9) & 2.12(36)\\
4.3 & 0.165 & 0.3319(2) & 0.0220(7) & 1.663(7) & 0.920(5) & 1.715(21)\\
4.3 & 0.175 & 0.3367(2) & 0.0293(8) & 1.463(7) & 0.696(5) & 1.546(22)\\
4.3 & 0.185 & 0.3440(2) & 0.0404(8) & 1.242(6) & 0.485(3) & 1.379(24)\\
4.3 & 0.205 & 0.3732(3) & 0.0736(8) & 0.647(8) & 0.094(3) & \\
4.3 & 0.21 & 0.4286(4) & 0.1369(12) & 0.755(74) & -.026(9) & \\
4.3 & 0.218 & 0.4661(4) & 0.1790(18) & 1.413(13) & -.083(9) & 1.784(55)\\
4.5 & 0.16 & 0.3524(2) & 0.0209(7) & 1.732(6) & 0.997(4) & 1.782(8)\\
4.5 & 0.17 & 0.3580(2) & 0.0282(6) & 1.520(6) & 0.760(4) & 1.595(1)\\
4.5 & 0.18 & 0.3656(2) & 0.0384(7) & 1.298(5) & 0.534(4) & 1.423(8)\\
4.5 & 0.195 & 0.3856(3) & 0.0590(9) & 0.882(11) & 0.201(4) & 1.145(38)\\
4.5 & 0.2 & 0.4007(6) & 0.0807(19) & 0.696(24) & 0.090(5) & \\
\hline
\end{tabular}
\caption{\baselineskip=.8cm
Results of the plaquette, the Polyakov loop, 
the pion screening mass, twice the quark mass, and the rho meson 
screening mass 
for $N_F=2$ obtained on an $8^2\times10\times4$ lattice.
Data marked with \dag \ are taken from 
Ref.~\protect\cite{ChNf2} 
obtained on an $8^2\times20\times4$ lattice. 
}
\protect\label{tab:resultsF2Nt4}
\end{center}
\end{table}
\clearpage
\addtocounter{table}{-1}
\begin{table}
\begin{center}
\begin{tabular}{lllllll}
\hline
$\beta$ & $K$ & plaquette & Polyakov & 
$m_\pi a$ & $2m_q a$ & $m_\rho a$\\
\hline
4.5 & 0.202 & 0.4591(3) & 0.1643(9) & 1.135(36) & -.072(8) & \\
4.5 & 0.205 & 0.4752(3) & 0.1809(17) & 1.421(18) & -.128(27) & 1.738(43)\\
4.5 & 0.2143 & 0.4949(3) & 0.2137(16) & 1.552(10) & -.034(8) & 1.746(19)\\
5 & 0.14 & 0.4095(2) & 0.0128(7) & 2.046(5) & 1.379(5) & 2.072(6)\\
5 & 0.15 & 0.4148(2) & 0.0217(7) & 1.801(12) & 1.093(1) & 1.828(19)\\
5 & 0.16 & 0.4215(2) & 0.0301(7) & 1.551(8) & 0.805(6) & 1.604(1)\\
5 & 0.17 & 0.4351(2) & 0.0409(8) & 1.279(6) & 0.522(5) & 1.394(12)\\
5 & 0.18 & 0.5174(2) & 0.2250(8) & 1.430(13) & -.086(9) & 1.637(11)\\
5 & 0.19 & 0.5378(1) & 0.2580(7) & 1.686(11) & -.096(8) & 1.861(8)\\
5 & 0.1982 & 0.5473(1) & 0.2789(6) & 1.717(5) & 0.029(5) & 1.837(7)\\
5.25 & 0.1 & 0.4426(2) & 0.0018(8) & 2.934(5) & 2.508(5) & 2.937(6)\\
5.25 & 0.11 & 0.4446(2) & 0.0041(5) & 2.681(10) & 2.191(8) & 2.687(11)\\
5.25 & 0.12 & 0.4472(2) & 0.0085(7) & 2.423(7) & 1.867(7) & 2.433(8)\\
5.25 & 0.13 & 0.4502(2) & 0.0140(7) & 2.184(8) & 1.563(6) & 2.200(9)\\
5.25 & 0.14 & 0.4556(2) & 0.0213(7) & 1.941(4) & 1.263(4) & 1.970(5)\\
5.25 & 0.15 & 0.4635(3) & 0.0305(9) & 1.657(12) & 0.939(9) & 1.709(14)\\
5.25 & 0.155 & 0.4746(3) & 0.0499(12) & 1.495(6) & 0.756(6) & 1.563(7)\\
5.25 & 0.16 & 0.4846(3) & 0.0678(11) & 1.324(9) & 0.570(7) & 1.397(12)\\
5.25 & 0.165 & 0.5307(2) & 0.2241(10) & 1.351(9) & 0.173(9) & 1.468(12)\\
5.25 & 0.175 & 0.5513(2) & 0.2695(8) & 1.531(12) & -.136(18) & 1.696(10)\\
5.25 & 0.18 & 0.5589(1) & 0.2861(8) & 1.696(6) & -.160(6) & 1.853(7)\\
5.5\dag & 0.15 & 0.5530(2) & 0.2413(7) & 1.486(6) & 0.512(1) & 1.528(8)\\
5.5\dag & 0.16 & 0.5662(2) & 0.2815(5) & 1.415(7) & 0.103(5) & 1.490(7)\\
5.5\dag & 0.1615 & 0.5677(2) & 0.2863(8) & 1.441(5) & 0.048(6) & 1.513(9)\\
5.5\dag & 0.163 & 0.5699(1) & 0.2905(8) & 1.438(8) & -.016(4) & 1.506(8)\\
6 & 0.15 & 0.6122(2) & 0.3456(10) & 1.469(7) & 0.233(5) & 1.510(8)\\
6 & 0.1524 & 0.6131(3) & 0.3478(16) & 1.467(7) & 0.142(8) & 1.514(7)\\
6 & 0.155 & 0.6157(2) & 0.3555(9) & 1.480(5) & 0.042(7) & 1.529(9)\\
6 & 0.16 & 0.6188(2) & 0.3607(9) & 1.534(6) & -.120(6) & 1.594(8)\\
10 & 0.13 & 0.7853(1) & 0.6126(11) & 1.496(6) & 0.447(2) & 1.491(6)\\
10 & 0.14 & 0.7865(1) & 0.6157(8) & 1.439(8) & -.010(4) & 1.437(9)\\
10 & 0.15 & 0.7873(1) & 0.6230(8) & 1.591(2) & -.427(6) & 1.598(3)\\
\hline
\end{tabular}
\caption{\baselineskip=.8cm
{\it Continued.}
}
\end{center}
\end{table}

\clearpage

\begin{table}
\begin{center}
\begin{tabular}{lllllll}
\hline
$\beta$ & $K$ & plaquette & Polyakov & 
$m_\pi a$ & $2m_q a$ & $m_\rho a$\\
\hline
4.2 & 0.2195 & 0.4410(3) & 0.0067(12) & 0.897(50) & -.082(11) & \\
4.3 & 0.2183 & 0.4593(2) & 0.0054(6) & 1.070(62) & -.111(21) & \\
4.4 & 0.2163 & 0.4742(1) & 0.0098(7) & 1.241(28) & -.105(8) & 1.677(52)\\
4.5 & 0.2143 & 0.4889(2) & 0.0071(10) & 1.371(11) & -.058(9) & 1.604(31)\\
5 & 0.1982 & 0.5455(1) & 0.0831(10) & 1.638(7) & 0.044(9) & 1.749(8)\\
5.02 & 0.16 & 0.4256(1) & 0.0023(4) & 1.542(5) & 0.800(4) & 1.605(6)\\
5.02 & 0.17 & 0.4384(1) & 0.0041(3) & 1.242(6) & 0.508(4) & 1.343(11)\\
5.02 & 0.18c & 0.4696(1) & 0.0102(5) & 0.710(10) & 0.149(5) & 0.986(29)\\
5.02 & 0.18d & 0.5180(2) & 0.0399(9) & 0.923(47) & -.164(18) & 1.40(16)\\
\hline
\end{tabular}
\caption{\baselineskip=.8cm
The same as Table~\protect\ref{tab:resultsF2Nt4} 
for $N_F=2$ on a $12^3\times6$ lattice.
}
\protect\label{tab:resultsF2Nt6}
\end{center}
\end{table}


\begin{table}
\begin{center}
\begin{tabular}{lllllll}
\hline
$\beta$ & $K$ & plaquette & Polyakov & 
$m_\pi a$ & $2m_q a$ & $m_\rho a$\\
\hline
4.5 & 0.16 & 0.3522(1) & 0.0010(7) & 1.731(7) & 0.999(5) & 1.779(20)\\
4.5 & 0.17 & 0.3574(1) & 0.0004(6) & 1.513(5) & 0.759(4) & 1.588(19)\\
4.5 & 0.18 & 0.3649(1) & 0.0023(6) & 1.281(5) & 0.529(3) & 1.398(18)\\
5.5\dag & 0.15 & 0.5377(3) & 0.0073(1) & 1.115(16) & 0.542(6) & 1.167(19)\\
5.5\dag & 0.155 & 0.5481(3) & 0.0081(2) & 0.807(35) & 0.308(14) & 0.874(39)\\
6 & 0.1524 & 0.6131(3) & 0.3478(16) & 0.837(19) & -.003(4) & 0.881(22)\\
\hline
\end{tabular}
\caption{\baselineskip=.8cm
The same as Table~\protect\ref{tab:resultsF2Nt4} 
for $N_F=2$ on an $8^3\times10$ lattice.
Data marked with \dag \ are taken from 
Ref.~\protect\cite{ChNf2} 
obtained on an $8^3\times20$ lattice. 
}
\protect\label{tab:resultsF2Nt8}
\end{center}
\end{table}
\clearpage

\clearpage

\begin{table}
\begin{center}
\begin{tabular}{lllllll}
\hline
$\beta$ & $K$ & plaquette & Polyakov & 
$m_\pi a$ & $2m_q a$ & $m_\rho a$\\
\hline
3 & 0.205 & 0.2402(2) & 0.0779(7) & 1.049(4) & 0.334(2) & 1.223(18)\\
3 & 0.215 & 0.2501(3) & 0.1002(12) & 0.820(4) & 0.180(1) & 1.247(67)\\
3 & 0.225 & 0.2635(6) & 0.1266(24) &  &  & \\
3 & 0.2352d & 0.3546(6) & 0.1718(16) & 0.988(30) & -.066(5) & \\
3.1 & 0.2341 & 0.3743(3) & 0.1812(12) & 1.084(21) & -.069(4) & 1.53(26)\\
3.2 & 0.2329 & 0.3889(2) & 0.1850(10) & 1.192(19) & -.077(7) & 1.52(22)\\
4 & 0.18 & 0.3176(2) & 0.0508(7) & 1.405(7) & 0.638(5) & 1.497(14)\\
4 & 0.19 & 0.3297(2) & 0.0669(7) & 1.179(8) & 0.424(5) & 1.346(17)\\
4 & 0.2 & 0.3486(3) & 0.0978(8) & 0.899(9) & 0.206(5) & 1.193(50)\\
4 & 0.205 & 0.4465(5) & 0.2102(11) & 1.313(4) & -.056(30) & 1.77(18)\\
4 & 0.21 & 0.4674(3) & 0.2341(11) & 1.542(10) & -.057(11) & 1.760(20)\\
4 & 0.2226 & 0.4944(2) & 0.2637(12) & 1.552(5) & 0.009(4) & 1.689(6)\\
4.5 & 0.16 & 0.3598(2) & 0.0334(7) & 1.717(5) & 0.979(5) & 1.768(6)\\
4.5 & 0.17 & 0.3691(2) & 0.0459(6) & 1.497(5) & 0.732(3) & 1.575(9)\\
4.5 & 0.18 & 0.3835(2) & 0.0641(7) & 1.250(6) & 0.478(4) & 1.385(14)\\
4.5 & 0.185 & 0.3954(2) & 0.0812(8) & 1.094(8) & 0.340(5) & 1.281(15)\\
4.5 & 0.186 & 0.4025(3) & 0.0927(10) & 1.070(9) & 0.299(5) & 1.267(23)\\
4.5 & 0.1875c & 0.4129(6) & 0.1094(15) & 1.023(6) & 0.250(8) & 1.287(34)\\
4.5 & 0.1875d & 0.4867(6) & 0.2343(17) & 1.394(44) & -.078(7) & 1.636(55)\\
4.5 & 0.189 & 0.4964(4) & 0.2492(15) & 1.502(19) & -.114(14) & 1.696(23)\\
4.5 & 0.19 & 0.5012(3) & 0.2560(11) & 1.580(10) & -.118(12) & 1.788(18)\\
4.5 & 0.2 & 0.5232(2) & 0.2852(11) & 1.693(5) & 0.010(6) & 1.814(8)\\
4.5 & 0.205 & 0.5318(3) & 0.2957(13) &  &  & \\
4.5 & 0.2143 & 0.5433(4) & 0.3183(27) &  &  & \\
\hline
\end{tabular}
\caption{\baselineskip=.8cm
The same as Table~\protect\ref{tab:resultsF2Nt4} 
for $N_F=3$ on an $8^2\times10\times4$ lattice.
}
\protect\label{tab:resultsF3Nt4Ns8}
\end{center}
\end{table}
\clearpage
\addtocounter{table}{-1}
\begin{table}
\begin{center}
\begin{tabular}{lllllll}
\hline
$\beta$ & $K$ & plaquette & Polyakov & 
$m_\pi a$ & $2m_q a$ & $m_\rho a$\\
\hline
5 & 0.13 & 0.4102(2) & 0.0138(6) & 2.257(10) & 1.647(76) & 2.271(10)\\
5 & 0.14 & 0.4163(2) & 0.0223(9) & 2.036(7) & 1.373(6) & 2.063(8)\\
5 & 0.15 & 0.4243(3) & 0.0319(9) & 1.790(7) & 1.077(6) & 1.830(10)\\
5 & 0.16 & 0.4382(3) & 0.0522(10) & 1.515(9) & 0.763(8) & 1.586(12)\\
5 & 0.165 & 0.4533(3) & 0.0798(10) & 1.359(12) & 0.574(12) & 1.444(16)\\
5 & 0.166 & 0.4697(4) & 0.1247(17) & 1.340(13) & 0.491(14) & 1.447(12)\\
5 & 0.167 & 0.5141(4) & 0.2369(17) & 1.379(11) & 0.185(17) & 1.488(17)\\
5 & 0.17 & 0.5297(3) & 0.2698(11) & 1.473(10) & 0.003(19) & 1.586(11)\\
5.5 & 0.1 & 0.5011(2) & 0.0185(7) & 2.846(5) & 2.413(5) & 2.849(5)\\
5.5 & 0.11 & 0.5052(3) & 0.0283(10) & 2.551(7) & 2.055(7) & 2.557(8)\\
5.5 & 0.12 & 0.5145(3) & 0.0516(15) & 2.253(9) & 1.679(8) & 2.262(10)\\
5.5 & 0.125 & 0.5216(3) & 0.0837(16) & 2.104(4) & 1.495(5) & 2.116(5)\\
5.5 & 0.1275 & 0.5276(3) & 0.1226(16) & 2.031(4) & 1.392(4) & 2.046(4)\\
5.5 & 0.13 & 0.5384(3) & 0.1872(13) & 1.950(5) & 1.254(4) & 1.967(6)\\
5.5 & 0.135 & 0.5453(4) & 0.2141(24) & 1.814(8) & 1.056(6) & 1.836(8)\\
5.5 & 0.14 & 0.5521(2) & 0.2413(13) & 1.672(4) & 0.843(4) & 1.696(5)\\
6 & 0.08 & 0.5963(2) & 0.2582(13) & 3.312(5) & 2.993(4) & 3.313(5)\\
6 & 0.09 & 0.5971(3) & 0.2745(15) & 2.982(5) & 2.591(4) & 2.984(5)\\
6 & 0.1 & 0.5984(2) & 0.2829(13) & 2.647(4) & 2.182(4) & 2.649(4)\\
6 & 0.11 & 0.5987(2) & 0.2874(15) & 2.344(5) & 1.799(4) & 2.347(5)\\
6 & 0.12 & 0.6024(2) & 0.3063(15) & 2.046(11) & 1.401(9) & 2.051(12)\\
6 & 0.135 & 0.6076(3) & 0.3346(16) &  &  & \\
\hline
\end{tabular}
\caption{\baselineskip=.8cm
{\it Continued.}
}
\end{center}
\end{table}

\clearpage

\begin{table}
\begin{center}
\begin{tabular}{lllllll}
\hline
$\beta$ & $K$ & plaquette & Polyakov & 
$m_\pi a$ & $2m_q a$ & $m_\rho a$\\
\hline
4 & 0.2 & 0.3479(2) & 0.0945(8) & 0.910(6) & 0.207(3) & 1.139(26)\\
4 & 0.202 & 0.3541(2) & 0.1053(6) & 0.859(4) & 0.174(2) & 1.129(25)\\
4 & 0.203 & 0.3580(2) & 0.1104(7) & 0.836(5) & 0.154(2) & 1.139(30)\\
4 & 0.204c & 0.3684(2) & 0.1270(8) & 0.809(9) & 0.119(2) & 1.077(79)\\
4 & 0.204d & 0.4378(3) & 0.2030(9) &  &  & \\
4 & 0.205 & 0.4486(2) & 0.2102(7) &  &  & \\
4 & 0.21 & 0.4679(2) & 0.2322(8) &  &  & \\
4.5 & 0.18 & 0.3828(1) & 0.0645(5) & 1.252(5) & 0.479(3) & 1.378(9)\\
4.5 & 0.186 & 0.4014(2) & 0.0921(7) & 1.070(7) & 0.304(4) & 1.266(12)\\
4.5 & 0.1875c & 0.4138(2) & 0.1115(4) & 1.032(6) & 0.244(3) & 1.257(14)\\
4.5 & 0.1875d & 0.4870(1) & 0.2353(3) & 1.430(10) & -.079(5) & 1.674(14)\\
4.5 & 0.189 & 0.4945(3) & 0.2457(10) & 1.556(7) & -.097(7) & 1.753(13)\\
4.5 & 0.19 & 0.5007(2) & 0.2525(7) & 1.586(9) & -.114(7) & 1.829(16)\\
4.7 & 0.17 & 0.3986(1) & 0.0536(4) & 1.417(6) & 0.647(5) & 1.505(8)\\
4.7 & 0.175 & 0.4076(2) & 0.0661(7) & 1.286(6) & 0.513(4) & 1.405(9)\\
4.7 & 0.178 & 0.4185(3) & 0.0814(7) & 1.190(8) & 0.408(4) & 1.328(13)\\
4.7 & 0.179 & 0.4234(2) & 0.0905(7) & 1.147(6) & 0.369(3) & 1.312(9)\\
4.7 & 0.1795c & 0.4275(1) & 0.0976(3) & 1.144(4) & 0.350(3) & 1.310(7)\\
4.7 & 0.1795d & 0.4968(1) & 0.2360(4) & 1.393(7) & -.004(7) & 1.597(9)\\
4.7 & 0.18 & 0.4995(3) & 0.2399(7) & 1.381(15) & 0.003(13) & 1.596(16)\\
5 & 0.165 & 0.4538(3) & 0.0786(10) & 1.357(4) & 0.569(4) & 1.446(5)\\
5 & 0.166 & 0.4630(2) & 0.1017(7) & 1.318(6) & 0.513(3) & 1.424(7)\\
5 & 0.16625 & 0.4791(2) & 0.1454(7) & 1.312(4) & 0.428(3) & 1.419(6)\\
5 & 0.1665 & 0.5031(2) & 0.2086(8) & 1.349(6) & 0.280(5) & 1.463(6)\\
5 & 0.167 & 0.5151(3) & 0.2377(8) & 1.384(7) & 0.178(8) & 1.495(9)\\
5 & 0.168 & 0.5193(2) & 0.2478(5) & 1.401(5) & 0.134(5) & 1.516(5)\\
5 & 0.169 & 0.5263(2) & 0.2609(10) & 1.421(17) & 0.062(8) & 1.537(19)\\
5 & 0.17 & 0.5294(2) & 0.2686(8) & 1.432(10) & 0.020(7) & 1.564(9)\\
\hline
\end{tabular}
\caption{\baselineskip=.8cm
The same as Table~\protect\ref{tab:resultsF2Nt4} 
for $N_F=3$ on a $12^3\times4$ lattice.
}
\protect\label{tab:resultsF3Nt4Ns12}
\end{center}
\end{table}

\clearpage

\begin{table}
\begin{center}
\begin{tabular}{lllllll}
\hline
$\beta$ & $K$ & plaquette & Polyakov & 
$m_\pi a$ & $2m_q a$ & $m_\rho a$\\
\hline
0 & 0.2 & 0.0286(5) & 0.1208(32) & 1.422(7) & 0.687(2) & 1.524(23)\\
0 & 0.21 & 0.0362(6) & 0.1567(20) & 1.248(4) & 0.514(4) & 1.384(27)\\
0 & 0.22 & 0.0412(6) & 0.1971(32) & 1.064(9) & 0.360(9) & 1.227(47)\\
0 & 0.235 & 0.0566(6) & 0.2632(24) & 0.777(5) & 0.160(1) & 1.157(86)\\
0.3 & 0.2485d & 0.2229(9) & 0.3109(19) & 1.040(5) & -.037(4) & \\
0.4 & 0.248 & 0.2364(12) & 0.3112(28) & 1.079(16) & -.042(5) & \\
0.5 & 0.2475 & 0.2540(8) & 0.3060(26) & 1.157(6) & -.063(3) & \\
1 & 0.2 & 0.0976(3) & 0.1313(14) & 1.364(4) & 0.615(4) & 1.494(7)\\
1 & 0.21 & 0.1075(3) & 0.1634(12) & 1.170(5) & 0.438(2) & 1.332(15)\\
1 & 0.22 & 0.1197(4) & 0.2099(13) & 0.984(3) & 0.281(2) & 1.217(19)\\
1 & 0.225 & 0.1251(2) & 0.2385(8) & 0.888(4) & 0.211(1) & 1.201(28)\\
1 & 0.23 & 0.1262(2) & 0.2595(7) & 0.797(3) & 0.154(1) & 1.218(47)\\
1 & 0.235 & 0.1633(6) & 0.3035(16) & 0.725(6) & 0.078(2) & 1.11(23)\\
1 & 0.24 & 0.2944(5) & 0.3032(47) & 1.261(7) & -.044(6) & 1.586(38)\\
1 & 0.245 & 0.3207(3) & 0.2988(15) & 1.314(12) & -.058(7) & 1.717(82)\\
4.5 & 0.15 & 0.3690(3) & 0.0493(17) & 1.909(11) & 1.200(8) & 1.945(34)\\
4.5 & 0.16 & 0.3879(5) & 0.0822(23) & 1.668(10) & 0.908(10) & 1.728(41)\\
4.5 & 0.165 & 0.4013(7) & 0.1083(33) & 1.551(7) & 0.765(3) & 1.641(36)\\
4.5 & 0.166 & 0.4171(13) & 0.1394(29) &  &  & \\
4.5 & 0.167 & 0.4177(4) & 0.1317(18) &  &  & \\
4.5 & 0.167 & 0.5024(3) & 0.3034(30) &  &  & \\
4.5 & 0.168 & 0.5156(6) & 0.3256(26) &  &  & \\
4.5 & 0.17 & 0.5295(14) & 0.3448(28) & 1.554(12) & -.041(23) & \\
4.5 & 0.18 & 0.5677(4) & 0.3964(27) & 1.799(5) & -.116(10) & \\
4.5 & 0.19 & 0.5889(5) & 0.4237(24) & 1.803(6) & 0.167(7) & \\
4.5 & 0.2143 & 0.6202(3) & 0.4666(19) & 1.616(6) & 0.156(10) & 1.641(8)\\
\hline
\end{tabular}
\caption{\baselineskip=.8cm
The same as Table~\protect\ref{tab:resultsF2Nt4} 
for $N_F=6$ on an $8^4\times10\times4$ lattice.
}
\protect\label{tab:resultsF6Nt4}
\end{center}
\end{table}

\clearpage

\begin{table}
\begin{center}
\begin{tabular}{llllllll}
\hline
$\beta$ & $K_{ud}$ & $K_s$ & plaquette & Polyakov & 
$m_\pi a$ & $m_\rho a$ & $m_\phi a$ \\
\hline
3.5d & 0.2295 & 0.2017 & 0.3909(2) & 0.178(1) & 0.991(19) 
&   & 1.432(33) \\
3.6 & 0.2281 & 0.2006 & 0.4119(6) & 0.189(1) & 1.182(21) &   &   \\
3.9d & 0.224 & 0.1677 & 0.4173(2) & 0.170(1) & 1.003(29) 
&   & 1.527(7) \\
4 & 0.2226 & 0.1669 & 0.4403(5) & 0.189(2) & 1.254(27) 
& 1.774(93) & 1.534(23) \\
4 & 0.2226 & 0.1964 & 0.4761(4) & 0.244(2) & 1.526(7) 
& 1.713(6) & 1.810(19) \\
4.3 & 0.218 & 0.1643 & 0.4902(4) & 0.246(2) & 1.535(8) 
& 1.699(12) & 1.520(10) \\
5.5 & 0.163 & 0.15 & 0.5801(2) & 0.328(1) & 1.487(10) 
& 1.569(12) & 1.532(9) \\
\hline
\end{tabular}
\caption{\baselineskip=.8cm
Results of the plaquette, the Polyakov loop, 
the pion screening mass, the rho meson screening mass, 
and the phi meson screening mass 
for $N_F=2+1$ obtained on an $8^2\times10\times4$ lattice.
}
\protect\label{tab:resultsF21Nt4}
\end{center}
\end{table}


\begin{table}
\begin{center}
\begin{tabular}{llllllll}
\hline
$\beta$ & $K_{ud}$ & $K_s$ & plaquette & Polyakov & 
$m_\pi a$ & $m_\rho a$ & $m_\phi a$ \\
\hline
3.9d & 0.224 & 0.1677 & 0.4180(2) & 0.169(1) & 1.078(29) 
&   & 1.518(8) \\
4 & 0.2226 & 0.1669 & 0.4407(1) & 0.190(1) & 1.270(9) 
& 1.702(54) & 1.509(6) \\
\hline
\end{tabular}
\caption{\baselineskip=.8cm
The same as Table~\protect\ref{tab:resultsF21Nt4}
for $N_F=2+1$ obtained on a $12^3\times4$ lattice.
}
\protect\label{tab:resultsF21Nt4Ns12}
\end{center}
\end{table}


\begin{table}
\begin{center}
\begin{tabular}{llllllll}
\hline
$\beta$ & $K_{ud}$ & $K_s$ & plaquette & Polyakov & 
$m_\pi a$ & $m_\rho a$ & $m_\phi a$ \\
\hline
3.5 & 0.195 & 0.2017 & 0.2814(2) & 0.001(1) & 1.150(4) 
& 1.305(8) & 1.198(10) \\
3.5 & 0.2 & 0.2017 & 0.2851(2) & 0.003(1) & 1.023(3) 
& 1.218(12) & 1.194(13) \\
3.5 & 0.205 & 0.2017 & 0.2891(2) & 0.003(1) & 0.899(4) 
& 1.114(18) & 1.172(15) \\
3.5 & 0.21 & 0.2017 & 0.2935(2) & 0.003(1) & 0.748(4) 
& 1.084(42) & 1.179(16) \\
\hline
\end{tabular}
\caption{\baselineskip=.8cm
The same as Table~\protect\ref{tab:resultsF21Nt4}
for $N_F=2+1$ obtained on an $8^3\times10$ lattice.
}
\protect\label{tab:resultsF21Nt8}
\end{center}
\end{table}

\clearpage

\begin{table}
\begin{center}
\begin{tabular}{ccccccc}
\hline
\multicolumn{3}{c}{$m_s \approx 150$MeV} & &
\multicolumn{3}{c}{$m_s \approx 400$MeV} \\
$\beta$ & $K_{ud}$ & $K_s$ & &
$\beta$ & $K_{ud}$ & $K_s$ \\
\hline
3.2 & 0.2329 & 0.2043 & & 3.7 & 0.2267 & 0.1692 \\
3.4 & 0.2306 & 0.2026 & & 3.8 & 0.2254 & 0.1684 \\
3.5 & 0.2295 & 0.2017 & & 3.9 & 0.2240 & 0.1677 \\
3.6 & 0.2281 & 0.2006 & & 4.0 & 0.2226 & 0.1669 \\
4.0 & 0.2226 & 0.1964 & & 4.3 & 0.2180 & 0.1643 \\
\hline
\end{tabular}
\caption{\baselineskip=.8cm
Hopping parameters for $N_F=2+1$ simulations performed 
on $8^2\times10\times4$ and $12^3\times4$ lattices. 
$K_{ud}$ for u and d quarks is set to be equal to $K_c$
and $K_s$ for s quark is chosen so that 
$m_s \approx 150$ MeV and 400 MeV in the left and right columns,
respectively.
\protect\label{tab:Ks}}
\end{center}
\end{table}

\clearpage

\clearpage

\clearpage

\clearpage

\clearpage

\clearpage

\clearpage

\clearpage

\clearpage

\clearpage

\clearpage

\clearpage

\clearpage

\clearpage

\clearpage

\clearpage

\clearpage

\clearpage

\clearpage

\clearpage

\clearpage

\clearpage

\clearpage

\end{document}